\documentclass[apj,numberedappendix,appendixfloats]{emulateapj}
\usepackage[utf8]{inputenc}
\usepackage{lipsum}
\usepackage{amsmath, amssymb, bm}
\usepackage{balance}
\usepackage{multirow}
\usepackage{color}
\usepackage{enumitem}

\usepackage[usenames,dvipsnames]{xcolor}
\usepackage{hyperref}
\hypersetup{
    colorlinks = true,
    citecolor = {MidnightBlue},
    linkcolor = {BrickRed},
    urlcolor = {BrickRed}
}

\newcommand{\aH}{{\cal H}}
\newcommand{\Tr}{\mathrm{Tr}}

\newcommand{\be}{\begin{equation}}
\newcommand{\ee}{\end{equation}}
\newcommand{\bea}{\begin{eqnarray}}
\newcommand{\eea}{\end{eqnarray}}
\newcommand{\nv}{\hat{\bf n}}

\newcommand{\egr}{\epsilon_{\rm GR}}
\newcommand{\ewl}{\epsilon_{\rm WL}}
\newcommand{\fnl}{f_{\rm NL}}
\newcommand{\fevo}{f_{\rm evo}}
\newcommand{\uJy}{\mu{\rm Jy}}

\begin{document}

\title{Ultra large-scale cosmology in next-generation experiments with single tracers}

\author{David Alonso}
%\email{david.alonso@astro.ox.ac.uk}
\affil{Astrophysics, University of Oxford, DWB, Keble Road, Oxford OX1 3RH, UK}

\author{Philip Bull}
%\email{p.j.bull@astro.uio.no}
\affil{Institute of Theoretical Astrophysics, University of Oslo, P.O. Box 1029 Blindern,
       N-0315 Oslo, Norway}

\author{Pedro G. Ferreira}
%\email{p.ferreira1@physics.ox.ac.uk}
\affil{Astrophysics, University of Oxford, DWB, Keble Road, Oxford OX1 3RH, UK}

\author{Roy Maartens}
%\email{roy.maartens@gmail.com}
\affil{Department of Physics, University of the Western Cape, Cape Town 7535, South Africa}
\affil{Institute of Cosmology \& Gravitation, University of Portsmouth, Portsmouth PO1 3FX, UK}

\author{M\'{a}rio G. Santos}
%\email{mgrsantos@gmail.com}
\affil{Department of Physics, University of the Western Cape, Cape Town 7535, South Africa}
\affil{SKA SA, The Park, Park Road, Pinelands 7405, South Africa}

%\maketitle

\begin{abstract}
Future surveys of large-scale structure will be able to measure perturbations 
on the scale of the cosmological horizon, and so could potentially probe a 
number of novel relativistic effects that are negligibly small on sub-horizon 
scales. These effects leave distinctive signatures in the power spectra of 
clustering observables and, if measurable, would open a new window on 
relativistic cosmology. We quantify the size and detectability of the effects 
for the most relevant future large-scale structure experiments: spectroscopic and 
photometric galaxy redshift surveys, intensity mapping surveys of neutral 
hydrogen, and radio continuum surveys. Our forecasts show that next-generation
experiments, reaching out to redshifts $z\simeq 4$, will not be able to detect
previously-undetected general-relativistic effects by using individual tracers
of the density field, although the contribution of weak lensing magnification
on large scales should be clearly detectable. We also perform a rigorous joint
forecast for the detection of primordial non-Gaussianity through the excess power
it produces in the clustering of biased tracers on large scales, finding that
uncertainties of $\sigma(\fnl)\sim1-2$ should be achievable. We study the level
of degeneracy of these large-scale effects with several tracer-dependent nuisance
parameters, quantifying the minimal priors on the latter that are needed for an
optimal measurement of the former. Finally, we discuss the systematic effects
that must be mitigated to achieve this level of sensitivity, and some alternative
approaches that should help to improve the constraints. The computational tools
developed to carry out this study, which requires the full-sky computation of the
theoretical angular power spectra for $\mathcal{O}(100)$ redshift bins, as well as
realistic models of the luminosity function, are publicly available at
\url{http://intensitymapping.physics.ox.ac.uk/codes.html}.
\end{abstract}

\section{Introduction}
The current success of relativistic cosmology is primarily based on the use of 
observations of large-scale structure to infer the properties of the Universe. 
The statistics of temperature and mass fluctuations, from maps of the cosmic 
microwave background (CMB) and galaxy surveys respectively, have been used to 
place remarkably tight constraints on the key cosmological parameters 
\citep{Ade:2015xua}, and we have now measured the spectral index of primordial 
fluctuations, the Hubble constant, and the present-day densities of dark matter, 
baryons, and relativistic species to exquisite precision. While many of these 
parameters can have similar effects on cosmological observations, by combining 
different observables at different times and length scales, we have been able 
to break the degeneracies between them.

While a variety of upcoming surveys will certainly improve the existing 
cosmological constraints, they will also allow us to probe an altogether new 
regime of large-scale structure -- perturbations that span the cosmological 
horizon. While such scales are routinely studied in the CMB, these 
measurements consist of a single two dimensional projection 
of the radiation density field and gravitational potentials at a fixed time. 
As such, the amount of information we can obtain from them is fundamentally 
limited by projection effects and cosmic variance. With 
three-dimensional maps of the matter density field, on the other hand, it should be 
possible to greatly refine our measurements of horizon-scale perturbations and, 
in doing so, explore a variety of new relativistic effects in cosmology. 

As pointed out in 
\cite{Bonvin:2005ps, 2009PhRvD..80h3514Y, 2010PhRvD..82h3508Y, 
2011PhRvD..84f3505B,2011PhRvD..84d3516C}, relativistic effects come into play 
through apparent distortions of the projected survey volume by lensing, the 
propagation of light through inhomogeneous potentials, and the large-scale 
effect of peculiar velocities at the source. While they are strongly 
sub-dominant on scales of order $100\,h^{-1}\,$Mpc, where current galaxy surveys 
are focused, they can substantially modify the power spectrum of (e.g.) the 
number density of galaxies from the usual Newtonian predictions on extremely 
large scales.

The familiar Kaiser redshift-space distortion is a 
sub-horizon approximation to a general relativistic redshift-space distortion, 
with the post-Kaiser terms becoming non-negligible only around the horizon 
scale. Terms accounting for lensing, time delays, and the 
Sachs-Wolfe and integrated Sachs-Wolfe effects are also present. 
\cite[For analysis of each of 
these effects in large-scale structure observables, see][]{Bruni:2011ta, Jeong:2011as,
Yoo:2012se, Bertacca:2012tp, 2013PhRvD..87f4026H, Lombriser:2013aj,
2013PhRvD..88b3502Y,2014JCAP...08..022R,Yoo:2013zga}.

Most of these additional effects can safely be ignored in standard analyses of 
galaxy clustering, simply because they are negligible for current surveys of 
limited volume. A partial exception is the lensing term, which contributes to 
the observed overdensity (modulated by the magnification bias), and which has 
occasionally been incorporated into clustering analysis -- it can become 
significant on sub-horizon scales, but only at high redshift. The remaining 
terms\footnote{While excluded from our list of ``GR terms'', we will also 
pay some attention to the lensing term, as it is a `hybrid' 
term that is only observable via clustering in very high volume surveys.} will 
be referred to as the ``GR terms'' for the purposes of this paper, which focuses 
on very large scale effects. We will define the GR terms more precisely in due 
course. The hope, then, 
is that the GR effects can be teased out of cosmological data on ultra-large 
scales, and used to test the standard relativistic model of cosmology.

Another effect that can come into play on large scales is a scale-dependent 
bias due to primordial non-Gaussianity \citep{2008PhRvD..77l3514D,2008ApJ...677L..77M}. 
If the primordial fluctuations are non-Gaussian, there will be a coupling 
between short and long wavelength modes such that the clustering of galaxies 
with respect to the underlying density field is enhanced on large scales. 
Specifically, the galaxy bias gains a scale-dependence proportional to 
$\sim k^{-2}$, where $k$ is the wavenumber of the mode being observed. The 
scale on which this effect comes into play is again of the order the 
cosmological horizon if the parameter quantifying the non-Gaussianity is 
$\fnl\sim 1$. Hence, in addition to detecting relativistic effects in structure 
formation, measurements of horizon-scale modes can also be used to constrain 
the statistics of primordial fluctuations, and thus the mechanism that seeded 
structure in the early Universe.

%There is also a general relativistic non-linear correction to the Poisson 
%constraint equation at early times that generates an effective non-Gaussianity 
%with $f^{\rm GR}_{\rm NL}=-2.2$, even for Gaussian initial conditions; see 
%\cite{2005JCAP...10..010B, 2009ApJ...706L..91V, Bruni:2013qta, Bruni:2014xma, 
%2014arXiv1412.4781C}.

As we can see, the scientific returns from measuring horizon-size fluctuations 
are legion. In this paper, we take a comprehensive view of future observations, 
and attempt to quantify how well both relativistic effects and primordial 
non-Gaussianity can be constrained with upcoming surveys. To do so, we examine 
four different types of survey: spectroscopic and photometric galaxy redshift 
surveys, continuum surveys of radio galaxies, and intensity mapping surveys of 
neutral hydrogen (HI). Each of these techniques will probe different redshifts 
with different sensitivities, but all will (in principle) be able to access 
horizon scale modes in the next decade. The surveys will also measure different 
combinations of the relativistic corrections, and will be sensitive to 
different systematic effects. We will therefore pay close attention to 
identifying the different obstacles for detecting large scale modes, and 
discussing possible methods for mitigating their effects.

The paper is structured as follows. In Section \ref{sec:observables} we 
present the various large-scale effects that we are targeting, focusing on the 
relativistic effects and primordial non-Gaussianity, and discuss how they might 
show up in galaxy surveys and intensity mapping experiments. We also discuss 
the nuisance parameters that determine the amplitudes of these effects, and 
their dependence on astrophysical uncertainties. In Section \ref{sec:fisher}, 
we briefly present the Fisher forecasting formalism we will be working with.
Then, in Section \ref{sec:forecasts}, we systematically present the forecasted
uncertainties in these observables for the different types of survey,
discussing the specific experiments that we consider, their sensitivity to
ultra large-scale observables, possible degeneracies
with nuisance parameters and their main sources of sysematic uncertainties.
The models used to describe the signal and noise for each experiment are
summarized in Appendix \ref{app:specs}. Finally, in Section \ref{sec:discussion}
we discuss our findings and draw conclusions for the prospects for detecting the
ultra large-scale effects.

%-------------------------------------------------------------------------------
\section{Ultra large-scale cosmological observables}
\label{sec:observables}

%-------------------------------------------------------------------------------
\subsection{Relativistic effects in large-scale structure observables}
\label{ssec:th_egr}

The aim of this section is to compute all the terms that give rise to 
fluctuations in the number of light-emitting sources measured by an observer in 
a given redshift interval $dz$ and observed solid angle $d\Omega_o$. The main 
result from this calculation is the following: neglecting all perturbations in 
the trajectory of the photons emitted by these sources, the observed 
perturbation in the number counts is simply given by the perturbation in the 
comoving number density of sources. The comoving 4-volume that we ascribe to 
the patch defined by $dz$ and $d\Omega_o$ depends on the direction and redshift 
of the photons received from these sources, however, and therefore any 
perturbation in their trajectory will induce additional contributions to the 
total fluctuation of source number counts.

Redshift space distortions (RSDs) are a perfect example of one of these 
contributions, where the observed redshifts are perturbed by the peculiar 
velocity of the source. While RSDs have been well understood for decades,
several other terms have only recently been rigorously quantified 
\citep{2011PhRvD..84d3516C, 2011PhRvD..84f3505B}. These terms are relativistic 
in nature, and affect the clustering spectrum only on extremely large scales, 
approaching the horizon size. Since these relativistic effects have been 
thoroughly discussed in the literature, the aim of this section is not to 
provide a derivation from first principles, but rather to give some physical 
intuition for the interpretation of each of these terms, as well as to 
establish the notation that will be used in what follows.

%-------------------------------------------------------------------------------
\subsubsection{Relativistic lightcone effects: number counts}
\label{sssec:rel_terms}

Consider a set of sources with comoving number density $n_s(\eta,{\bf x})$ (as 
measured in their own rest frame), and 4-velocity $u^\mu_s$. These sources emit 
photons with a wave vector $k^\mu\equiv dx^\mu/d\lambda$ ($\lambda$ is an 
affine parameter of the photon geodesic), and rest-frame energy $k_\mu u_s^\mu$. 
During an interval $d\lambda$ of the affine parameter, the photons cover a 
volume $dA_e\,(k_\mu\,u_s^\mu)d\lambda$, where $dA_e$ is the invariant area of 
the wavefront corresponding to the observed solid angle $d\Omega_o$. 
Throughout, we have labelled quantities measured in the emitter's and 
observer's frames with subscripts $e$ and $o$ respectively.

The total number count in a redshift interval $dz$ corresponding to $d\lambda$ 
is therefore
\begin{equation}\label{eq:snc_1}
  \frac{dN}{dz\,d\Omega_o}=n_s\,\frac{dA_e}{d\Omega_o}(k_\mu\,u_s^\mu)
  \frac{d\lambda}{dz}.
\end{equation}
Each of these terms is straightforward to compute in the absence of perturbations:
\begin{align}\nonumber
 n_s(\eta,{\bf x})=\bar{n}_s(\eta(z)),&\hspace{10pt}
   \frac{dA_e}{d\Omega_o}=a^2(\eta(z))\,r^2(z),\\\nonumber
 (k_\mu\,u_s^\mu)\frac{d\lambda}{dz}&=\frac{a(\eta(z))}{H(\eta(z))}.
\end{align}
Here $\eta(z)$ is the background conformal time at redshift $z$, $r(z)$ is the 
background comoving angular diameter distance, $a(\eta(z))=1/(1+z)$ is the 
scale factor, and $H\equiv\dot{a}/a$ is the expansion rate. For the rest of 
this work we will assume a flat background cosmology, so that radial ($\chi$) 
and angular distances are the same.

In the presence of inhomogeneities, all of these quantities are perturbed with 
respect to their background values at redshift $z$, and in general we can write:
\begin{align}
  &\eta(z,\nv)\equiv\eta(z)+\delta\eta,\\\label{eq:ns}
  &n_s(z,\nv)=\bar{n}_s(\eta(z))\left[1+\delta_n+
             \frac{\partial\ln\bar{n}_s}{\partial\eta}\delta\eta\right],\\
  &\frac{dA_e}{d\Omega_o}\equiv a^2(\eta(z))\chi^2(z)\,[1+2\delta_\perp],\\
  &(k_\mu\,u_s^\mu)\frac{d\lambda}{dz}\equiv\frac{a(\eta(z))}{H(\eta(z))}[1+\delta_\parallel],
\end{align}
where $\delta_n$ is the perturbation to $n_s$, and we have defined the
perturbations to the conformal time, $\delta\eta$, transverse distance, 
$\delta_\perp$, and radial distance, $\delta_\parallel$.

One extra detail must be taken into account: not all sources are equally 
bright, and will generally be distributed according to a particular luminosity 
function, $n_s(\eta,\ln L,{\bf x})$, which we define as the density of sources 
in a logarithmic interval of luminosity:
\begin{equation}
 n_s\equiv\frac{d({\rm \#\,sources)}}{dVd\ln L}.
\end{equation}
Only sources with a flux (observed power per unit detector area) above a given
detection cut, $F_{\rm cut}$, will be detected. Flux and luminosity are related 
by an inverse-square law in angular distance, so perturbations to the 
angular diameter distance will affect the measured flux. Linearizing with 
respect to these perturbations gives
\begin{equation}
 F(z,\nv)=\frac{L}{4\pi(1+z)^4\,a^2(\eta(z))\,\chi^2(z)}[1-2\delta_\perp].
\end{equation}
At a given redshift and flux cut, we will only observe sources with luminosities 
above a threshold $L_{\rm cut}$, related to $F_{\rm cut}$ by the previous 
equation. In order to take this into account, we must replace $n_s$ by the 
cumulative luminosity function,
\begin{equation}\label{eq:def_cumlum}
 \mathcal{N}(\eta,{\bf x},>\ln L)\equiv\int_{\ln L}^\infty d\ln L'\,
 n_s(\eta,{\bf x},\ln L'),
\end{equation}
so that Eq.~(\ref{eq:ns}) becomes
\begin{equation}\nonumber
 \mathcal{N}(z,{\bf n},F_{\rm cut})=\bar{\mathcal{N}}\left[1+\delta_{\mathcal{N}}+
 \frac{\partial\ln\bar{\mathcal{N}}}{\partial\eta}\,\delta\eta-
 2\frac{\bar{n}_s}{\bar{\mathcal{N}}}\,\delta_\perp\right].
\end{equation}
We have shortened our notation such that
\begin{equation}
\bar{\mathcal{N}}\equiv\bar{\mathcal{N}}(\eta(z),>\ln \bar{L}_{\rm cut}(z,F_{\rm cut}))
\end{equation}
(and likewise for $\bar{n}_s$), and have overlined ($\,\bar{\,}\,$) all 
quantities evaluated in the background.

The full linear expression for the source number counts can finally be written as
\begin{equation}
  \frac{dN}{dzd\Omega_o}=\bar{\mathcal{N}}\,\frac{a^3(z)}{H(z)}\,\chi^2(z)\,
  [1+\Delta_N(z,\nv)],
\end{equation}
where the perturbation is given by
\begin{equation}\label{eq:total_snc}
 \Delta_N(z,\nv)=\delta_{\mathcal{N}}+
 \frac{\partial\ln\bar{\mathcal{N}}}{\partial\eta}\,\delta\eta+\delta_\parallel+
 2\delta_\perp\left[1-\frac{\bar{n}_s}{\bar{\mathcal{N}}}\right].
\end{equation}
In order to simplify the notation, from now on we will refer to the observed 
background number of sources found per unit redshift and solid angle simply as 
$\bar{N}(z)$, i.e.
\begin{equation}
 \bar{N}(z)\equiv\bar{\mathcal{N}}\,\frac{a^3(z)}{H(z)}\,\chi^2(z).
\end{equation}

The terms $\delta\eta$, $\delta_\parallel$, and $\delta_\perp$ can be related 
to the metric, density, and velocity perturbations by solving the geodesic 
equation for photons in any gauge. In the conformal Newtonian gauge, defined
by the line element
\begin{equation}
  ds^2=-a^2(\eta)\left[(1+2\psi)\,d\eta^2-(1-2\phi)\,\delta_{ij}dx^idx^j\right],
\end{equation}
these perturbations read
\begin{align}
 \aH\delta\eta=&-\psi+\int(\phi'+\psi')\,d\eta+v_r\\
 \delta_\parallel=&\left[1-\frac{\aH'}{\aH^2}\right]\aH\delta\eta+\psi+v_r\\
 &+\frac{1}{\aH}\left[-\frac{d\psi}{d\eta}+\phi'+\psi'+\frac{dv_r}{d\eta}\right]\\
 \delta_\perp=&\aH\delta\eta-\frac{1}{\chi}\left[\delta\eta+\int(\phi+\psi)\,d\eta\right]
 -\phi-\kappa\\
 \kappa\equiv&\frac{1}{2}\int_0^\chi\frac{\chi-\chi'}{\chi\chi'}
 \nabla_\Omega^2(\phi+\psi)\,d\chi',
\end{align}
where $v_r\equiv\nv\cdot{\bf v}_s$ is the radial peculiar velocity of the 
sources, $\nabla_\Omega^2$ is the Laplacian on the unit sphere, and $\kappa$ is 
the lensing convergence. Note that we have denoted all partial derivatives with 
respect to conformal time as $\partial_\eta b \equiv b'$ (and $\aH\equiv a'/a$), 
and that the operator $d/d\eta$ denotes a total lightcone derivative along the 
unperturbed photon trajectory,
\begin{equation}
 \frac{db}{d\eta}\equiv\frac{d}{d\eta}[b(\eta,{\bf x}=(\eta_0-\eta)\nv)],
\end{equation}
where $\eta_0$ is the age of the Universe. Likewise, all integrals shown in the
equations above must be understood as lightcone integrals along the same 
trajectory.

%-------------------------------------------------------------------------------
\subsubsection{Relativistic lightcone effects: intensity mapping}
\label{sssec:imap_th}

Besides source number counts, another promising observational tool for studying 
large-scale structure is a technique known as \emph{intensity mapping}. The 
technical details of this method are discussed in Section \ref{sec:imapping}, 
but we will describe the relevant relativistic effects here 
\citep[see also][]{2013PhRvD..87f4026H}.

In intensity mapping, the observable used to trace the matter density is the 
intensity received from a line-emitting medium integrated over a patch of the 
sky (i.e. the total power measured in a frequency interval per unit detector 
area and observed solid angle). We assume that this line emission is caused by 
some well-defined transition line, and can therefore be used to recover the 
redshift of the source by comparing the observed frequency with the known 
rest frame one. In the rest frame of a set of line-emitting sources, the 
emissivity is
\begin{equation}
 \frac{dE_e}{dt_ed\nu_ed\Omega_edV_e}=
 \frac{\hbar}{2}A_{21}\nu_e\varphi(\nu_e)\frac{x_2\,\rho_a}{m_a},
\end{equation}
where $A_{21}$ is the Einstein coefficient for the transition, $\rho_{a}$ is 
the comoving density of the emitting gas, $m_a$ is its atomic mass, $x_2$ is 
the (number) fraction of the gas in the excited state, and $\varphi(\nu)$ is 
the line profile (normalized to unity when integrated over all frequencies).

As shown in Section \ref{sssec:rel_terms}, the volume covered by the emitted 
photons in an affine parameter interval $d\lambda$ is
\begin{equation}
 dV_e=dA_e\,(k_\mu\,u^\mu_s)d\lambda.
\end{equation}
Assuming that no absorption or scattering of the emitted photons occurs, the 
emissivity can be related to the measured intensity by
\begin{align}\nonumber
  I(\nu_o,\nv)&\equiv\frac{dE_o}{dt_odA_od\Omega_od\nu_o}\\\nonumber
  &=\frac{\hbar\nu_oA_{21}x_2\rho_a}{2m_a}\varphi(\nu_e)(k_\mu u_s^\mu)d\lambda
  \frac{dA_ed\Omega_e}{dA_o\Omega_o}\frac{d\nu_edt_e}{d\nu_odt_o}.
\end{align}
The frequencies and time intervals in both frames are directly related through 
the redshift $z$, as are angles and invariant areas (from Etherington's 
reciprocity relation),
\begin{equation}
  \frac{dt_ed\nu_e}{dt_od\nu_o}=1,\hspace{12pt}
  \frac{dA_ed\Omega_e}{dA_od\Omega_o}=\frac{1}{(1+z)^2}.
\end{equation}
Using these relations, and assuming that observations will take place on 
frequency intervals $\Delta\nu_o$ much larger than the line width, we finally 
obtain the relation
\begin{equation}
 I(\nu_o,\nv)=\frac{\hbar A_{21}\nu_{21}x_2}{2m_a(1+z)^2}\,\rho_a\,(k_\mu\,u_s^\mu)\,
 \frac{d\lambda}{dz},
\end{equation}
where $\nu_{21}$ is the rest-frame line frequency. We can see that this is 
equivalent to Eq.~(\ref{eq:snc_1}) for number counts, except for the factor of 
the angular diameter distance, $dA_e/d\Omega_o$. This is because the observable 
in intensity mapping is not the total number of objects in a given patch of 
sky, but the combined emitted light from the same patch. Since luminosities and 
angular distances are affected in the same way by lightcone effects, these 
cancel exactly for intensity mapping.

Expanding both $\rho_a$ and $(k_\mu\,u_s^\mu)\,\frac{d\lambda}{dz}$ to linear 
order, we can therefore compute all of the linear perturbations to the observed 
intensity mapping signal,
\begin{align}\nonumber
 I(\nu,\nv)&=\bar{I}(\nu)\,[1+\Delta_I]\\\label{eq:total_im}
 &\equiv\bar{I}(\nu)\left[1+\delta_a+\frac{\partial\ln\bar{\rho}_a}{d\eta}\delta\eta+
 \delta_\parallel\right],
\end{align}
where $\delta_a$ is the intrinsic perturbation of the emitting gas density. By
comparing this with Eq.~(\ref{eq:total_snc}), we can see that the linear 
perturbation for intensity mapping is equivalent to the perturbation of the 
number counts for a population of sources with a particular form of the 
luminosity function, such that the number of sources observed above a given 
flux is proportional to the luminosity associated with that flux,
\begin{equation}
 \bar{\mathcal{N}}_{\rm IM}(>L)\propto L.
\end{equation}
The somewhat unfortunate consequence of this result is that there are no linear 
perturbations to angular distances for intensity mapping,\footnote{Note that 
this is equivalent to the result that lensing of the CMB is a second-order 
effect.} which could potentially reduce the amount of cosmological information 
that can be extracted from this probe.

The background term in Eq.~(\ref{eq:total_im}) is commonly expressed in terms of 
antenna temperature, defined through the Rayleigh-Jeans relation for a 
black-body emitter $T(\nu)=I(\nu)\,c^2/(2k_B\nu^2)$, where $k_B$ is the 
Boltzmann constant. In terms of background quantities, the homogeneous 
intensity mapping signal is
\begin{equation}\label{eq:im_bg}
  \bar{T}(z)=\frac{3\hbar A_{21}x_2c^2}{32\pi G k_B m_a \nu_{21}^2}
  \frac{H_0^2\,\Omega_{b,0}\,x_a(z)(1+z)^2}{H(z)},
\end{equation}
where $x_a(z)$ is the fraction of baryons made up by the line-emitting species 
under study.

%-------------------------------------------------------------------------------
\subsubsection{Clustering bias}

Until now, we have not related the intrinsic perturbation in the number density 
of sources to the perturbations of the energy-momentum tensor. Assuming that 
galaxies form in dark matter haloes, which themselves form preferentially in 
high density regions, one would expect the halo (or galaxy) number overdensity 
to trace the fluctuations in the overall matter density on large (linear) 
scales with a simple linear bias factor, $\delta_{\rm halo}\simeq
b_{\rm halo}\,\delta_M$. This bias is a central piece of the halo model of 
structure formation \citep{1996MNRAS.282..347M,2000MNRAS.318.1144P}, the 
validity of which has been extensively tested against numerical simulations 
\citep{2008MNRAS.383..546C}. Although the linear bias is expected to be 
scale-dependent on non-linear and mildly non-linear scales, where non-linear 
and stochastic bias terms could also be important, we are mainly interested in 
large-scale observables in this paper, where the approximation of a linear,
scale-independent bias should be valid (although see Section \ref{ssec:th_fnl}). 
This bias will depend on redshift and luminosity, however (e.g. more luminous, 
and therefore rarer, objects are expected to be more highly biased).

Since it is not possible to unambiguously define the matter overdensity 
$\delta_M$ in a gauge-invariant way in a general-relativistic context, a subtle 
point is the choice of overdensity field on which the bias relation is applied. 
In this work we take the point of view of \cite{2011PhRvD..84d3516C, 
2011JCAP...10..031B, Jeong:2011as, Bruni:2011ta}, and argue that, since the 
process of galaxy formation is due to local physics, and since we expect our 
sources to follow the same velocity field as the dark matter, the bias relation 
should be applied in the synchronous comoving gauge. Note that it is also the 
comoving gauge perturbation that appears in the Poisson equation. A more 
complete discussion of this argument, which can also be extended to the case 
of primordial non-Gaussianity, can be found in \cite{2011JCAP...10..031B}.

It follows that the intrinsic perturbation to the number density of sources, 
$\delta_N$, in the Newtonian gauge -- our choice for this work -- is related to 
the matter overdensity in the synchronous comoving gauge, $\delta_{M,{\rm syn}}$, 
through
\begin{equation}
  \delta_{\rm N}=b(L,z,k)\,\delta_{M,{\rm syn}}+
  \frac{\partial\ln\bar{\mathcal{N}}}{d\eta}\frac{v}{k},
\end{equation}
where $v$ is the peculiar velocity in Newtonian gauge, and we have allowed the 
bias, $b$, to be scale dependent, in anticipation of the discussion in Section 
\ref{ssec:th_fnl}.

%-------------------------------------------------------------------------------
\subsubsection{Magnification and evolution bias}

The amplitudes of the perturbations to the conformal time and transverse 
distances depend explicitly on the derivatives of the luminosity function of 
the source population with respect to luminosity and time (see Eq.~\ref{eq:total_snc}).
It has become common to express these derivatives in terms 
of the so-called \emph{magnification bias}, $s(\eta)$, and 
\emph{evolution bias}, $\fevo(\eta)$, defined as
\begin{align}\label{eq:s_def}
 s(\eta)&\equiv\frac{5}{2}\frac{\bar{n}_s(\eta,\ln \bar{L}_{\rm cut})}
 {\bar{\mathcal{N}}(\eta,>\ln\bar{L}_{\rm cut})},\\\label{eq:fevo_def}
 \fevo(\eta)&\equiv\frac{\partial\ln[a^3\bar{\mathcal{N}}(\eta,>\ln\bar{L}_{\rm cut})]}
 {\partial\ln a}.
\end{align}
As with the clustering bias, the values of $s$ and $\fevo$ depend on the source
population under study, so must be modelled correctly in order to maximise the
information that can be extracted from any clustering analysis. While $b(z)$ 
must be determined directly from clustering statistics, it is possible to 
estimate $s$ and $\fevo$ directly from the overall number counts of sources as 
a function of redshift and magnitude. Let $\bar{N}(z,<m_*)$ be the cumulative 
number of sources with magnitude $m$ brighter than $m_*$, per unit solid angle 
and redshift interval. $\bar{N}$ is related to the luminosity function, 
$\bar{n}_s$, through
\begin{equation}\label{eq:nz_def}
  \bar{N}(z,<m_*)=\frac{c\,\chi^2(z)}{(1+z)^3H(z)}
  \int_{\ln L_*}^\infty\bar{n}_s(\eta,\ln L)\,d\ln L,
\end{equation}
where the threshold luminosity $L_*$ is $L_*=4\pi(1+z)^2\chi^2(z)F_*$, and 
fluxes and magnitudes are related through
\begin{equation}
 m=-\frac{5}{2}\log_{10}\left[\frac{F}{F_0}\right].
\end{equation}
Note that we have neglected evolution and $k$-corrections.

Using the definitions of $\fevo$ and $s$ in Eqs.~(\ref{eq:s_def}) and 
(\ref{eq:fevo_def}), these quantities can be related to the derivatives of 
$\bar{N}$ with respect to $z$ and $m_*$ by
\begin{align}
 &\frac{\partial\log_{10}\bar{N}}{\partial m_*}=s,\\
 &\frac{\partial\log_{10}\bar{N}}{\partial\log_{10}(1+z)}=\frac{(2-5s)}{\chi\,aH}-5s
 +\frac{H'}{aH^2}-\fevo.
\end{align}
Note that, in order to use these relations to estimate $s$ and $\fevo$, it is 
necessary to have full redshift information about the source distribution. 
While this is available by default for spectroscopic surveys, determining the 
redshift distribution becomes more involved for photometric and radio continuum 
surveys. This is nevertheless a necessary task if these probes are to be usable 
for cosmological studies, where (e.g.) the redshift and photometric redshift 
distributions must be correctly modelled. In any case, the uncertainties on 
$s$ and $\fevo$ will tend to grow towards large $z$, and must therefore be 
taken into account in any cosmological analysis.

As we described in Section \ref{sssec:imap_th}, the case of intensity mapping 
is slightly different. In this case, perturbations to the angular distance 
cancel, which is equivalent to setting the magnification bias to the critical 
value $s_{\rm IM}=2/5$. $\fevo$ can be determined directly from the redshift 
dependence of the background brightness temperature, $\bar{T}(z)$ 
(Eq.~\ref{eq:im_bg}).

%-------------------------------------------------------------------------------
\subsubsection{Power spectra}

\begin{figure}
 \centering
 \includegraphics[width=0.49\textwidth]{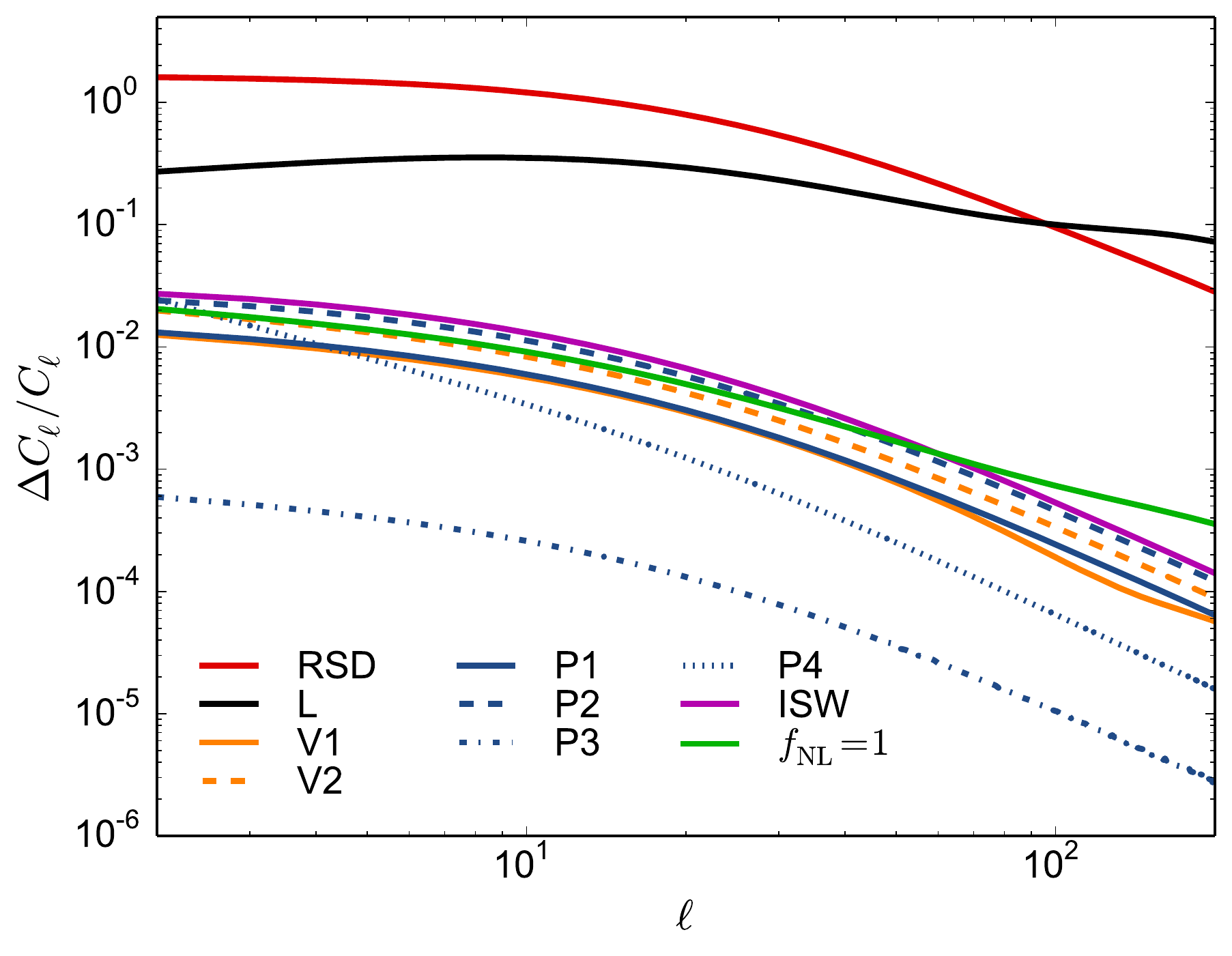}
 \caption{Amplitude of the different terms listed in Eqs.~(\ref{eq:terms0}-\ref{eq:isw})
          relative to the amplitude of $\Delta^{\rm D}_\ell$ in the power spectrum. The
          $C_\ell$s were calculated for a redshift bin at $z=2$ with width
          $\Delta z=0.12$ and for constant bias functions ($b(z)=1.5$, $s(z)=1$,
          $\fevo(z)=1$). We have also included in green the contribution of primordial
          non-Gaussianity for $\fnl=1$.}
 \label{fig:cl_terms}
\end{figure}
The most informative observable regarding the clustering of astrophysical 
sources is their two-point correlation, $\langle\Delta_{\mathcal{N}}(z_1,\nv_1)
\Delta_{\mathcal{N}}(z_2,\nv_2)\rangle$. The perturbation $\Delta_{\mathcal{N}}$ 
can be expressed in terms of spherical harmonic coefficients,
\begin{equation}
a_{\ell m}(z)\equiv\int d\nv\,\Delta_{\mathcal{N}}(z,\nv)\,Y_{\ell m}(\nv),
\end{equation}
where $Y_{\ell m}(\nv)$ are the spherical harmonics. The clustering of number 
counts can then be studied through the angular power spectrum, defined by the 
correlation
\begin{equation}
  \langle a_{\ell m}(z_1)a_{\ell' m'}^*(z_2)\rangle\equiv \delta_{\ell \ell'}\delta_{mm'}
  \,C_\ell(z_1,z_2),
\end{equation}
where angle brackets denote an ensemble average.

In practice, $\Delta(z,\nv)$ is not measured in infinitesimal intervals of $z$, 
but by averaging over a set of finite radial bins, which we will label here by 
a Latin index, $i$. The observed anisotropy in bin $i$ is
\begin{equation}\label{eq:project}
  a^i_{\ell m}\equiv\int dz\,W_i(z)\,\Delta(z,\nv),
\end{equation}
where the window function $W_i$ is normalized to $1$ when integrated over 
redshift. The shape of $W_i$ is determined both by the background redshift 
distribution of observed sources, $\bar{N}(z)$, and the probability that a 
source at redshift $z$ will be included in the $i$-th bin $p_i(z)$, so that
\begin{equation}
 W_i(z)\propto \bar{N}(z)\,p_i(z).
\end{equation}
Using this, one can show that the cross-spectrum between two bins can be written 
as \citep{2013JCAP...11..044D}
\begin{equation}\label{eq:cls_def}
  C^{ij}_\ell=4\pi\int_0^\infty\frac{dk}{k}\mathcal{P}(k)\Delta_\ell^i(k)\Delta_\ell^j(k),
\end{equation}
where $\mathcal{P}(k)$ is the dimensionless primordial power spectrum, which 
is assumed to take the form $\mathcal{P}(k)=A_s\,(k/k_0)^{n_s-1}$, and 
$\Delta_\ell^i(k)$ contains the transfer functions of the terms that 
contribute to the anisotropy in bin $i$, in Fourier space and projected on the 
sky. Expanding the various contributions to Eq.~(\ref{eq:total_snc}), the 
functions $\Delta^i_\ell(k)$ can be written as a sum of 10 terms corresponding 
to different physical effects \citep{2013JCAP...11..044D}:
\begin{widetext}
\begin{align}\label{eq:terms0}
  &\Delta^{{\rm D},i}_\ell(k)\equiv\int d\eta \,b\,\tilde{W}_i\,\delta_{M,{\rm syn}}(k,\eta)\,
  j_\ell(k\chi(\eta)),\hspace{12pt}
  ~~~~~~~~\Delta^{{\rm RSD},i}_\ell(k)\equiv\int d\eta\,(aH)^{-1}\tilde{W}_i(\eta)\,\theta(k,\eta)\,
  j_\ell''(k\chi(\eta)),\\
  &\Delta^{{\rm L},i}_\ell(k)\equiv \ell(\ell+1) \int d\eta\,\tilde{W}^{\rm L}_i(\eta)\,
  (\phi+\psi)(k,\eta)\,j_\ell(k\chi(\eta)),\hspace{12pt}
  \Delta^{{\rm V1},i}_\ell(k)\equiv \int d\eta\,(\fevo-3)\,aH\,\tilde{W}_i(\eta)\,
  \frac{\theta(k,\eta)}{k^2}\,j_\ell(k\chi(\eta)),\\
  &\Delta^{{\rm V2},i}_\ell(k)\equiv \int d\eta\,
  \left(1+\frac{H'}{aH^2}+\frac{2-5s}{\chi\,aH}+5s-\fevo\right)
  \tilde{W}_i(\eta)\,\frac{\theta(k,\eta)}{k}\,j_\ell'(k\chi(\eta)),\\
  &\Delta^{{\rm P1},i}_\ell(k)\equiv \int d\eta\,
  \left(2+\frac{H'}{aH^2}+\frac{2-5s}{\chi\,aH}+5s-\fevo\right)
  \tilde{W}_i(\eta)\,\psi(k,\eta)\,j_\ell(k\chi(\eta)),\\
  &\Delta^{{\rm P2},i}_\ell(k)\equiv \int d\eta\,
  (5s-2)\tilde{W}_i(\eta)\,\phi(k,\eta)\,j_\ell(k\chi(\eta)),\hspace{12pt}
  \Delta^{{\rm P3},i}_\ell(k)\equiv \int d\eta\,
  (aH)^{-1}\tilde{W}_i(\eta)\,\phi'(k,\eta)\,j_\ell(k\chi(\eta)),\\\label{eq:isw}
  &\Delta^{{\rm P4},i}_\ell(k)\equiv \int d\eta\,
  \tilde{W}^{\rm P4}_i(\eta)\,(\phi+\psi)(k,\eta)\,j_\ell(k\chi(\eta)),\hspace{12pt}
  \Delta^{{\rm ISW},i}_\ell(k)\equiv \int d\eta\,
  \tilde{W}^{\rm ISW}_i(\eta)\,(\phi+\psi)'(k,\eta)\,j_\ell(k\chi(\eta)),
\end{align}
where we have defined the window functions
\begin{align}\label{eq:windows}
  &\tilde{W}_i(\eta(z))\equiv W_i(z)\left(\frac{d\eta}{dz}\right)^{-1},\hspace{12pt}
  \tilde{W}^{\rm L}_i(\eta)\equiv\int_0^\eta d\eta'\tilde{W}_i(\eta')\frac{2-5s(\eta')}{2}
  \frac{\chi(\eta)-\chi(\eta')}{\chi(\eta)\chi(\eta')},\\
  &\tilde{W}^{\rm P4}_i(\eta)\equiv\int_0^\eta d\eta'\tilde{W}_i(\eta')
  \frac{2-5s}{\chi},\hspace{12pt}
  \tilde{W}^{\rm ISW}_i(\eta)\equiv\int_0^\eta d\eta'\tilde{W}_i(\eta')
  \left(1+\frac{H'}{aH^2}+\frac{2-5s}{\chi\,aH}+5s-\fevo\right)_{\eta'}.
\end{align}
\end{widetext}
In these equations, the quantities $\delta_{M,{\rm syn}}(k,\eta)$, 
$\theta(k,\eta)$, $\psi(k,\eta)$, and $\phi(k,\eta)$ are the transfer functions 
for the synchronous comoving gauge matter density perturbation, the divergence 
of the peculiar velocity, and the two metric potentials respectively.

Each term is sourced by a different physical effect. $\Delta_\ell^{\rm D}$ 
corresponds to the intrinsic perturbation in the comoving number density of 
sources, which is the dominant contribution in most cases, and is the only term 
that has traditionally been taken into account when RSDs and lensing can be 
neglected. $\Delta_\ell^{\rm RSD}$ is the usual RSD term corresponding to
the Kaiser effect, due to the deformation of the Lagrangian volume in redshift 
space. $\Delta_\ell^{\rm L}$ is the lensing convergence term, caused by the 
deformation of the Lagrangian volume in the transverse directions due to weak 
lensing. The terms $\Delta_\ell^{\rm V1}$ and $\Delta_\ell^{\rm V2}$ are extra 
RSD contributions that come from evaluating the background terms at a redshift 
perturbed by the Doppler effect. The remaining terms correspond to to the same 
effect, but for redshift perturbations caused by gravitational redshifting 
instead of peculiar velocities. In particular, $\Delta_\ell^{\rm ISW}$
is the analogue of the integrated Sachs-Wolfe (ISW) effect 
\citep{1967ApJ...147...73S} for number counts.

Of these terms, the first three give the largest contribution to the total 
clustering anisotropy, so only these have traditionally been included in 
clustering analyses. The remaining terms are mainly relevant on super-horizon 
scales at the position of the sources, and even on those scales their amplitude 
is significantly smaller than the first three (see Fig. \ref{fig:cl_terms}).
Nevertheless, these terms contain useful information that could potentially be 
used, for example, to constrain different theories of gravity 
\citep{Lombriser:2013aj, BakerBull}. One of the aims of this paper is to 
forecast the detectability of these terms by future experiments. In order to do 
so, we have defined an effective parameter, $\egr$, which multiplies the terms 
$\Delta_\ell^{\rm V1,2},\,\Delta_\ell^{\rm P1-4}\,{\rm and}\,\Delta_\ell^{\rm ISW}$ 
and has a fiducial value of $1$. $\egr$ therefore parametrizes the amplitude of 
the relativistic corrections to the clustering of sources.

Even though the origin of the lensing convergence term ($\Delta_\ell^{\rm L}$)
is clearly General-Relativistic, we have not included it under the umbrella
of $\egr$ for two main reasons: first, we would like to focus on yet-undetected
effects, and lensing magnification has already been detected by cross-correlating
pairs of distant tracers \citep{2005ApJ...633..589S, 2009A&A...507..683H}. Secondly,
in this work we aim to identify possibly-detectable observables on
ultra-large scales, but the lensing term has a non-negligible effect on small angular
scales. Nevertheless, for completeness we have also forecasted for the detectability
of lensing magnification by defining an effective amplitude, $\ewl$, multiplying
$\Delta_\ell^{\rm L}$. In keeping with the main aim of this paper, we will
only produce forecasts for this parameter based on its effects on the largest
angular scales (lowest multipoles), however.

Thus, to clarify the terminology used here, we will refer to the terms 
$\Delta^{\rm V1,2}$, $\Delta^{\rm P1-4}$ and $\Delta^{\rm ISW}$ as ``GR effects''
or ``GR terms'' and to $\Delta^{\rm L}$ as the ``lensing term'', even if the
nature of the latter is clearly relativistic.

%-------------------------------------------------------------------------------
\subsection{Primordial non-Gaussianity}
\label{ssec:th_fnl}

A fundamental assumption of current theories of large-scale structure is that 
the primordial seed fluctuations can be described as a multivariate Gaussian 
random process, uniquely characterised by the primordial power spectrum 
\citep{Baumann:2009ds}. For many years, the Gaussianity of primordial 
fluctuations was one of the main predictions of inflationary theories of the 
early Universe. More recently, the possibility of non-Gaussian primordial 
fluctuations has been revisited for two main reasons. On the one hand, a 
battery of statistical techniques have been developed to quantify primordial 
non-Gaussianity, primarily from the CMB, but also adapted to large-scale 
structure data \citep{Yadav:2007yy,Fergusson:2008ra,Ade:2013ydc}. There is a
hope that these techniques will uncover something that will enrich our 
understanding of the early Universe, above simple one- and two-point 
statistics. On the other hand, the Effective Field Theory approach to 
inflation \citep{Cheung:2007st,Weinberg:2008hq} can be used to systematically 
quantify all possible deviations from the quadratic action of linear 
perturbation theory around quasi-de Sitter space. These deviations, in the form 
of higher-order terms, will lead to non-trivial Gaussian signatures that are 
directly related to the fundamental parameters (and more importantly, the 
fundamental structure) of the theory of the early Universe.

A useful (though not universal) way of describing deviations from primordial
Gaussianity \citep{Komatsu:2001rj} is to assume that the gravitational 
potential $\Phi$ can be decomposed into a quadratic polynomial in a Gaussian 
random field $\phi$, of the form
\begin{eqnarray}
\Phi=\phi+\fnl(\phi^2-\langle\phi^2\rangle), \nonumber
\end{eqnarray}
where $\fnl$ is, in the simplest scenarios, a constant. $\fnl$ has 
been calculated for a family of inflationary models. In the local 
(`squeezed') configuration, it is expected to be of order 
$|n_s-1|\sim{\cal O}(10^{-2})$, where $n_s$ is the spectral scalar index of 
primordial fluctuations \citep[although some non-canonical models can lead to 
$\fnl\sim 1$;][]{Verde:1999ij,Liguori:2005rj,Smith:2006ud}. Current 
constraints from higher order statistics of CMB maps place the constraint at
$|\fnl|\lesssim 7$ \citep{2015arXiv150201592P}.

A novel proposal is to look for the effects of non-Gaussian initial 
fluctuations at the level of the power spectrum. It has been shown that 
primordial non-Gaussianity induces a scale- and redshift-dependence for any 
biased tracer, $X$, of the overall density field 
\citep{2008ApJ...677L..77M, 2008PhRvD..77l3514D}. If the density contrast of 
$X$ has a linear, Gaussian bias, $b_X^G$, then the non-Gaussianity of the 
distribution will induce a correction of the form
\begin{eqnarray}\label{eq:db_fnl}
\Delta b_X(z,k)=3\fnl\frac{[b_X^G(z)-1]\Omega_MH_0^2\delta_c}{(T(k)D(z)k^2)}.\label{NGPK}
\end{eqnarray} 
Here, $\Omega_M=\Omega_b+\Omega_{C}$ is the fractional energy density in matter 
(i.e. baryons plus dark matter), $H_0$ is the Hubble constant, 
$\delta_c\simeq1.686$ is the critical density contrast of matter from the 
spherical collapse model, $T(k)$ is the matter transfer function, and $D(z)$ 
is the linear growth function for density perturbations.

As can be seen from Eq.~(\ref{NGPK}), there will be a substantial enhancement of 
the tracer power spectrum on large scales, with a $\sim1/k^2$ scale dependence 
(since $T \sim 1$ on large scales). A rough estimate is that the transition
scale is of order $k_{NG}\sim \fnl H_0$, i.e. we expect the 
scale-dependent signature to kick in close to the horizon scale for 
$\fnl\sim 1$. There have already been some attempts to constrain 
$\fnl$ from the scale-dependent galaxy bias, although they have been 
severely hampered by systematic effects on extremely large scales 
\citep{Giannantonio:2013uqa}. In parallel, there have also been attempts to 
forecast the possibility of measuring $\fnl\sim 1$ with future surveys
\citep[e.g.][]{2011PhRvD..83l3514N, 2012MNRAS.422.2854G, 2013PhRvL.111q1302C, 
2014MNRAS.442.2511F, 2015MNRAS.448.1035C, 2015JCAP...01..042R}.

Given the nature of this signature -- the fact that it arises on large scales 
and has a $1/k^2$ scale dependence -- it has been argued that non-Gaussianity 
may be degenerate with the relativistic effects we are studying in this paper 
\citep{Bruni:2011ta,Jeong:2011as,Bertacca:2012tp}. We will therefore include 
the effect of $\fnl$ throughout our analysis and, in the process, present the 
most up-to-date and conservative forecasts for its detectability with future 
surveys. As illustrated in Fig. \ref{fig:cl_terms}, the extra power induced by 
a value of $\fnl\sim1$ on large scales is typically similar to the amplitude 
of the relativistic corrections presented in the previous section.

Including our three main observables ($\fnl$, $\egr$, and $\ewl$) the total
perturbation to the number counts is:
\begin{widetext}
\begin{align}
 \Delta^i_\ell=\Delta^{{\rm D},i}_\ell(\fnl)+\Delta^{{\rm RSD},i}_\ell+
 \ewl\,\Delta^{{\rm L},i}_\ell+\egr\left[\Delta^{{\rm V1},i}_\ell+
 \Delta^{{\rm V2},i}_\ell+\Delta^{{\rm P1},i}_\ell+\Delta^{{\rm P2},i}_\ell
 +\Delta^{{\rm P3},i}_\ell+\Delta^{{\rm P4},i}_\ell+\Delta^{{\rm ISW},i}_\ell\right],
\end{align}
\end{widetext}
where there is an implied scale- and time-dependence in all of these terms.

%--------------------------------------------------------------------------------------------
\section{Forecasting formalism}
\label{sec:fisher}

The spherical harmonic coefficients of the fluctuation in the observed number 
counts in the $i$-th redshift bin, $a_{\ell m}^i$, contain most of the 
information about the clustering of sources. Assuming that they are 
Gaussian-distributed with a variance given by the cross-power spectra 
$C_\ell^{ij}$ (Eq.~\ref{eq:cls_def}), it is straightforward to show that the 
log-likelihood for a given realization of the harmonic coefficients is
\begin{equation}\label{eq:loglhood}
  \ln \mathcal{L}=-\frac{1}{2}\left[\sum_{\ell,m}{\bf a}^T_{\ell m}
  \mathsf{C}^{-1}_\ell{\bf a}_{\ell m}-
  \ln\left(\det[\mathsf{C}_\ell]\right)\right]+\text{const.},
\end{equation}
where we have written $a^i_{\ell m}$ for each $\ell$ and $m$ as a vector 
$[{\bf a}_{\ell m}]_i \equiv a^i_{\ell m}$, and the set of cross-spectra 
$C^{ij}_\ell$ as a matrix $[\mathsf{C}_\ell]_{ij} \equiv C^{ij}_\ell$. Our aim 
here is to forecast the precision with which different experiments will be able 
to measure a certain set of parameters, $\{\theta_\alpha\}$,
which boils down to predicting the parameter covariance matrix. An efficient 
way of doing this is to use the Fisher matrix formalism, wherein the likelihood 
is approximated by a Gaussian expansion of Eq.~(\ref{eq:loglhood}) around a 
fiducial set of parameters,
\begin{equation}
  \ln \mathcal{L} = -\frac{1}{2}
  \sum_{\alpha,\beta}(\theta_\alpha-\bar{\theta}_\alpha)F_{\alpha\beta}
  (\theta_\beta-\bar{\theta}_\beta)+\mathcal{O}(\theta^3),
\end{equation}
where we have defined the Fisher matrix
$F_{\alpha\beta}\equiv\langle\partial^2\ln \mathcal{L}/
\partial\theta_\alpha\partial\theta_\beta\rangle$.
The covariance matrix $C_{\alpha\beta}\equiv\langle(\theta_\alpha-\bar{\theta}_\alpha)
(\theta_\beta-\bar{\theta}_\beta)\rangle$ can then be approximated as the 
inverse of $F_{\alpha\beta}$. For our likelihood, one can show that
\begin{equation}\label{eq:fisher}
  F_{\alpha\beta}=
  f_{\rm sky}\sum_{\ell=2}^{\ell_{\rm max}}\frac{(2\ell+1)}{2}\,
  \Tr\left[(\partial_\alpha\mathsf{C}_\ell)\,
  \mathsf{C}^{-1}_\ell\,(\partial_\beta\mathsf{C}_\ell)\,\mathsf{C}^{-1}_\ell\right],
\end{equation}
where $\partial_\alpha\equiv\partial/\partial\theta_\alpha$.

We model the observable ${\bf a}_{\ell m}$ as the sum of two 
contributions: ${\bf a}_{\ell m}={\bf a}^S_{\ell m}+{\bf a}^N_{\ell m}$, 
corresponding to signal (i.e. cosmological anisotropies) and noise 
(non-cosmological fluctuations due to instrumental or shot noise). We will also 
assume that both contributions are uncorrelated, so that 
$\mathsf{C}_\ell=\mathsf{C}^S_\ell+\mathsf{N}_\ell$. Here, $\mathsf{C}^S_\ell$ 
is given by Eq.~(\ref{eq:cls_def}), and $\mathsf{N}_\ell$ is the noise power 
spectrum, the exact form of which will depend on the particular type of 
experiment.

The theoretical power spectra $\mathsf{C}^S_\ell$ were computed using a 
modified version of the public {\tt CLASS} code \citep{2011arXiv1104.2932L,
2013JCAP...11..044D}. Our modifications are documented in detail in Appendix 
\ref{app:CLASS}. For our fiducial cosmology, we adopted a model consistent with 
the best-fit flat $\Lambda$CDM parameters from Planck \citep{2014A&A...571A..16P}, 
given by $(\Omega_M,f_b,h,w,A_s,n_s)=
(0.315,0.156,0.67,-1,2.46\times10^{-9},0.96)$, where 
$f_b\equiv\Omega_b/\Omega_M$ is the baryon fraction. We further set the fiducial
value of $\fnl$ to $0$, the value for Gaussian initial conditions, and $\egr=\ewl=1$.

While we are primarily interested in forecasting for $\fnl$ and $\egr$, we must
also marginalize over other parameters that could be correlated with them, 
which includes the six other cosmological parameters listed above, as well as 
the bias nuisance parameters described below. When forecasting for the 
uncertainty on $\fnl$, we do not consider $\egr$ as an extra free parameter, and 
fix it to its fiducial value of 1. Conversely, for the $\egr$ forecasts we 
fix all but $\fnl$ and $\egr$, assuming that only a possible degeneracy with 
$\fnl$ could hamper a detection of the GR effects (all other parameters would 
simply change the shape of the GR correction `template').

The derivatives required by Eq.~(\ref{eq:fisher}) were computed using central 
finite differences,
\begin{equation}\nonumber
 \partial_\alpha f=\frac{f(\theta_\alpha+\delta\theta_\alpha)-
 f(\theta_\alpha-\delta\theta_\alpha)}{2\delta\theta_\alpha}+\mathcal{O}(\delta\theta^3),
\end{equation}
where we chose intervals $\delta\theta_\alpha$ such that the estimated 
derivatives converged to the required numerical accuracy.

We also need to impose priors on certain parameters (e.g. the bias functions) 
to mitigate degeneracies. These are straightforward to incorporate into the 
Fisher matrix formalism: a Gaussian prior on $\{\theta_\alpha\}$ with a 
covariance $C^p_{\alpha\beta}$ can be added directly to the Fisher matrix as
\begin{equation}
 F_{\alpha\beta}\longrightarrow F_{\alpha\beta}+\left(C^p\right)^{-1}_{\alpha\beta}.
\end{equation}
For the cosmological parameters in particular, it is useful to add a Planck CMB 
prior, which we construct by estimating their covariance matrix directly from 
the corresponding Planck MCMC chains \citep{2014A&A...571A..15P}.

\begin{figure}
 \centering
 \hspace{-0.8em}\includegraphics[width=0.49\textwidth]{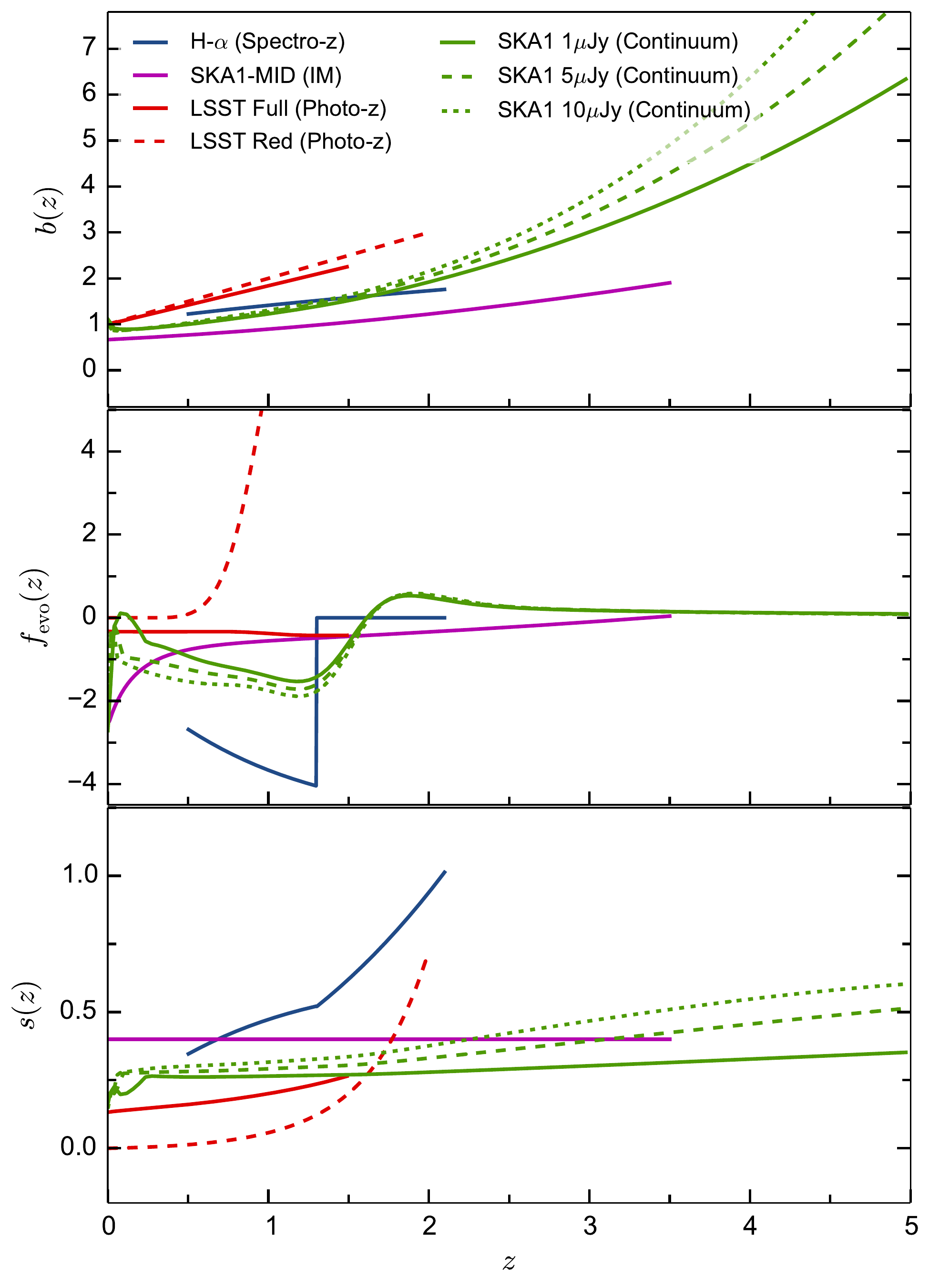}
 \caption{Linear clustering bias (top panel), evolution bias (middle panel) and 
 magnification bias (bottom panel) for the different experiments considered 
 here.}\label{fig:bias}
\end{figure}

Finally, the constraints that any experiment will be able to yield will depend 
crucially upon the smallest and largest scales that can be used. For angular 
scales, this is explicitly taken into account in Eq.~(\ref{eq:fisher}) as the 
maximum multipole, $\ell_{\rm max}$, that we sum up to. This cutoff is 
determined by either the angular resolution of the experiment (e.g. the beam 
size for intensity mapping), or by the non-linear scale, beyond which 
the theoretical predictions become unreliable and the modes must be discarded 
from the analysis. The smallest radial scale corresponds to the comoving width 
of the redshift bins used, and is also determined by either instrumental 
effects (e.g. the redshift resolution in continuum and photometric surveys) or 
the non-linear scale. In any case, since the effects we aim to study in this 
work are dominant on large scales, most of the information about them is 
concentrated on scales well inside the linear regime, and our final results 
are fairly insensitive to the choice of a minimum scale. By default we assume
$\ell_{\rm max}=500$ for all of the probes considered here, and address the 
redshift binning for each case individually. In terms of the maximum angular 
scales, this is set by the sky area surveyed by each telescope. In our 
analysis, all surveys can probe down to $\ell=2$. In later sections we analyse 
the effect of changing this minimum $\ell$. The largest radial scale is set by 
the available redshift range, or the maximum scale that is free from foreground
contamination in the case of intensity mapping.

As described in Section \ref{ssec:th_egr}, the amplitude of the number count 
spectra also depends on three redshift-dependent parameters: the clustering 
bias, $b(z)$, the magnification bias, $s(z)$, and the evolution bias, 
$\fevo(z)$. We will refer to these three parameters collectively as the 
``bias functions'' in what follows. The value of $b(z)$ can only be determined 
from the clustering statistics of the sample under study, and must therefore 
be marginalized over in the cosmological analysis. As discussed previously, 
$s(z)$ and $\fevo(z)$ can be estimated from the redshift-magnitude distribution 
of the sources, although these estimates will inevitably be uncertain and, to 
some extent, model-dependent. Properly accounting for this uncertainty is 
vital, as the behaviour of the bias functions can strongly affect the 
detectability of the signal. In the absence of strong prior measurements from
observations or simulations, these parameters must therefore also be 
marginalized over.

\begin{figure}
 \centering
 \includegraphics[width=0.49\textwidth]{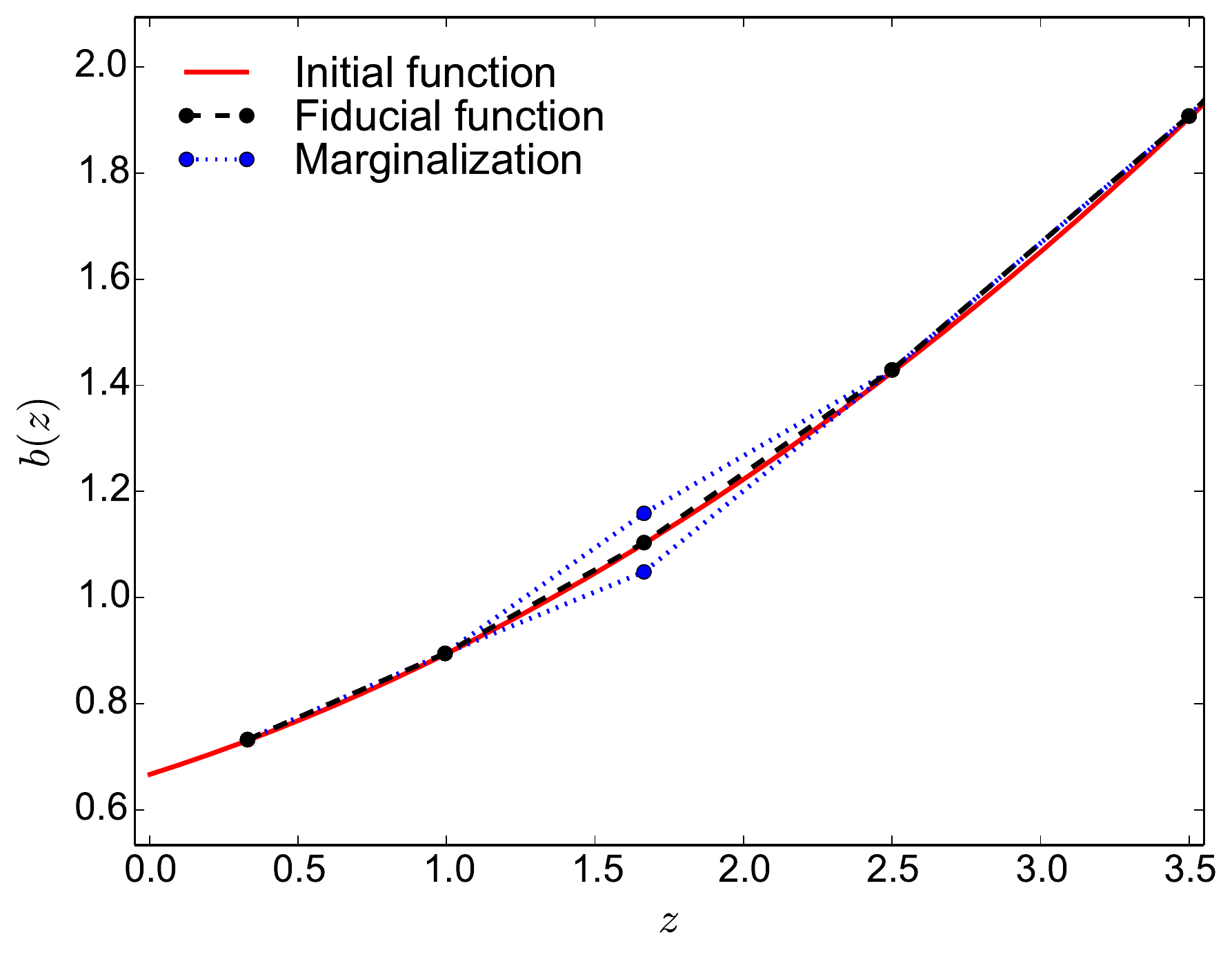}
 \caption{Illustration of the procedure used in this work to 
 marginalize over the bias functions in the case of the HI clustering bias: 
 first the theoretical function is calculated (solid red line). Then, the mean 
 values of the function in the redshift bins listed in Table 
 \ref{tbl:nuisancebins} are computed (black circles), and the linear 
 interpolation between these values (black dashed line) is used as the fiducial 
 function in the computation of the power spectra. Finally, in order to 
 marginalize with respect to this bias function, the mean values mentioned 
 above are used as free parameters, and varied to compute the numerical 
 derivatives in Eq.~(\ref{eq:fisher}) (blue circles and dotted blue lines).}
 \label{fig:nuisance}\vspace{1em}
\end{figure}

\begin{table}[b]
%\vspace{-2em}
\centering{
 \begin{tabular}{|l|l|l|}
 \hline
 {\bf Survey} & {\bf \# bins} & {\bf Bin edges} \\
 \hline
 Int. map. (SKA1-MID) & 5 & 0.0, 0.7,  1.3, 2.0,  3.0, 4.0\\
 Cont. survey (SKA)      & 5 & 0.0, 0.5,  1.0, 2.0,  3.0, 5.0\\
 Spectro. (H$\alpha$ survey) & 4 & 0.5, 0.75, 1.0, 1.3,  2.1\\
 Photo. (LSST), red      & 4 & 0.0, 0.35, 0.7, 1.05, 1.4\\
 Photo. (LSST), all      & 5 & 0.0, 0.5,  1.0, 1.5,  2.2, 3.0\\
 \hline
 \end{tabular}}
 \caption{Redshift bins used for the bias function nuisance parameters, for each survey.}
 \label{tbl:nuisancebins}
\end{table}

In order to do this, we defined a small number of redshift bins that sample the 
bias functions well enough for each survey. The mean values of the bias 
functions were computed in each bin, and a linear interpolation between those 
values was used to define each fiducial bias function. We then marginalized 
over the functions by treating each of the mean values as an additional free 
parameter that was included in the computation of the Fisher matrix (see 
Figure \ref{fig:nuisance}). We also explored other strategies, such as higher 
order interpolation and local modifications to the fiducial functions in each 
bin, but the method described above was found to be the best compromise between 
simplicity and stability to variations in each parameter. We also confirmed 
that the final results do not change significantly for the different methods. 
The redshift bins used for the bias functions for each of the four survey 
types are given in Table \ref{tbl:nuisancebins}, and the input bias functions 
for the surveys\footnote{The codes used to estimate the bias functions for the 
models detailed in Sects. \ref{sec:imapping}--\ref{sec:photometric} can be 
found at \url{http://intensitymapping.physics.ox.ac.uk/codes.html}.} are shown
in Figure \ref{fig:bias}.

\section{Forecasts}\label{sec:forecasts}
\begin{table*}[t]
\centering{
{\renewcommand{\arraystretch}{1.6}
% \begin{tabular}{|p{3cm}|p{2.3cm}|c|c|c|}
 \begin{tabular}{|c|c|c|c|c|c|c|}
 \hline
 {\bf Experiment type}& {\bf Experiment}     & $[z_0,z_f]$ & $\langle S/N\rangle$ 
 & $\langle b\rangle$ & $\langle s\rangle$ & $\langle f_{\rm evo}\rangle$ \\
 \hline
 Intensity mapping    & SKA1-MID             & $[0.1,3.5]$    & 6.7 & 1.2  & 0.4 & -0.4  \\
 \hline
 Continuum survey     & $S_{\rm cut}=10\uJy$ & $[0,3]$        & 8.3 & 1.9  & 0.4 & -0.6  \\
                      & $S_{\rm cut}=5\uJy$  & $[0,4]$        & 13  & 2.5  & 0.3 & -0.3  \\
                      & $S_{\rm cut}=1\uJy$  & $[0,5]$        & 32  & 2.9  & 0.3 & -0.2  \\
 \hline
 Spectroscopic survey & H$\alpha$ survey     & $[0.65, 2.05]$ & 3.6 & 1.5  & 0.55 & -2.0 \\
 \hline
 Photometric survey   & LSST-red             & $[0,1.4]$      & 25  & 1.75 & 0.04 & 3.5  \\
                      & LSST-full            & $[0,2.5]$      & 210 & 2.1  & 0.35 & -0.7 \\
 \hline
 \end{tabular}} }
 \caption{Properties of the experiments under consideration. The quantities shown in
 columns 2-6 are, in order: the
 approximate redshift range probed by each experiment, the average signal-to-noise
 ratio (computed by averaging the ratio of the signal and noise power spectra
 over redshift and angular scale), and the clustering, magnification and
 evolution biases averaged over redshift.}
 \label{tbl:properties}
\end{table*}
The aim of this section is to present the forecasted uncertainties on our
main observables ($\fnl,\,\ewl$ and $\egr$) for the main four types of 
cosmological surveys that will be used to measure them in the future:
intensity mapping, radio continuum surveys, spectroscopic redshift surveys
and photometric redshift surveys. For each of them we will also discuss the
main sources of systematic uncertainties that could prevent these measurements.
The signal and noise models assumed for each of these experiments are presented
in detail in Appendix \ref{app:specs}.

\subsection{HI intensity mapping}\label{sec:imapping}
Intensity mapping (IM) is a relatively new technique, but has a 
number of potential advantages for the study of ultra-large scales 
\citep{Battye:2004re, 2008MNRAS.383..606W,Chang:2007xk}. The idea is to observe
the unresolved emission integrated over many line-emitting sources that are 
assumed to trace the large-scale matter distribution, sacrificing angular 
resolution for survey speed. For source populations with sufficiently narrow, 
isolated emission lines, high redshift resolution can nevertheless be obtained 
with a suitable spectroscopic instrument, allowing the 3D redshift-space 
matter distribution to be reconstructed. Thus, for a given pointing on the sky, 
one measures the combined emission from all the sources in it essentially as a 
continuous field, much like the CMB, rather than as a set of 
separately-identifiable objects. There is therefore no need to set a flux 
threshold that rejects most of the peaks in the signal for being insufficiently 
far above the noise level; the whole of the signal can be used, but different 
modes will be recovered with more or less noise.
This leads to significant gains in survey speed, making it possible to cover 
extremely large volumes -- and thus constrain ultra-large scales -- in a 
relatively short time.

The most developed technique to date is HI intensity mapping, which uses the 
redshifted 21cm radio emission line of neutral hydrogen (HI) as its tracer. HI 
is ubiquitous even in the post-reionisation Universe, where it is mostly 
confined to dense, self-shielded Damped Lyman-$\alpha$ systems. Forthcoming 
large, high-sensitivity, wide-bandwidth radio arrays such as the Square 
Kilometre Array (SKA), and purpose-built IM experiments like CHIME, are 
expected to be able to detect fluctuations in the cosmological HI signal over 
$\sim 75\%$ of the sky, for a wide range of redshifts \citep{2015ApJ...803...21B}. 
We have focused our analysis on Phase I of the SKA, which has the large survey 
area and extremely wide frequency/redshift coverage needed to access ultra-large 
scales (although we have also computed the constraints for a cosmic 
variance-limited experiment).

The SKA is a proposed and partially-funded multi-science radio facility that 
will be able to survey a large fraction of the sky in the frequency range from 
$\sim50\,{\rm MHz}$ to $\sim20\,{\rm GHz}$. It will comprise two different 
instruments, built at separate sites in the South African Karoo region and 
Western Australia's Murchinson region:
\begin{itemize}
 \item SKA1-MID: an array of $\sim\! 200$ single-pixel, 15m dishes to be installed 
       in South Africa. It will cover the frequency range 350-1760 MHz 
       ($z\lesssim3$) in two separate bands.
 \item SKA1-LOW: a set of about 455 aperture array stations each with 35m diameter. 
       It will cover the frequency range 50-350 MHz ($3\lesssim z\lesssim20$).
\end{itemize}
Further information regarding the baseline design for the SKA can be found in 
\cite{2009IEEEP..97.1482D} and \cite{braun15}.

As discussed in \cite{2015ApJ...803...21B} and \cite{2015arXiv150103989S}, the 
most efficient use of SKA1 for intensity mapping at late times ($z\lesssim3$) 
would be to use SKA1-MID as an auto-correlation experiment. If used in 
interferometric mode, the number of short baselines -- needed to resolve the BAO 
scale and larger -- is not large enough. This fact is all the more important 
for ultra-large scales, and so we have focused on the SKA1-MID, single-dish case.

%-------------------------------------------------------------------------------
\subsubsection{Constraints on relativistic effects}

As we have argued above, the two main sources of extra power on ultra-large 
scales are primordial non-Gaussianity and relativistic corrections. In order to 
study the detectability of the latter, we can therefore treat only $\fnl$ and 
$\egr$ as free parameters, and fix the rest to their fiducial values. While 
this procedure would clearly yield an optimistic prediction of the actual 
constraint on $\egr$, it mimics what a survey attempting a first detection of 
any new effect would do: fix all non-degenerate parameters to their best-fit 
values, and fit for the amplitude of the terms related to the new effect. If, 
in doing this, the SNR on the amplitude of the effect is smaller than unity, 
then there is no point in even considering the covariance with other 
parameters.

\begin{figure}[t]
 \centering
 \includegraphics[width=0.49\textwidth]{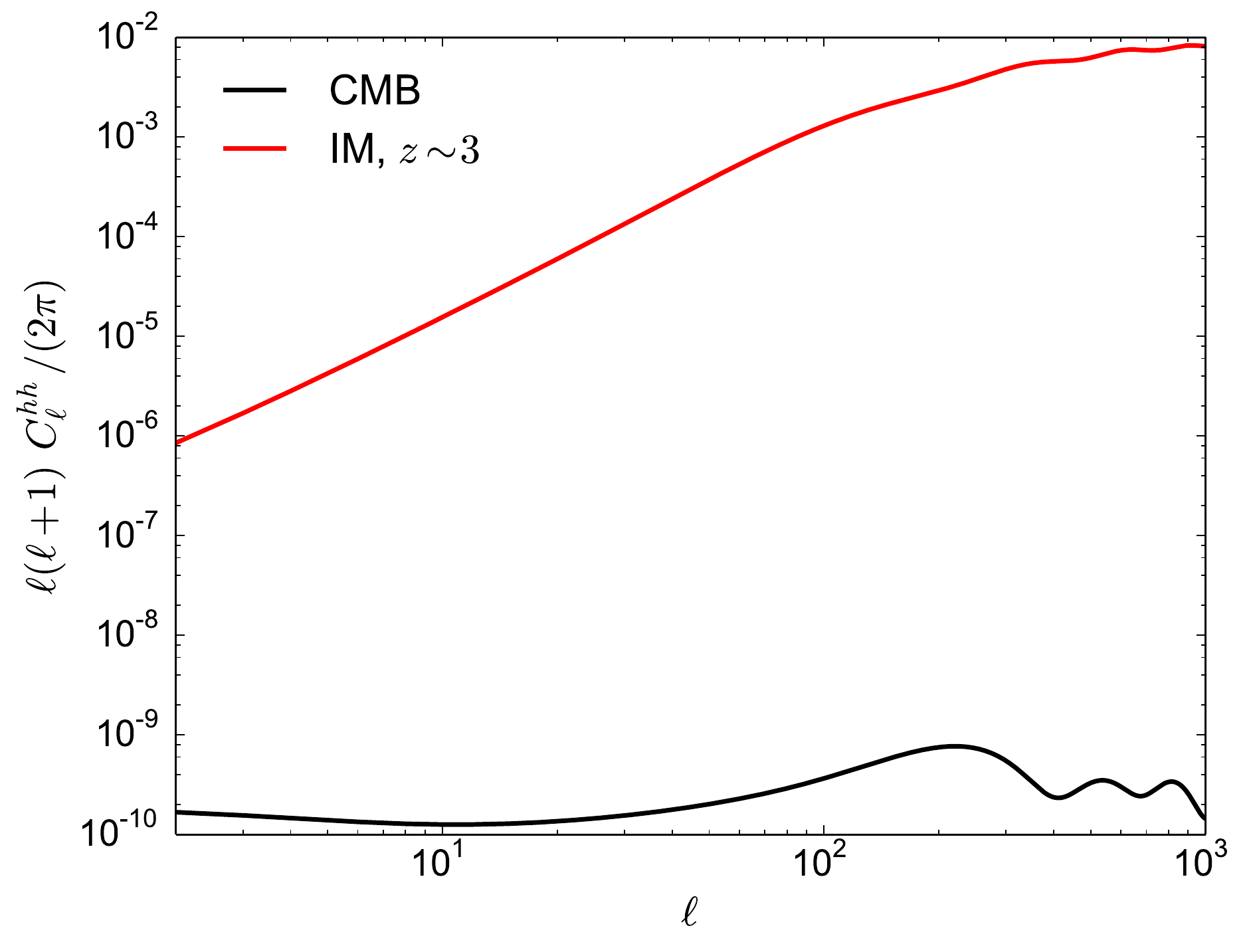}
 \caption{Dimensionless angular power spectrum for a HI intensity map at 
          redshift $z\sim3$ with width $\Delta z\sim0.5$ (orange) and 
          dimensionless ${\rm TT}$ angular power spectrum of the CMB (blue). 
          The much larger amplitude of perturbations at lower redshifts for 
          intensity mapping can explain the difficulty in detecting
          the effect of relativistic corrections in LSS.}\label{fig:im_vs_cmb}
\end{figure}

We applied this procedure for both SKA1-MID and a cosmic variance-limited 
survey ($f_{\rm sky}=1,\,N^{ij}_\ell=0$), obtaining the following result:
\begin{align}
 \text{SKA1-MID}        &\longrightarrow \sigma(\egr)=2.75,\\ 
 \text{CV-limited} &\longrightarrow \sigma(\egr)=1.97.
\end{align}
Thus, even in the best-case scenario, it is not possible to detect the effect 
of the relativistic corrections on the brightness temperature fluctuations.

This may seem like a striking result, as one of the relativistic effects is 
the equivalent of the CMB ISW effect for large-scale structure 
($\Delta^{\rm ISW}$ in Eq.~\ref{eq:isw}), and the CMB ISW has already been 
detected above $4\sigma$ by cross-correlating CMB maps with different LSS 
datasets \citep{2008PhRvD..77l3520G}. In simplistic terms, intensity mapping 
surveys can be thought of as a set of uncorrelated ``CMB'' maps at different
redshifts, so it is legitimate to ask why a similarly significant measurement 
is not possible in this case.

\begin{figure*}
 \centering
 \includegraphics[width=0.49\textwidth]{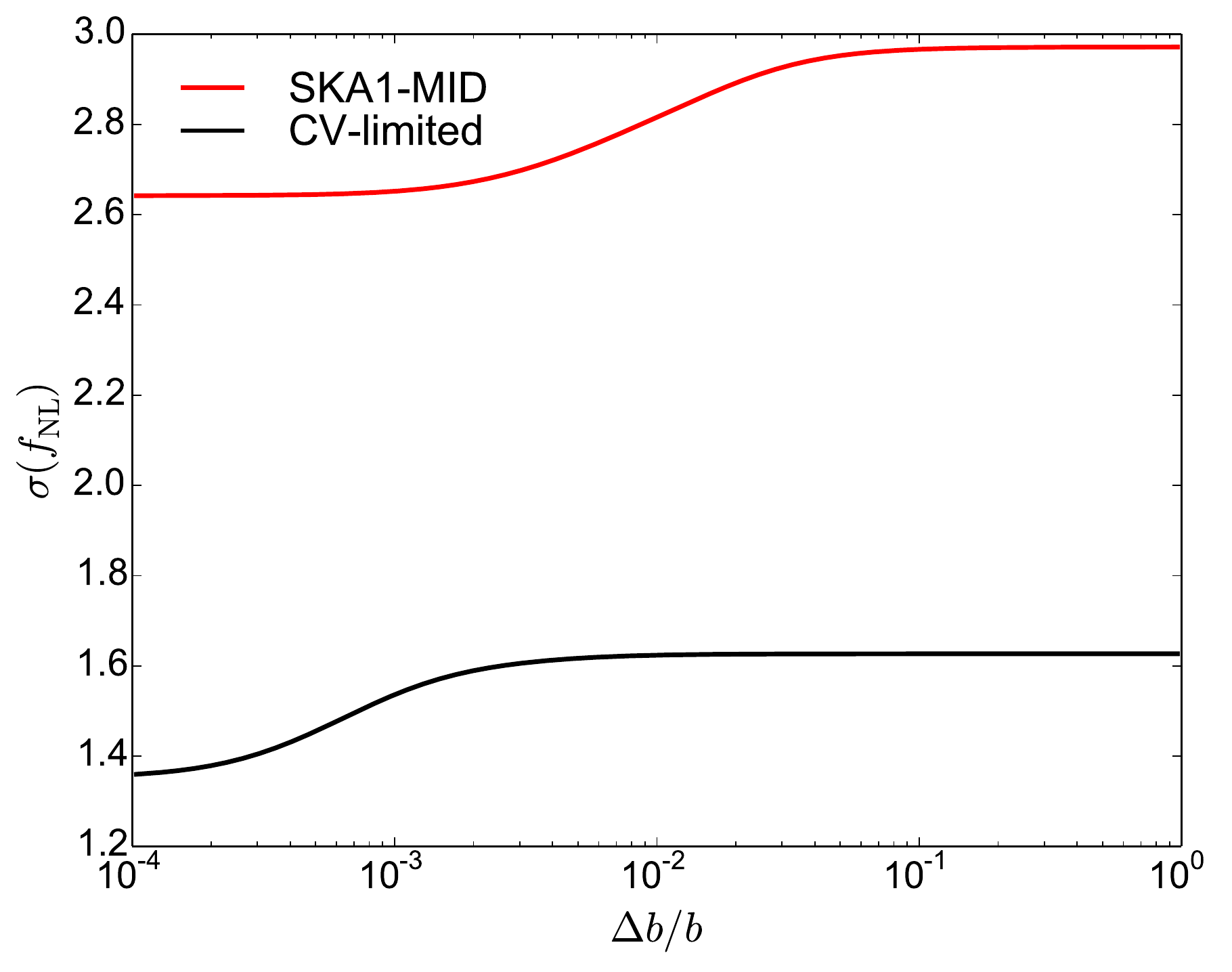}
 \includegraphics[width=0.49\textwidth]{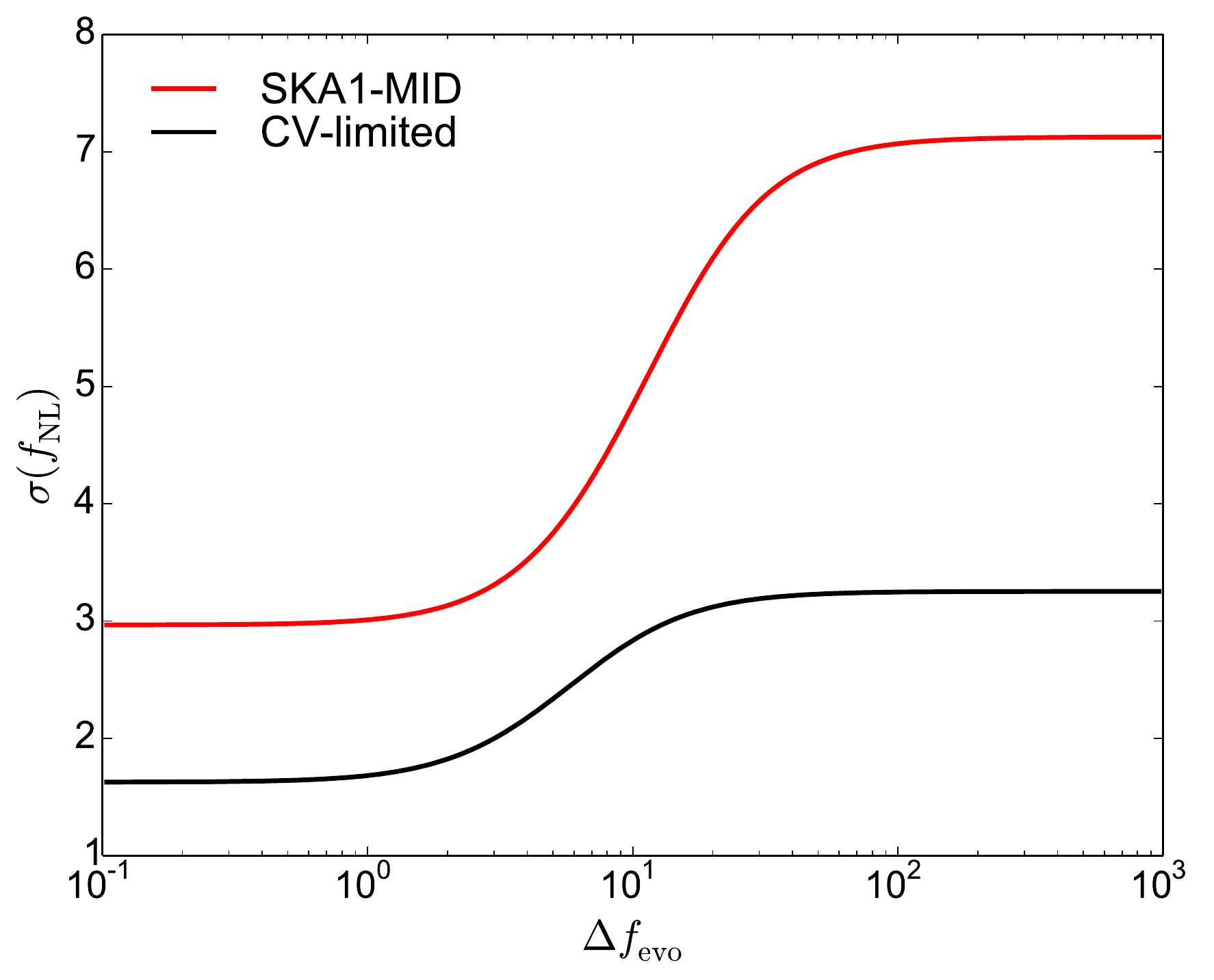}
 \caption{{\sl Left panel}: uncertainty in $\fnl$ as a function of the prior 
          on $b(z)$. A constant Gaussian relative prior was imposed over the 
          whole redshift range. {\sl Right panel}: assuming a prior of $10\%$ 
          on $b(z)$ (e.g., $10\%$ uncertainty). This plot shows the dependence 
          of $\sigma(\fnl)$ on the prior for $\fevo$.}\label{fig:sfnl_vs_sfevo}
\end{figure*}

This can be explained in terms of the clustering variance of both datasets. 
Consider an attempt to measure the ISW effect by cross-correlating two 
datasets, one at high redshift (e.g. the CMB or a high-$z$ HI intensity map), 
which we label here by a superscript $h$, and another at low redshift (e.g. a 
galaxy survey or low-$z$ intensity map), which we label by $g$. Assuming that 
the ISW is the only term that could give rise to a significant cross-correlation 
between both datasets, the signal would be given by the cross-power spectrum,
\begin{equation}
 S_{\rm ISW}=C^{hg}_\ell.
\end{equation}
Neglecting any instrumental or shot noise, and assuming full-sky coverage and 
Gaussian statistics, the noise is purely given by the sample variance,
\begin{equation}
 N_{\rm ISW}\simeq\sqrt{\frac{2}{2\ell+1}\,C^{gg}_\ell\,C^{hh}_\ell},
\end{equation}
where we have assumed that $C^{hg}_\ell\ll \sqrt{C^{gg}_\ell\,C^{hh}_\ell}$. 
Except for factors of order unity, the amplitude of the signal depends 
only on the value of $\phi'+\psi'$ at the redshift of $g$, and not on the 
nature of the high-redshift sample, so it will be roughly the same for both 
a high-$z$ intensity mapping bin and the CMB. The difference in SNR between 
the two cases must therefore depend primarily on the amplitude of the noise, 
which differs by the ratio of $C^{hh}_\ell$ for the two cases. Since 
perturbations have grown significantly since $z_{\rm CMB}\sim1100$, we can 
expect the intensity mapping power spectrum to have a much larger amplitude, 
$C^{hh, {\rm IM}}_\ell \gg C^{hh, {\rm CMB}}_\ell$, which would explain the 
difficulty of achieving a good SNR in this case. This is explicitly shown in 
Figure~\ref{fig:im_vs_cmb}: even at the highest redshift we considered, the 
intensity mapping power spectrum is 4 orders of magnitude larger than that of 
the CMB.

Another effect conspiring against the detection of the relativistic terms in an 
intensity mapping survey is the fact that the perturbations on transverse 
scales cancel exactly (i.e. $s(z)\equiv2/5$), as previously discussed. This 
further suppresses the overall amplitude of the relativistic effects, and is
the reason why we do not present forecasts for $\ewl$ in this case.

%-------------------------------------------------------------------------------
\subsubsection{Constraints on primordial non-Gaussianity}
\label{sssec:fnl_im}

The uncertainty on the level of primordial non-Gaussianity measured by an 
intensity mapping experiment will depend on our prior knowledge of the free 
parameters of the model. Here we have imposed a (non-diagonal) Gaussian prior on 
the cosmological parameters $\{\Omega_M,f_b,h,w_0,n_s,A_s\}$ using a prior 
covariance matrix estimated from the appropriate Planck 2015 MCMC chains for 
our set of parameters \citep{Ade:2015xua}. As we have argued, it is reasonable 
to assume that, by the time the SKA attempts to measure $\fnl$, prior 
information will be available regarding the HI clustering bias (e.g. from 
experiments on smaller scales) and evolution bias (e.g. from external 
measurements of $T_{\rm HI}(z)$).

Before we assume any specific priors for these parameters, it is worth studying 
their effect on $\sigma(\fnl)$ in order to quantify how good prior measurements 
will need to be in order to optimize the constraints on $\fnl$. We first 
studied the level of degeneracy between $\fnl$ and $b(z)$ by estimating the 
value of $\sigma(\fnl)$ assuming a relative Gaussian prior on $b(z)$, constant 
across the whole redshift range. The result is shown in the left panel of 
Fig.~\ref{fig:sfnl_vs_sfevo}. We observe that a mild improvement on 
$\sigma(\fnl)$ ($\sim10\%$ for SKA1-MID and $\sim20\%$ for a CV-limited survey) 
can be achieved only for extremely accurate prior measurements of the 
clustering bias ($\Delta b/b\lesssim10^-3$). Since it would not be realistic to 
expect such a tremendous accuracy, we adopted a fiducial prior on $b$ of 
$10\%$, more compatible with current measurements of the bias of neutral 
hydrogen \citep{2013ApJ...763L..20M}.

Using this fiducial prior, we then explored the degeneracy between $\fnl$ and 
$\fevo$ by studying the dependence of $\sigma(\fnl)$ on the prior uncertainty, 
$\Delta\fevo$. The result is shown in the right panel of 
Fig.~\ref{fig:sfnl_vs_sfevo}. While $\sigma(\fnl)$ increases only slightly when 
factoring in the uncertainties on the clustering bias, we observe a much 
larger increase (e.g. by a factor $2.4$ for SKA1-MID) when we assume no 
knowledge about the evolution bias of the sample at all. This suggests a much 
stronger degeneracy between $\fnl$ and $\fevo$, which could affect any attempt 
at measuring $\fnl$ with LSS probes \cite[note that a similar issue was reported
by][]{2015MNRAS.448.1035C}.

The source of the degeneracy can be understood by comparing the scale 
dependence of the terms in Eqs.~(\ref{eq:terms0}--\ref{eq:isw}) that are 
proportional to $\fevo$ and $\fnl$ respectively. Primordial non-Gaussianity 
introduces a term, included in $\Delta^{\rm D}_\ell$, with a $k$-dependence of 
the form $\propto \delta({\bf k})j_\ell(k\chi)/(k^2\,T(k)) \simeq 
\delta({\bf k})j_\ell(k\chi)/k^2$, where the second equality holds on 
ultra-large scales. The evolution bias, on the other hand, multiplies four 
different terms:
\begin{enumerate}
 \item $\Delta^{\rm P1}_\ell$, proportional to $\psi({\bf k})j_\ell(k\chi)
       \propto\delta({\bf k})j_\ell(k\chi)/k^2$.
 \item $\Delta^{\rm V1}_\ell$, which is proportional to
       $\theta({\bf k})j_\ell(k\chi)/k^2\propto\delta({\bf k})j_\ell(k\chi)/k^2$.
 \item $\Delta^{\rm V2}_\ell$, proportional to $\theta({\bf k})j_\ell'(k\chi)/k$. 
       For sufficiently large $\ell$ ($\gtrsim5$), this term is also proportional 
       to $\delta({\bf k})j_\ell(k\chi)/k^2$.
 \item $\Delta^{\rm ISW}_\ell$, the ISW term, which involves a much wider 
       window function covering the full photon path from the source.
\end{enumerate}
Thus, on large scales, three out of the four terms involving $\fevo$ have the 
same scale dependence as the $\fnl$ term, which explains the result found 
above. Fortunately, as can be seen in Fig.~\ref{fig:sfnl_vs_sfevo}, the effect 
of this degeneracy disappears if we can assume a relatively loose prior on
$\fevo$ of $\Delta\fevo\lesssim1$. I.e. if we parametrize the evolution of
background density of HI in the Universe as $\rho_{\rm HI}(a)\propto a^\alpha$,
the slope $\alpha$ should then be measured with an error $\Delta\alpha\lesssim1$.
In what follows we assume that such an accuracy will be available from external 
measurements of $T_{\rm HI}(z)$, and impose a Gaussian prior of $\Delta\fevo=1$.

For our fiducial set of priors (Planck CMB priors for the cosmological 
parameters, $\Delta b/b=0.1$, and $\Delta\fevo=1$), the final constraints on 
$\fnl$ for SKA1-MID and for a cosmic variance-limited survey are given in 
Table \ref{tbl:constraints_summary}. We also include results for an IM survey 
with a higher fiducial bias, which we discuss in Section \ref{sec:discussion}.

%-------------------------------------------------------------------------------
\subsubsection{Systematic uncertainties}

One of the most important observational challenges for intensity mapping is the 
presence of galactic and extragalactic radio foregrounds (e.g. galactic 
synchrotron emission and extragalactic continuum radio sources) with 
amplitudes several orders of magnitude larger than the cosmological HI signal. 
The potential bias and extra variance induced in the measured signal by the 
process of foreground removal must be correctly taken into account in any 
analysis.

\begin{figure}[t]
 \centering
 \includegraphics[width=0.49\textwidth]{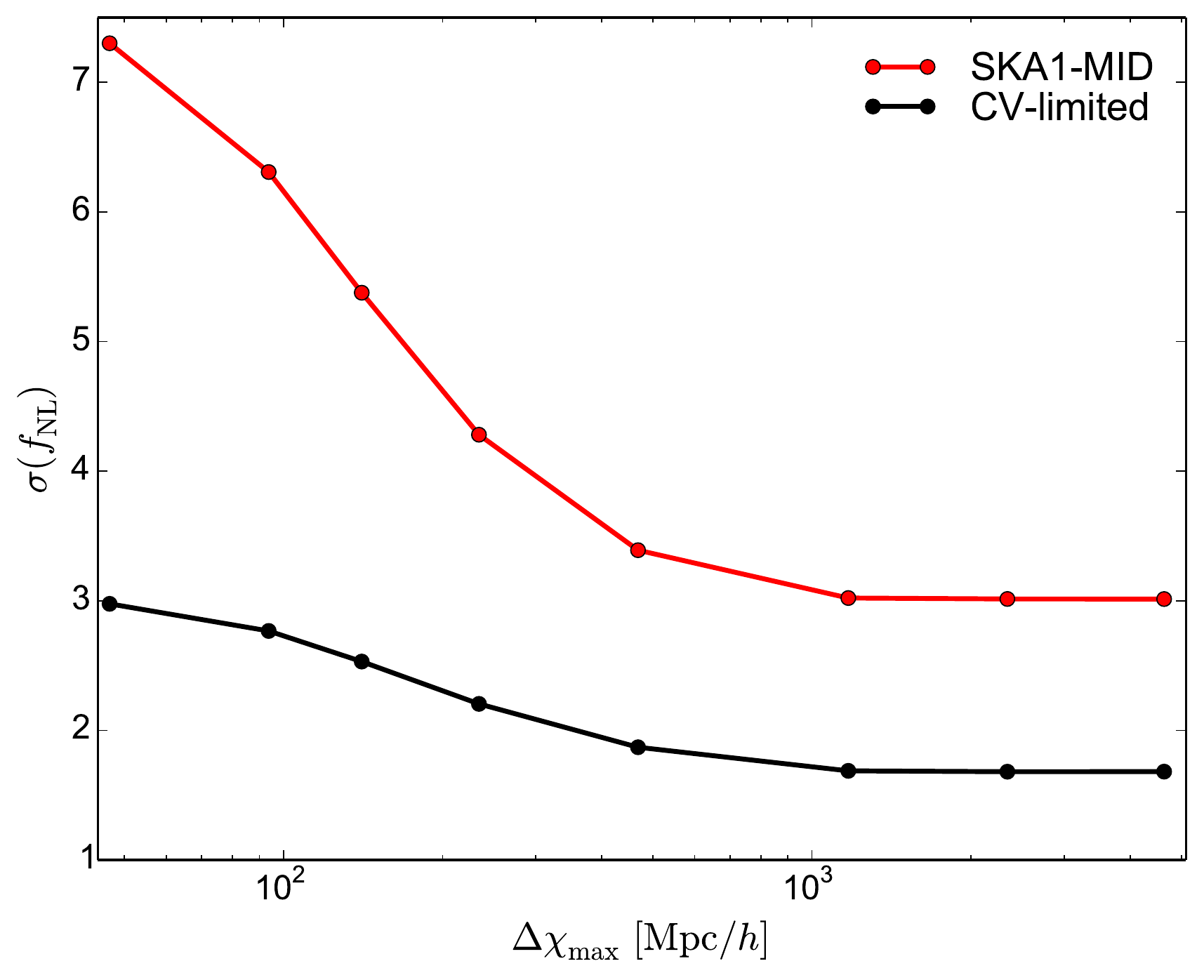}
 \caption{Constraint on $\fnl$ as a function of the maximum comoving separation 
          between pairs of redshift bins included in the analysis (at 
          $z\sim 1.7$, 1 Gpc corresponds to $\sim 100$ MHz).}
          \label{fig:sfnl_vs_dchi}
\end{figure}

The strategy underlying most foreground cleaning methods is to exploit the very 
different frequency structures of signal and foregrounds. Most foreground 
signals have a very smooth frequency dependence, while the cosmological signal 
traces the stochastic fluctuations in the matter density, and is therefore much 
``noisier'' in the radial (frequency) direction. Broadly speaking, most 
cleaning methods try to remove the foregrounds by fitting and subtracting a set 
of smooth functions of the frequency from the combined foreground $+$ 
cosmological signal. See \cite{2015MNRAS.447..400A} for a description and 
comparison of different methods.

Since foregrounds are smooth in frequency, and frequency is a proxy for radial 
distances for an IM experiment, we can expect the foreground-cleaned maps to be 
dominated by systematics on large radial scales. These scales must then be 
omitted from the analysis, which reduces the sensitivity of an experiment to 
$\fnl$. In order to understand the importance of this effect, we have studied 
the dependence of $\sigma(\fnl)$ on the maximum radial separation between 
redshift bins included in the computation of the Fisher matrix, i.e. we set 
to zero the off-diagonal elements of $\mathsf{C}_\ell$ corresponding to pairs 
of bins separated by more than some radial separation $\Delta\chi$.

The results are shown in Figure \ref{fig:sfnl_vs_dchi} for SKA1-MID and for a 
cosmic variance-limited survey. Reducing the range of the cross-correlations 
included in the analysis can degrade the sensitivity to $\fnl$ significantly, 
enlarging the errors by up to a factor $\sim3.3$ in the case of 
SKA1-MID.

\begin{figure}[t]
 \centering
 \hspace{-1em}\includegraphics[width=0.49\textwidth]{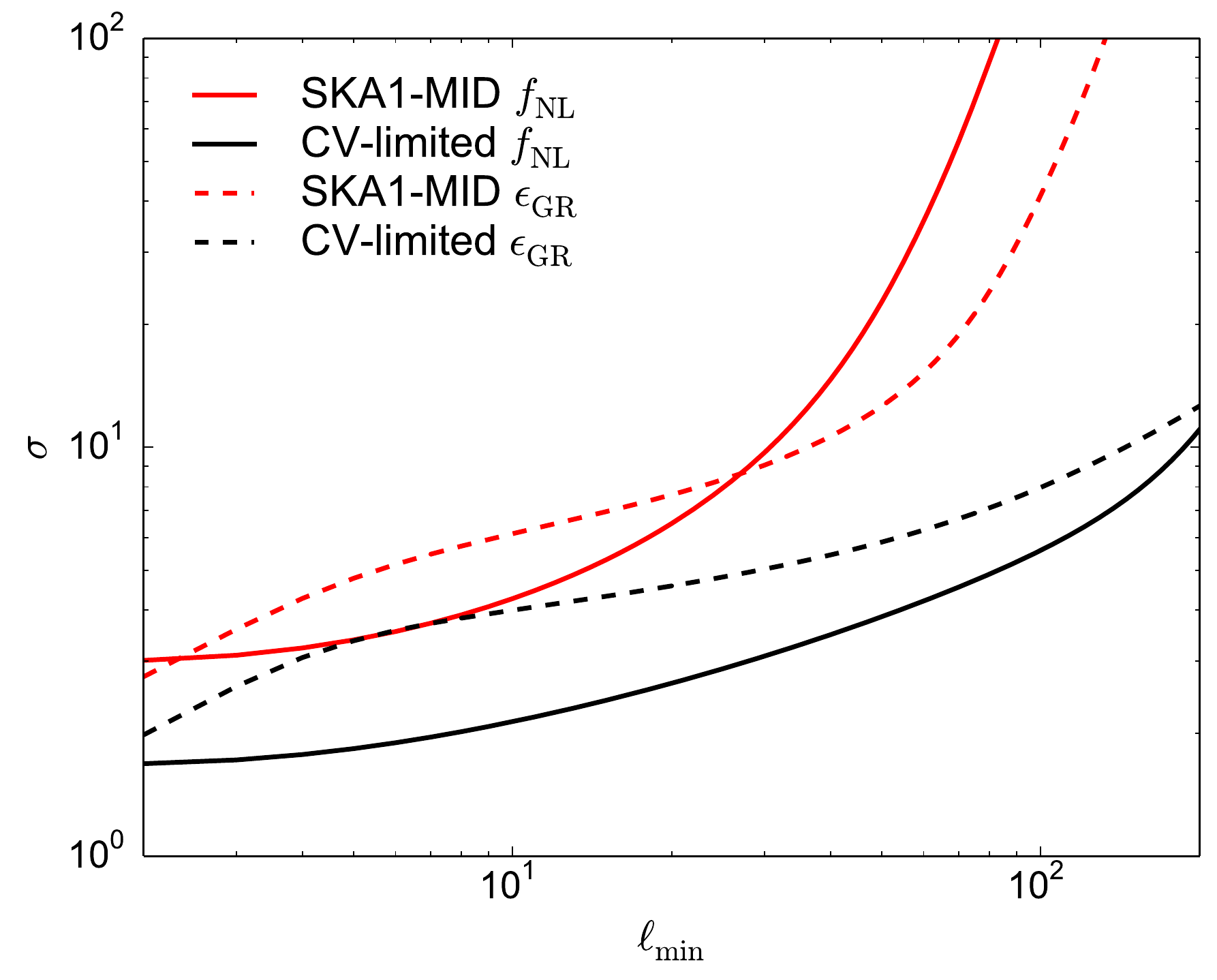}
 \caption{Uncertainty on $\fnl$ as a function of the largest angular scale 
          (minimum $\ell$) included in the Fisher matrix calculation. 
          Systematic effects and incomplete sky coverage can prevent recovery 
          of the lowest multipoles (in this plot, sky coverage, $f_{\rm sky}$, 
          is fixed to the survey specifications).}
          \label{fig:sigma_llim_im}
\end{figure}

Given that measurements of ultra-large scale modes will typically need to be 
done in single dish (autocorrelation) mode, one also needs to worry about the 
stability of the instrument and the observational strategy. Much as in CMB 
experiments, long term noise drifts (the ``$1/f$'' noise) will lead to 
striping in the maps, i.e. a coherent set of large angle features that have 
been artificially projected on the sky. A sensible choice of scan strategy 
that leads to appreciable cross-linking between the scans can mitigate the 
effect, but there will always be a residual large-angle contaminant. Again, we 
can model this effect by not including the very large angular modes in the 
analysis (tantamount to assuming that they are filtered out by the destriping 
process). Fig.~\ref{fig:sigma_llim_im} gives some idea of the impact of this 
effect on the constraints: a severe cut in the large angle data significantly 
degrades any attempt to detect large-scale features.

%-------------------------------------------------------------------------------
\subsection{Radio continuum surveys}
\label{sec:continuum}

Radio continuum surveys measure the integrated emission of radio sources in one 
broad frequency band (or a small number of them). At radio wavelengths, the 
spectral energy distributions (SEDs) of most sources are generally smooth and 
featureless, except for a few radio lines such as the aforementioned 21cm 
signal (which can only be resolved for individual sources after a long 
integration time). For this reason, there is little distance information to be 
gained by integrating the flux density over more than a single, wide band. In 
turn, using a wide band significantly increases sensitivity, allowing much 
fainter sources to be observed than it would be possible to detect otherwise. 
Continuum surveys can therefore cover extremely large volumes, with the caveat
that essentially all information on radial scales (even the ultra-large ones) 
is completely inaccessible.

That continuum surveys have the potential to constrain the level of primordial 
non-Gaussianity has already been shown in the literature 
\citep{2014MNRAS.442.2511F, 2015JCAP...01..042R}, especially if the survey can 
be split into several different populations so that the multi-tracer technique 
\citep{2009PhRvL.102b1302S} can be used. We attempt to reproduce this result 
here, as well as exploring the possible degeneracies of $\fnl$ with the 
relativistic corrections, and the possibility of detecting the latter. As with 
the other probes, we treat the continuum survey as being single-tracer only, 
and will not study the potential of the multi-tracer technique here. (This is 
left for future work, in which all the possible cross-correlations will be 
considered systematically.)

Our forecasts focus on a continuum survey with Phase 1 of the SKA, since this
would correspond to the widest proposed survey area to date in the $\sim\uJy$ 
regime. SKA1 should be able to detect radio sources out to $z\sim5$ over an 
area of about $3\pi$ steradians. This survey would be carried out using the 
SKA1-MID facility, integrating the source flux in the band 350-1050 MHz with 
an rms noise of $S_{\rm rms}\simeq1\uJy$ (see \cite{2015arXiv150103825J} for 
details).
\begin{figure*}
 \centering
 \includegraphics[width=0.49\textwidth]{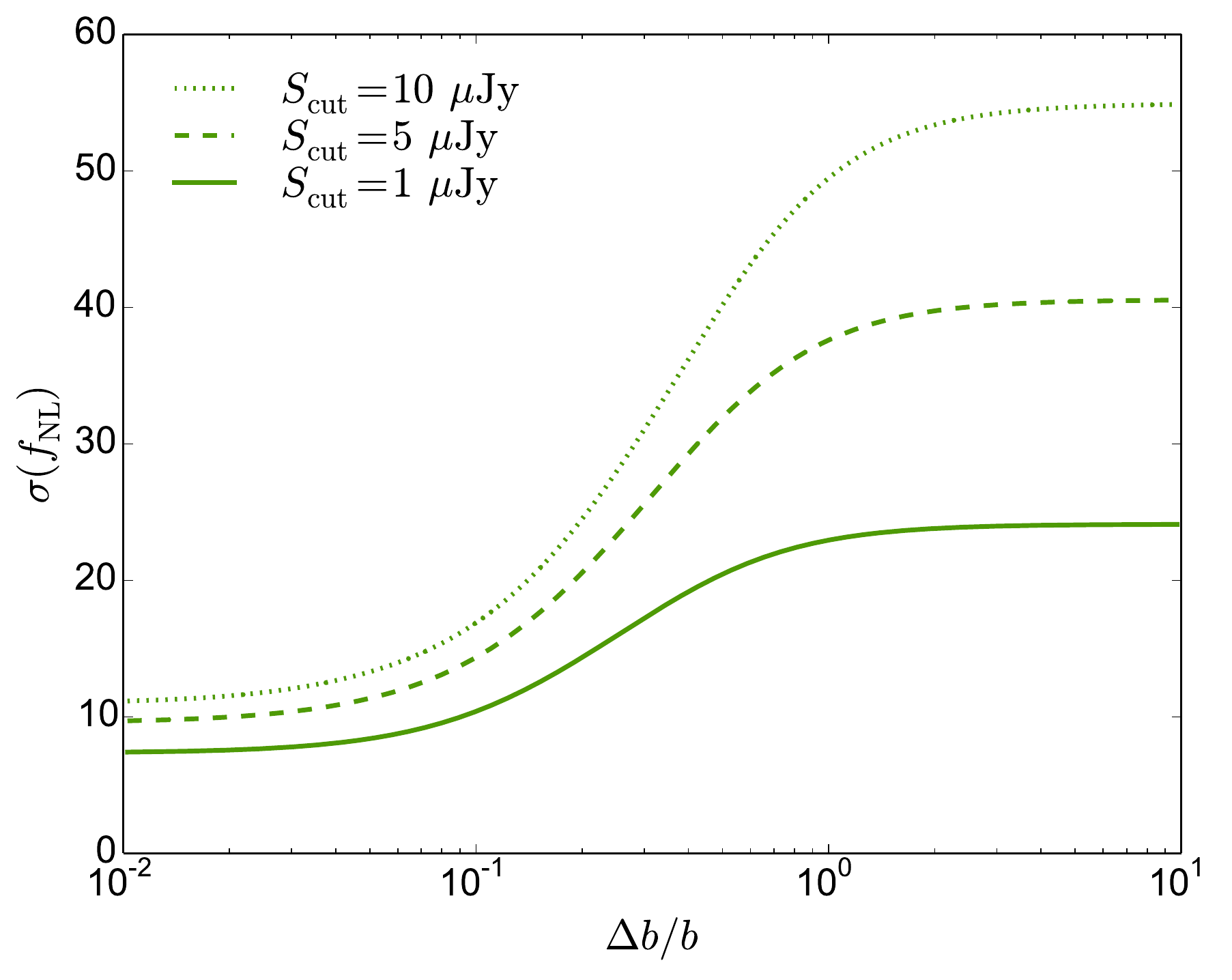}\\
 \includegraphics[width=0.49\textwidth]{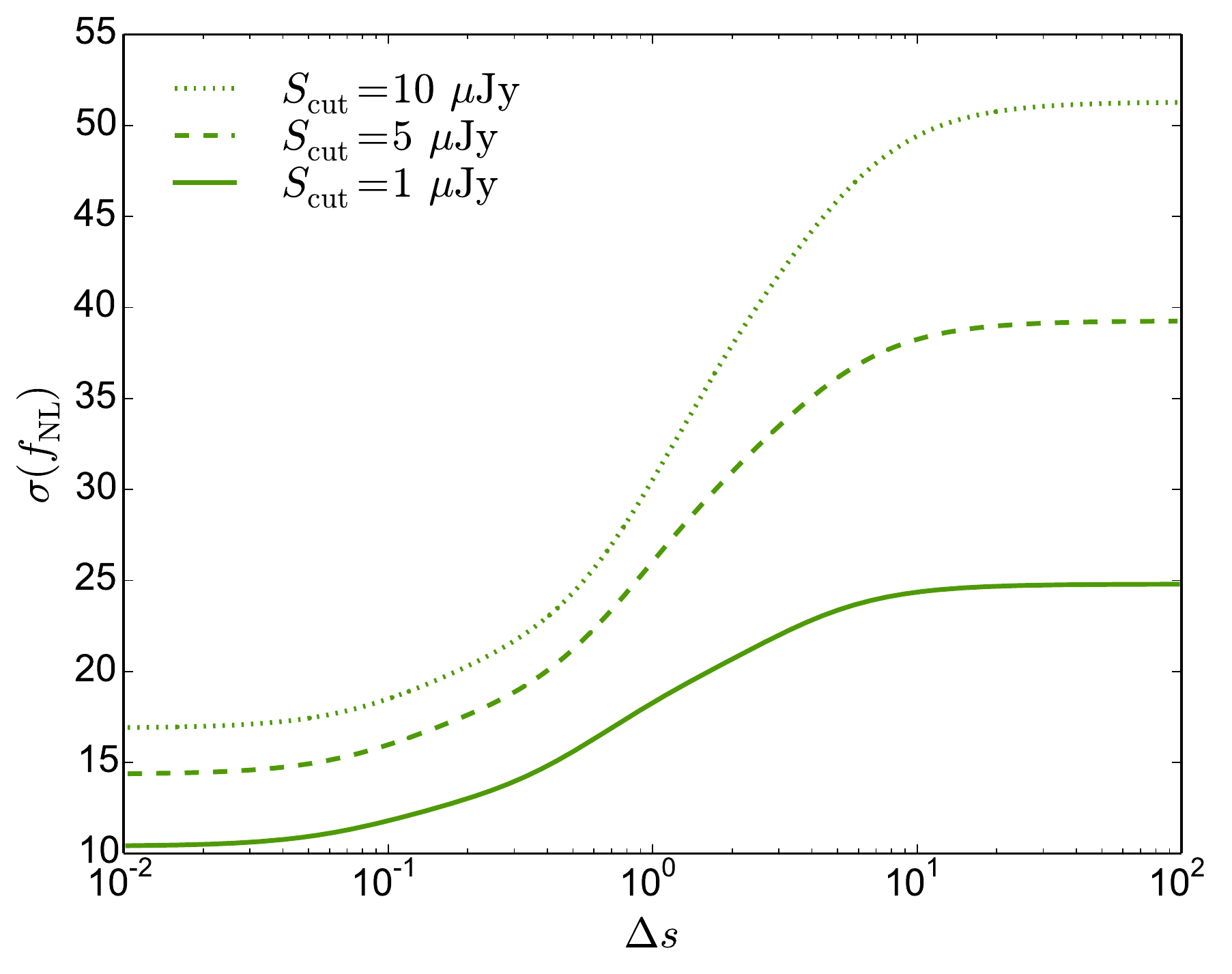}
 \includegraphics[width=0.49\textwidth]{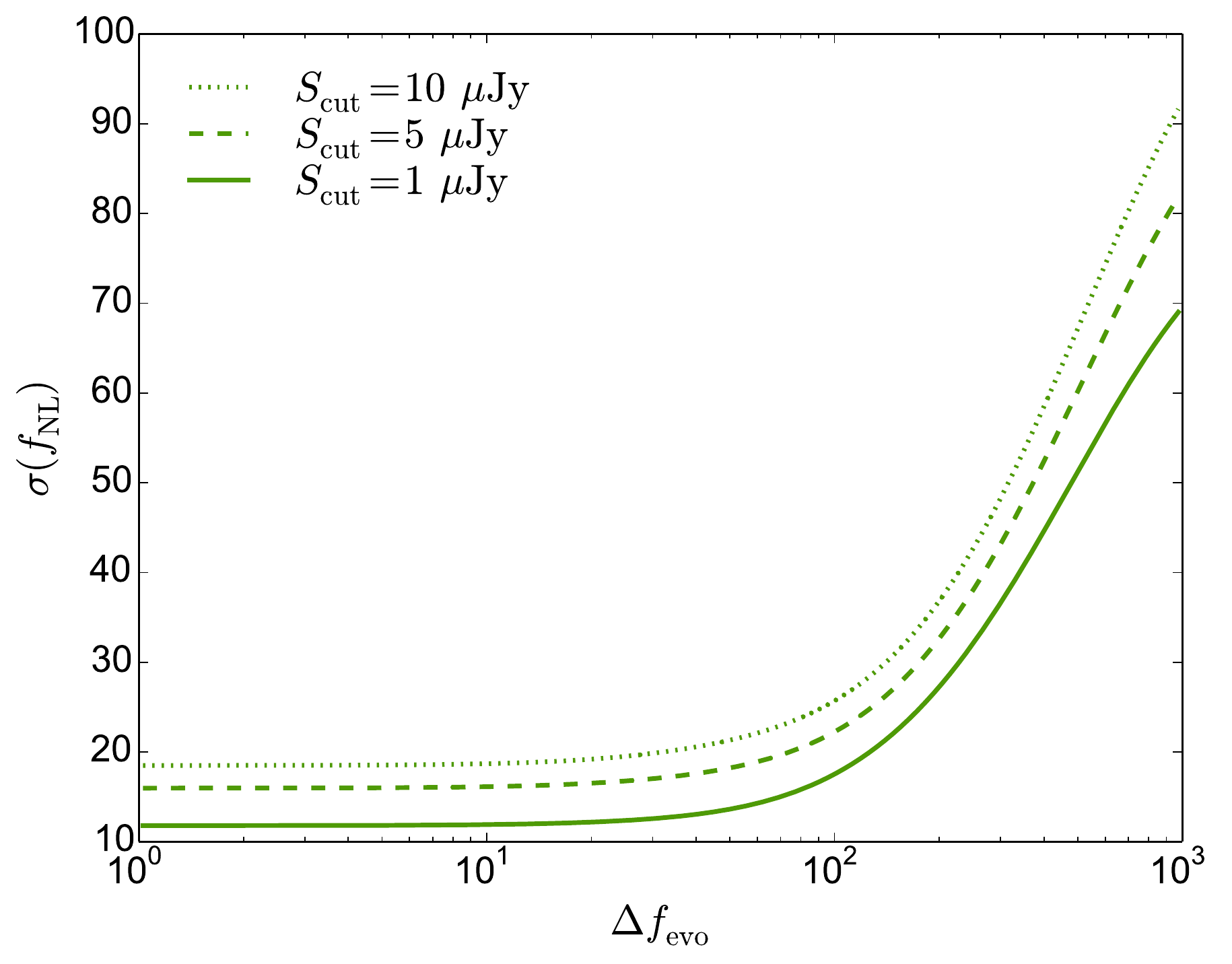}
 \caption{Dependence of the uncertainty on $\fnl$ on the prior assumed for
          the clustering bias (top right panel, in terms of a relative prior)
          and the magnification and evolution biases (bottom left and right panels,
          shown in terms of an absolute prior).}\label{fig:sfnl_vs_spar_cont}
\end{figure*}

It is common to define the source detection limit $S_{\rm cut}$ to be several 
times higher than the noise level of the experiment (usually by a factor of 5 
or 10, depending on the intended use of the sample). Assuming an rms 
instrumental noise for SKA1-MID of $S_{\rm rms}=1\uJy$, we have defined a 
fiducial $5\sigma$ detection limit (i.e. $S_{\rm cut}=5\uJy$). In order to 
explore the dependence on the survey depth, we have also produced forecasts for
$S_{\rm cut}=10\uJy$ and $S_{\rm cut}=1\uJy$.

\subsubsection{Constraints on relativistic effects}

Relativistic effects in the continuum angular power spectrum were considered in 
\cite{2013JCAP...02..044M} and \cite{2015PhRvD..91d3507C}.
As in the previous section, we first present the most optimistic forecasts for 
$\egr$ by marginalizing only over $\fnl$. If the uncertainty on $\egr$ is 
larger than unity (i.e. no detection of GR effects) in this optimistic case, 
there is no point in exploring more realistic scenarios.

There are two main differences between intensity mapping and continuum surveys 
in terms of the quantities that can affect these forecasts. First of all, we 
can expect the lack of radial information in continuum surveys to considerably 
degrade the constraints on most parameters. Perturbations to transverse scales 
will affect the observed clustering of radio sources in this case, however 
(i.e. $s(z)\ne2/5$, unlike for intensity mapping), which will enhance the 
amplitude of the relativistic terms.

The constraints on $\egr$ found for the three flux limits that we considered 
are shown in Table \ref{tbl:constraints_summary}. Due to the lack of radial 
information, there is no hope of detecting GR effects using only a 
single-tracer analysis, in spite of the enormous volume probed; 
$\sigma(\egr) = 17.1$ even for the deepest ($1\mu$Jy) survey.

We have also produced forecasts for the detectability of the weak lensing term
on large scales for continuum surveys, following the same logic used in the
case of $\egr$ (i.e. we keep all other parameters fixed, except for $\fnl$). In
order to pick up only the large-scale lensing contribution we also used a more
stringent value of $\ell_{\rm max}=100$. The results are summarized in Table
\ref{tbl:constraints_summary}: a continuum survey with a flux limit of
$1\uJy$ would be able to clearly detect the large-scale lensing effect
above $\sim4\sigma$, although the level of this detection would be below
$2\sigma$ for $S_{\rm cut}=5\uJy$. No detection would be possible for
a flux cut of $10\uJy$.

%-------------------------------------------------------------------------------
\subsubsection{Constraints on primordial non-Gaussianity}
\label{sec:continuum_fnl}

In a continuum survey, the available information is compressed into only a 
small amount of data -- the angular clustering statistics of radio sources -- 
due to the lack of any sensitivity to radial modes. We can therefore expect an
even larger degree of degeneracy between the various cosmological and nuisance 
parameters than for intensity mapping. As before, we assume Planck CMB priors 
on all cosmological parameters (except $\fnl$), and start our discussion of the 
primordial non-Gaussianity forecasts by exploring the effect of prior 
information about the bias functions on the $\fnl$ constraints. The results of 
this analysis are shown in Figure \ref{fig:sfnl_vs_spar_cont}.

The top left panel of Fig.~\ref{fig:sfnl_vs_spar_cont} shows the forecast 
uncertainty on $f_{\rm NL}$ as a function of a constant relative Gaussian prior 
for $b$. In contrast with the situation for intensity mapping, the constraints 
are much more sensitive to the prior on the clustering bias. The main reason 
for this is that, for IM experiments, the availability of redshift information 
helps to break the degeneracy between the parameters through the scale 
dependence of the $\fnl$ term along the line of sight. We find that a 
$\sim10\%$ error on $b(z)$ would be sufficient to minimize the uncertainty on
$\fnl$. Since this is compatible with previous smaller-scale observations
\citep[c.f.][]{2014MNRAS.440.2322L}, we chose this value as our fiducial prior on 
the clustering bias.

The effect of a prior on the magnification bias is shown in the bottom left 
panel of Fig.~\ref{fig:sfnl_vs_spar_cont}. We observe a similar degeneracy 
with $\fnl$, again mainly due to the lack of redshift information, which can 
only be mitigated by prior measurements of $s(z)$ with an error better than 
$\Delta s\approx\pm0.1$. We have assumed that such an accuracy would be achievable 
using the magnitude-redshift distribution of sources, although this could be an 
optimistic assumption at the highest redshifts. Finally, as in the case of 
intensity mapping, we observe a significant degradation in the uncertainty on 
$\fnl$ if we assume no knowledge about the evolution bias of the sample (bottom 
right panel in Fig.~\ref{fig:sfnl_vs_spar_cont}). This is again due to the 
degeneracy in the scale dependence of the terms corresponding to both 
quantities (see Sect. \ref{sssec:fnl_im}), and can be mitigated by measuring 
$\fevo$ to an accuracy of better than $\Delta\fevo\lesssim10$.

\begin{figure}[t]
 \centering
 \hspace{-1em}\includegraphics[width=0.49\textwidth]{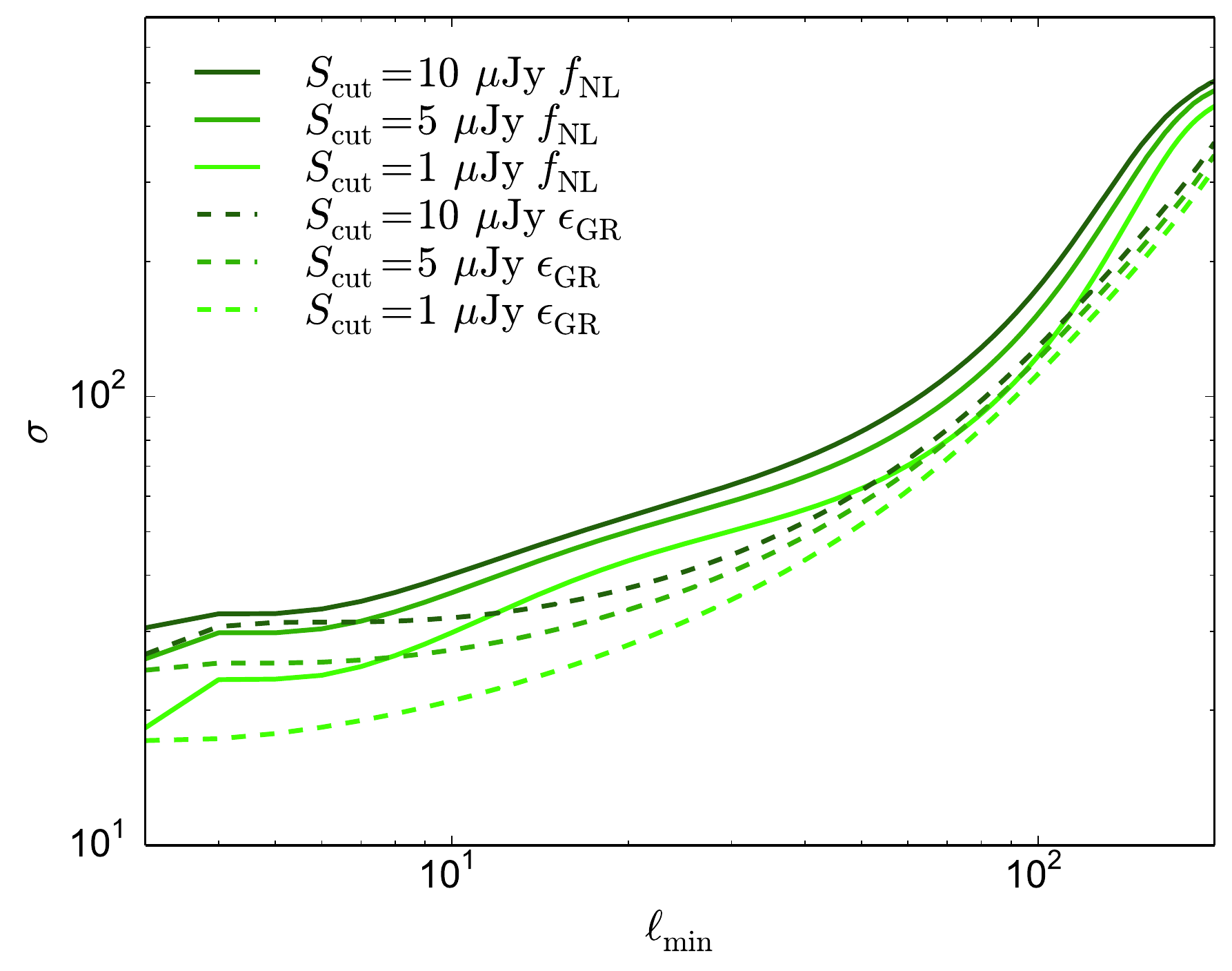}
 \caption{Dependence of the uncertainty on $\fnl$ (solid lines) and $\egr$ 
          (dashed lines) on the minimum scale probed by the survey for 
          different flux limits.}
 \label{fig:sigma_llim_cont}
\end{figure}

In view of this analysis, our final fiducial set of priors on the bias 
functions is $\Delta b/b=0.1$, $\Delta s=0.1$, and $\Delta\fevo=1$. Forecasts 
for $\fnl$ with these priors are listed in Table \ref{tbl:constraints_summary}  
and, with $\sigma(\fnl) =16$ for the 5$\mu$Jy sample, are compatible with the 
results of \cite{2014MNRAS.442.2511F} for their combined sample. This is 
far worse than the constraints possible with multiple tracers, however.

%-------------------------------------------------------------------------------
\subsubsection{Systematic uncertainties}

The different frequency range and observational techniques involved in radio 
astronomy give rise to potentially very different sources of systematics for 
continuum surveys compared with optical and near-infrared surveys. To begin 
with, the diffuse nature of galactic synchrotron emission (the largest galactic 
foreground at radio frequencies) makes it virtually transparent to the long 
interferometer baselines needed for a survey aiming to resolve individual 
sources, and hence the problem of galactic foregrounds is greatly ameliorated. 
On the other hand, in order to produce a full-sky catalogue, mosaicing of the 
individual pointings must be implemented. If the mosaicing pattern and 
correlations in the noise properties between pointings are not fully understood, 
they could introduce systematic deviations on large angular scales. Ionospheric 
effects will also be relevant at low frequencies, although this should not be a 
problem in the SKA1-MID frequency range. Bright point sources would also need 
to be masked in a non-trivial way, due to dynamical range issues causing 
increased noise in the far beam sidelobes. The extent to which this would 
affect ultra-large scales is instrument-dependent, however. Avoiding these 
systematics might again entail removing the smallest multipoles of the power 
spectrum from the analysis; we show how the constraints on $\fnl$ depend on the 
minimum multipole $\ell_{\rm min}$ in Fig.~\ref{fig:sigma_llim_cont}.

%-------------------------------------------------------------------------------
\subsection{Spectroscopic redshift surveys}
\label{sec:spectroscopic}

Spectroscopic galaxy redshift surveys in the optical and near-infrared 
represent the current state-of-the-art in large-scale structure observations 
\citep{Percival:2013awa}. The premise is simple: to detect redshifted emission 
lines from as many resolved sources as possible, over as large a spatial volume 
as possible, and then to reconstruct the 3D distribution of sources in 
redshift-space. Assuming that the source population traces the underlying 
matter density field, and samples it sufficiently well, one can then constrain 
the statistical properties of the cosmological matter distribution as a 
function of scale and redshift. Because the positions and redshifts of the 
sources can both be measured with high precision, information about the matter
distribution is retained in the angular {\it and} radial directions, unlike 
with more lossy techniques like intensity mapping and photometric redshifts. 
The downside is that taking high-resolution spectra for a large number of 
sources is extremely time-consuming.

For ultra-large scales, we are primarily interested in beating-down sample 
variance by increasing the effective survey volume. This requires a wide survey 
area, broad redshift coverage (preferentially at higher redshifts, where longer 
wavelength Fourier modes are accessible), and a sufficiently high signal 
intensity/source density that noise will be sub-dominant. Of the methods 
considered in this paper, spectroscopic galaxy surveys appear the least 
well-suited to the task of surveying extremely large volumes. Spectroscopy is 
time-consuming, and while precision redshift information is extremely useful 
for studying baryon acoustic oscillations and other smaller-scale phenomena, it 
is less necessary for the largest scales, apart from allowing the survey to be 
split into a larger number of redshift bins (c.f. intensity mapping). 
Nevertheless, spectroscopic galaxy surveys are the most developed of the 
methods, and so have comparatively well-understood systematic effects. The 
additional small-scale information also allows them to measure basic 
cosmological parameters more accurately, which helps to break parameter 
degeneracies.

We base the specifications of our reference spectroscopic survey on a large 
emission-line galaxy (ELG) survey along the same lines as Euclid 
\citep{2011arXiv1110.3193L}, a satellite mission with a near-infrared 
spectrograph that will detect $\sim 6 \times 10^7$ H$\alpha$-emitting galaxies 
over 15,000 deg$^2$ in the redshift range $0.65 \lesssim z \lesssim 2.05$. A 
similar ELG survey will be performed by DESI (formerly BigBOSS), which will 
target [OII] galaxies out to $z = 1.7$ over 14,000 deg$^2$ 
\citep{2011arXiv1106.1706S}. These are the largest planned spectroscopic 
surveys,\footnote{A proposed HI galaxy survey with Phase 2 of the SKA would 
detect $\sim 10^9$ galaxies over $\sim$30,000 deg$^2$ for redshifts 
$0 \le z \lesssim 2$ \citep{Yahya:2014yva}, but is considerably more 
futuristic.} but have the smallest area of the experiments considered here, 
and the lowest maximum redshift apart from LSST-red (see Section 
\ref{sec:photometric}).

%-------------------------------------------------------------------------------
\subsubsection{Constraints on relativistic effects}
\label{sec:spectroz:rel}

The most optimistic forecast for $\egr$, marginalizing over $\fnl$ only, yields 
$\sigma(\egr) = 2.6$ for the H$\alpha$ survey's narrowest redshift binning 
($\Delta z = 0.025$). This result is insensitive to bin width, increasing only 
slightly to $2.7$ for the widest binning ($\Delta z = 0.1$); the additional 
information gained by decreasing the bin width is mostly confined to small 
scales, where the relativistic effects are essentially negligible. The 
correlation between $\egr$ and $\fnl$ is very weak, and there is no change in 
the constraint whether $\fnl$ is marginalized or fixed. The survey is also 
quite close to its ideal (sample variance-limited) performance, with 
$\sigma(\egr)$ improving only slightly to 2.2 in the limit $N_\ell \to 0$ for
$\Delta z = 0.025$. These results, together with the constraints corresponding 
to a cosmic variance-limited results (assuming $N_\ell = 0$ and $f_{\rm sky}=1$) 
are summarized in Table~\ref{tbl:constraints_summary}, and have a qualitatively 
similar behaviour.

As with the two previous surveys, then, the relativistic effects are 
undetectable. This is despite the relatively high magnification bias of the 
H$\alpha$ galaxies, which boosts the size of some of the relativistic correction 
terms. The H$\alpha$ survey's sky coverage and maximum redshift are smaller 
than for the other surveys though, which weakens its constraining power. An 
experiment with the same specifications as the H$\alpha$ survey but covering 
twice the area (30,000 deg$^2$) would give $\sigma(\egr) = 1.8$ (compared with 
1.4 in the CV-limited case), which is still not enough to gain a detection.

Nevertheless, the H$\alpha$ surveys's forecast constraint of $\sigma(\egr)=2.6$ 
is the best so far, and the CV-limited figure of 1.36 is markedly better than 
SKA1-MID's value of 1.97, despite the IM survey having a significantly wider 
redshift range. The enhanced performance of the spectroscopic survey over
intensity mapping is primarily due to the different behaviour of the bias 
functions, particularly $s(z)$, which caused many of the relativistic effects 
to cancel for the IM survey.

As before, we also forecast for the detectability of the large-scale lensing
effect, parametrized by $\ewl$. After restricting to modes $\ell \le 100$ and
marginalizing only over $\fnl$, we find $\sigma(\ewl) = 0.19$ for all three
choices of redshift bin width -- a strong detection (see Table
\ref{tbl:constraints_summary}).
\begin{figure}[t]
 \centering
 \includegraphics[width=\columnwidth]{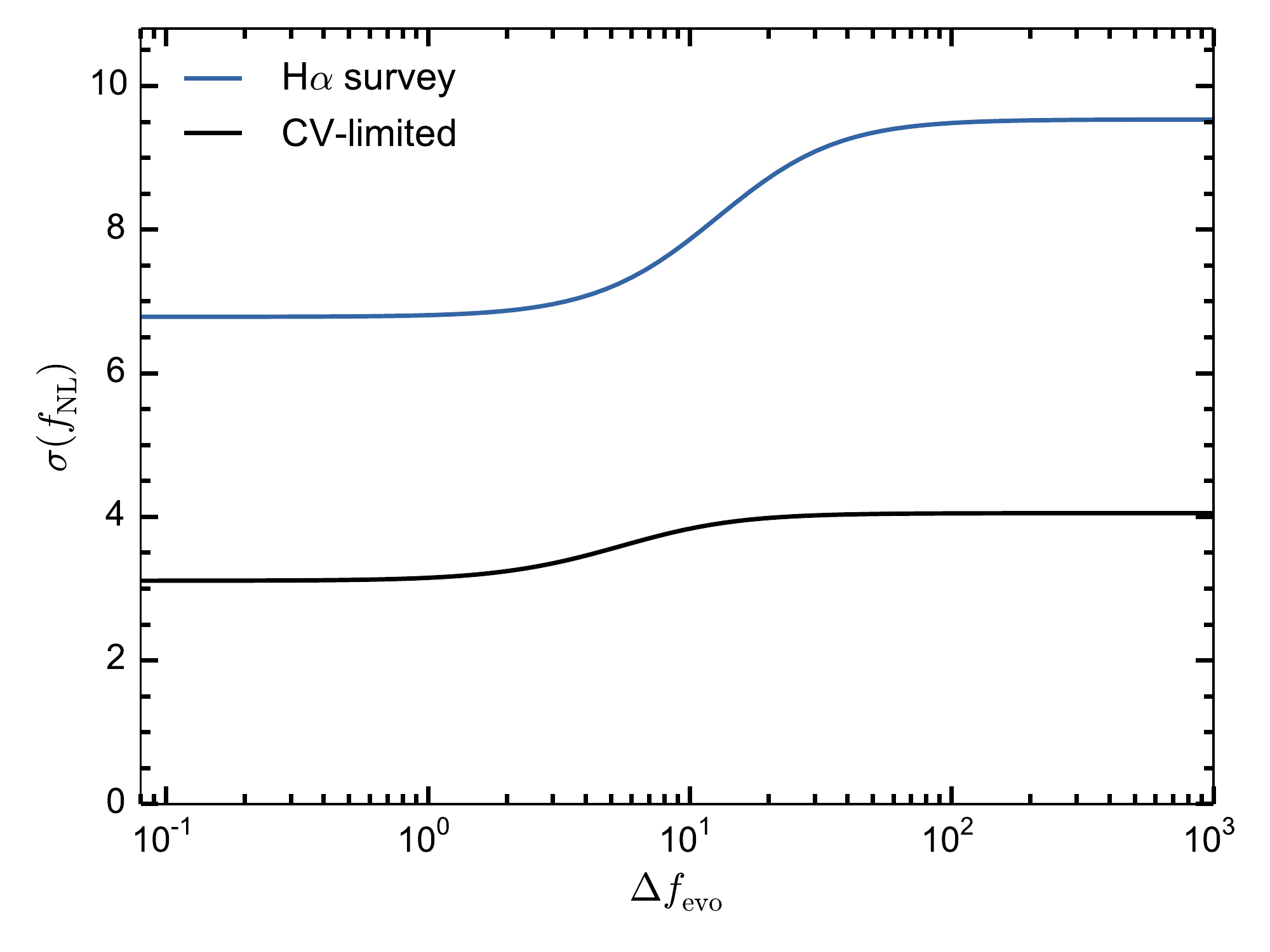}
 \caption{1D marginal errors on $\fnl$ for a H$\alpha$ spectroscopic survey as 
          a function of the prior on $f_{\rm evo}$. A Planck prior on the 
          cosmological parameters is also included.}\vspace{1em}
 \label{fig:spectro_fevo}
\end{figure}

%-------------------------------------------------------------------------------
\subsubsection{Constraints on primordial non-Gaussianity}
 \label{sec:spectroz:fnl}

As with the previous two probes, our $\fnl$ forecasts include a Planck CMB 
prior and priors on the bias functions ($\Delta b / b = 0.1$, $\Delta s = 1$, 
$\Delta f_{\rm evo} = 1$). The results are shown in Table 
\ref{tbl:constraints_summary} for the $\Delta z = 0.025$ redshift binning.

The forecast constraint from the H$\alpha$ survey is $\sigma(\fnl) = 6.8$, which 
is worse than the intensity mapping survey by a factor of $\sim 2$. While the 
H$\alpha$ survey has a consistently higher bias (which enhances the non-Gaussian 
bias signal, $\propto b - 1$), it covers a narrower redshift range and smaller 
area than the IM survey, so ultimately loses out when the higher-redshift bins 
of the IM survey are taken into account (see Fig.~\ref{fig:sfnl_z}, below). The 
difference in performance remains in the CV-limited case, again mostly due to 
the wider redshift range of the IM survey.

The H$\alpha$ constraint degrades only slightly to $\sigma(\fnl) = 7.2$ for the 
widest redshift binning, $\Delta z = 0.1$. Similarly, it is only weakly 
sensitive to $\ell_{\rm max}$, improving from $\sigma(\fnl) = 7.2$ for 
$\ell_{\rm max} = 200$ to $6.6$ for $\ell_{\rm max} = 1000$ (both for 
$\Delta z = 0.025$). The addition of significantly more small-scale information 
in both the radial and transverse directions is therefore only mildly 
beneficial.

As with the IM survey, there is a reasonably strong degeneracy between $\fnl$ 
and $f_{\rm evo}$, predominantly for the highest-redshift nuisance parameter 
bin. Fig.~\ref{fig:spectro_fevo} shows the effect of changing the prior on 
$f_{\rm evo}$ -- an $\mathcal{O}(1)$ prior is sufficient to completely break 
the degeneracy. The results are insensitive to the prior on the magnification 
bias, and there is no gain to be had from tightening the bias prior until a 
very low level of $\Delta b / b \lesssim 1\%$ is reached.

%-------------------------------------------------------------------------------
\subsubsection{Systematic uncertainties}
\label{sec:spectro:systematics}

Spectroscopic surveys are prone to systematic effects on large angular scales. 
Redshift surveys commonly consist of samples from several non-contiguous 
fields, surveyed during different observing seasons and possibly even with 
different instruments, which can make it tricky to patch them together into a 
single coherent survey volume. They may also suffer from the problem of not 
having a homogeneous magnitude limit, i.e. the magnitude cuts vary, and cannot 
easily be mapped onto a substantial and complete 3D volume of the sky.

\begin{figure}[t]
 \centering
 \vspace{-0.4em}\includegraphics[width=1.04\columnwidth]{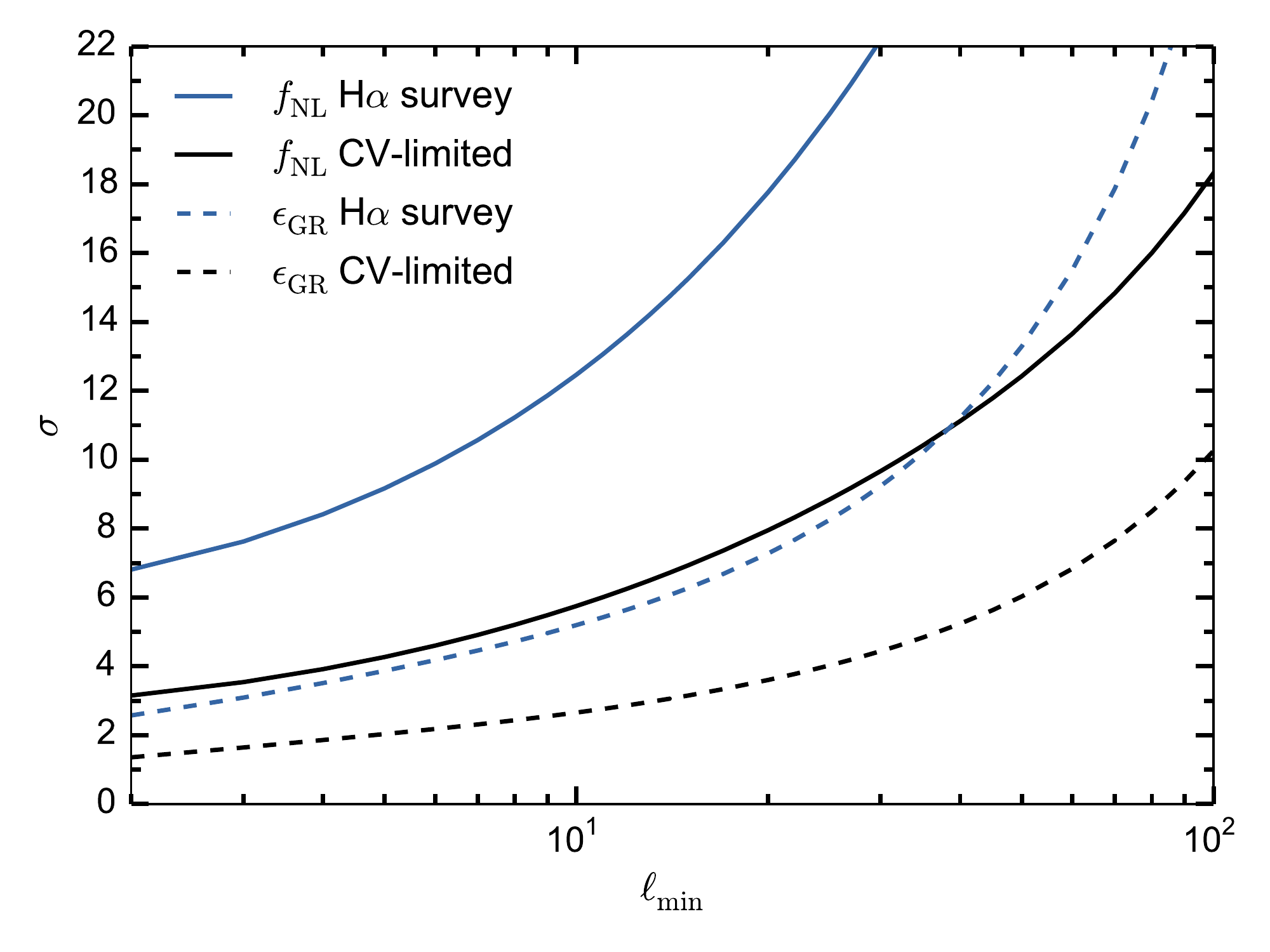}
 \caption{1D marginal errors on $\fnl$ and $\egr$ as a function of the minimum 
          usable spherical harmonic mode, $\ell_{\rm min}$, for the H$\alpha$ 
          survey $(\Delta z = 0.025)$. For the $\fnl$ results, we have added a 
          Planck prior to break degeneracies with other cosmological 
          parameters, as well as priors on the bias functions 
          ($\Delta b / b = 0.1$, $\Delta s = 1$, $\Delta f_{\rm evo} = 1$).}\vspace{1em}
 \label{fig:spectro_lmin_lmax}
\end{figure}

Galactic extinction is a dominant source of systematic error on large angular 
scales. Dust in our galaxy changes the overall true flux cut of the survey, and 
introduces number density fluctuations that vary with the shape of the Galaxy. 
This must be corrected for (e.g. by fitting an extinction template) to avoid 
biasing the inferred large-scale power. Variations in airmass and seeing also 
affect the number of photons reaching the detector in a way that is correlated 
with the elevation of the telescope. This induces additional dispersion in the 
magnitude of the measured galaxies, and the ability to distinguish them from 
stars.

Stars themselves are problematic. For example, a small fraction of the observed 
sources may in fact be misidentified stars that contaminate the galaxy sample. 
Stars also obscure regions of the sky of order the size of the point-spread 
function, which reduces the observed density of galaxies nearby. While this 
should be a small effect for an individual star, which will mask an area of 
$\sim 10^{-6}$ of a degree, the total obscured area can be significant given 
the large stellar density (which grows towards the galactic plane). In fact, 
stellar contamination was found to be a dominant source of systematic error in 
the recent analysis of the BOSS data \citep{Ross:2011cz}, where it introduced a 
significant bias in the measurement of the correlation function on large 
angular scales if left uncorrected. This bias was well above the statistical 
uncertainty, to the extent that the correlation functions measured in the 
Northern and Southern Galactic hemispheres of the survey were inconsistent 
with one other.

While many of these effects can lead to fluctuations in the number density of 
galaxies as a function of redshift, inducing systematics along the radial 
direction, the dominant effect is on large angular scales, reducing the 
effective area of the survey and hampering accurate recovery of the lowest 
$\ell$ modes. Fig.~\ref{fig:spectro_lmin_lmax} shows how the forecast 
constraints on $\egr$ and $\fnl$ depend on the minimum recoverable $\ell$ mode 
of the survey. There is a rapid loss of information on both parameters as 
$\ell_{\rm min}$ increases, with $\sigma(\fnl)$ doubling from $6.8$ for 
$\ell_{\rm min} = 2$ to around $13$ at $\ell_{\rm min} = 10$. The degradation 
is similar for $\sigma(\egr)$, which also doubles in the same range. As such, 
future spectroscopic surveys will likely need excellent control over large 
scale systematics if they are to be used to constrain $\fnl$.

%-------------------------------------------------------------------------------
\subsection{Photometric redshift surveys}
\label{sec:photometric}

One of the main drawbacks of spectroscopic surveys is the long integration 
times needed to resolve galaxy spectra sufficiently well to yield a good 
redshift estimate. Because of this, the number of targets selected for 
spectroscopic follow-up is usually much smaller than the total imaged sample, 
which significantly limits the survey depth and number density that can be 
achieved.

In a photometric redshift survey, each galaxy that is detected with a 
sufficiently high signal-to-noise is imaged in a small number of wide frequency 
bands. This provides a very coarse measurement of the galaxy's spectrum, 
convolved with the bandpass of each band, which can be used to statistically 
infer its redshift. These photometric redshifts (usually abbreviated 
``photo-$z$'s'') have much larger uncertainties than their spectroscopic 
counterparts, and most of the information about gravitational clustering on 
radial scales is lost. Photo-$z$s can be recovered for much fainter galaxies 
than spectroscopic redshifts, however, and so photometric surveys have the 
ability to cover significantly larger volumes. This potentially makes them more 
suitable for constraining cosmological observables on ultra-large scales.

The first wide-area, deep photometric surveys are already underway 
\citep{2002SPIE.4836..154K,2005astro.ph.10346T}. Their results will pave the 
way for the Large Synoptic Survey Telescope \citep[LSST,][]{2009arXiv0912.0201L}, 
which will surpass them in terms of area, depth, and angular resolution 
(although the latter is not a critical factor for this work). We have thus 
chosen to produce Fisher forecasts for LSST as the best photometric survey that 
we will have access to in the foreseeable future.

We have assumed that the LSST will observe two separate galaxy populations:
early-type (``red'') galaxies, and late-type (``blue'') galaxies. Although this 
is a simplistic picture, it allows us to study the effect that different 
properties of the sample will have on the final constraints:
\begin{itemize}[leftmargin=1em]
 \item Early-type galaxies form preferentially in high-density regions, and are 
       associated with high-mass haloes. They are therefore more highly biased 
       than blue galaxies, which is desirable for measuring $\fnl$.
       
       The number density of red galaxies decays very fast beyond redshift 
       $z\sim1$. On the one hand, the drop should be associated with a larger 
       evolution bias, which could enhance the amplitude of the relativistic 
       terms. On the other, it limits the largest scales that the red sample 
       can probe.
       Finally, the spectra of red galaxies show prominent features, most 
       importantly the 4000\AA~ Balmer break. These features are easy to 
       locate, even using only photometric information, and therefore 
       photo-$z$'s for red galaxies are more accurate on average.
       
 \item Blue galaxies are found in lower density regions and correspond to 
       lower-mass haloes. They are therefore more faithful tracers of the 
       total matter density field, and have a lower bias, which impacts their 
       usefulness for measuring $\fnl$. LSST should be able to observe a 
       significant number of blue galaxies up to much higher redshifts than the 
       red population ($z\sim3$) however, so their large-scale clustering 
       properties can be studied more accurately. Finally, photometric 
       redshifts for blue galaxies will be more uncertain than those for red 
       galaxies, as discussed above.
\end{itemize}
We will provide forecasts for two samples: a sample of red galaxies only, 
labeled ``red'', and a sample containing all of the galaxies observed by LSST 
(red $+$ blue), which we will call the ``full'' sample. Although LSST will be 
able to detect galaxies down to a magnitude limit of 27.5 in the $r$-band, it 
is not clear that the photometric redshift requirements will be satisfied for 
this survey depth. We have therefore adopted a more stringent magnitude cut of 
$i<25.3$, corresponding to the so-called LSST ``gold'' sample 
\citep{2009arXiv0912.0201L}. We have further assumed that LSST will cover the 
whole southern hemisphere ($f_{\rm sky}=0.5$).

%-------------------------------------------------------------------------------
\subsubsection{Constraints on relativistic effects}

As with the previous experiments, we start by exploring the possibility of 
detecting the contribution of the relativistic terms to the clustering of LSST 
galaxies in the best-case scenario, by marginalizing over only $\fnl$ whilst 
keeping all other cosmological parameters fixed to their fiducial values. Two 
main differences with respect to the previous tracers give some hope for 
detecting $\egr$ with LSST. First of all, the sharp decay in the number density 
of red galaxies can enhance the amplitude of the relativistic terms thanks to 
the large value of $\fevo$. Also, LSST covers a wider survey area and redshift 
range than spectroscopic surveys, so has access to larger scales.

\begin{figure}
 \centering
 \includegraphics[width=0.49\textwidth]{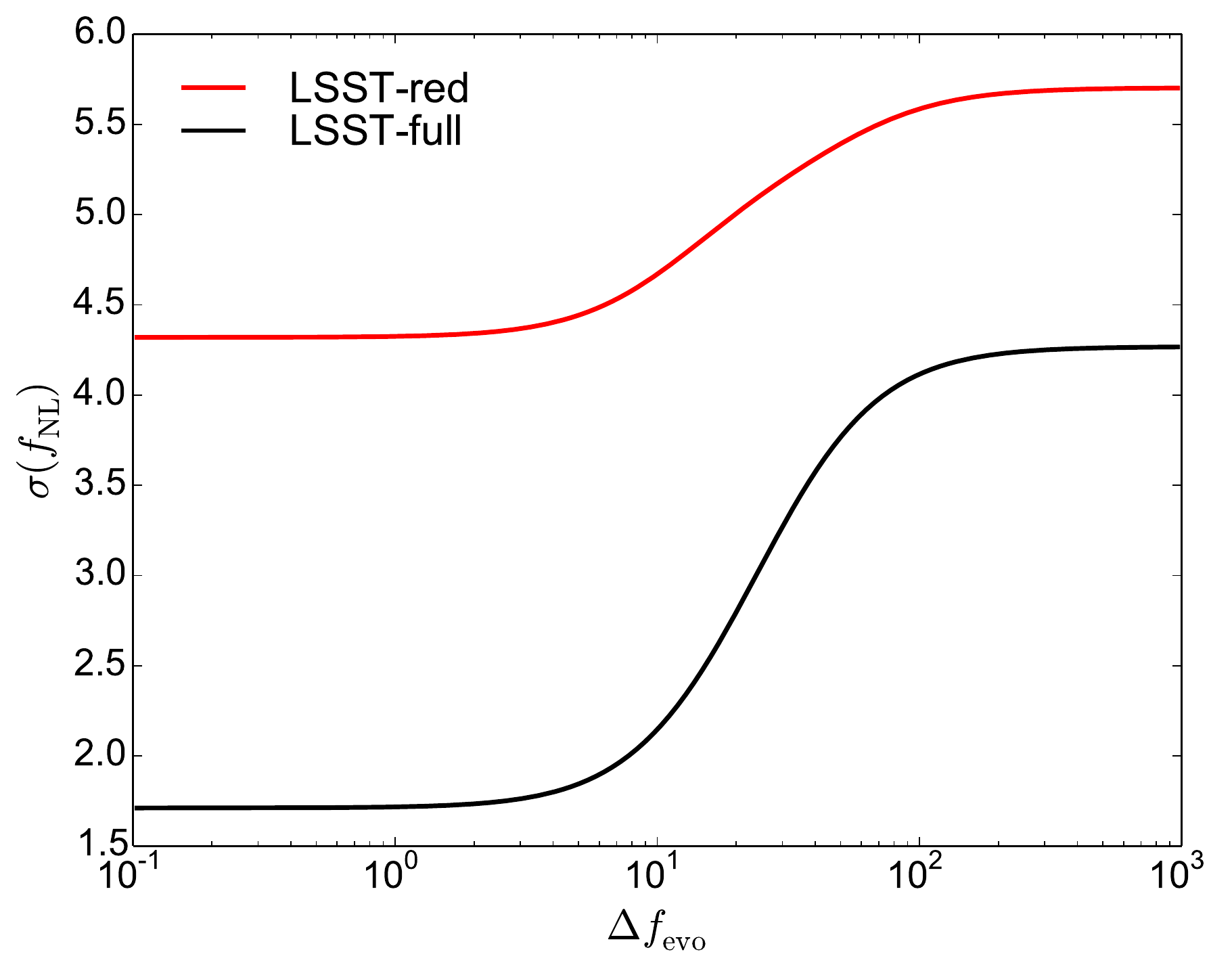}
 \caption{Uncertainty on $\fnl$ as a function of the size of the prior assumed 
          for $\fevo$ for the ``red'' and ``full'' samples.}
 \label{fig:sfnl_vs_sfevo_lsst}
\end{figure}

The results are summarized in Table \ref{tbl:constraints_summary}. Even though 
the higher value of $\fevo$ for red galaxies helps to decrease the forecast 
uncertainty on $\egr$, it is still impossible to detect relativistic effects 
using either sample; $\sigma(\egr)= 1.4$ and $2.3$ for the red and full samples 
respectively. The red sample nevertheless produces the best constraint on 
$\egr$ of any of the surveys considered above.

As before, we predict the detectability of the large-scale magnification
lensing by marginalizing only over $\fnl$ and using a small-scale cutoff
$\ell_{\rm max}=100$. Our results (see Table \ref{tbl:constraints_summary})
show that this effect should be clearly detectable (well above $5\sigma$)
for both the ``red'' and ``full'' samples.

%-------------------------------------------------------------------------------
\subsubsection{Constraints on primordial non-Gaussianity}

Even though photometric redshifts erase most of the clustering signal on all 
but the largest radial scales, they are still sufficient to enable a 
tomographic analysis of galaxy clustering to be performed. This helps immensely 
in breaking many of the degeneracies reported for continuum surveys (which 
retain essentially no radial information).

We studied the importance of breaking these degeneracies by again calculating 
$\sigma(\fnl)$ as a function of the priors on the bias parameters, finding that 
the forecasted uncertainty is almost completely insensitive to any priors on 
the clustering and magnification biases, $b(z)$ and $s(z)$. For the reasons 
outlined in Section \ref{sssec:fnl_im}, this behaviour does not follow for the 
evolution bias, so it is useful to explore the prior constraints on $\fevo$ 
that are required in order to optimize the measurement of $\fnl$. Figure 
\ref{fig:sfnl_vs_sfevo_lsst} shows the dependence of $\sigma(\fnl)$ on a 
constant Gaussian prior imposed on $\fevo$. The degeneracy between the 
parameters can be largely mitigated by measuring the evolution bias with an 
accuracy of $\Delta\fevo\lesssim1$.

As in the previous cases, we produced our final forecasts for $\fnl$ by 
assuming Planck CMB priors for the cosmological parameters, a $10\%$ 
uncertainty on the clustering bias, and priors of $\Delta s=1$ and 
$\Delta\fevo=1$. The final results are summarised in Table 
\ref{tbl:constraints_summary}. LSST should be able to impose very tight
constraints of $\sigma(\fnl)\simeq1.7$ using galaxy clustering autocorrelations 
(single-tracer) alone.

%-------------------------------------------------------------------------------
\subsubsection{Systematic uncertainties}

\begin{figure}
 \centering
 \includegraphics[width=0.49\textwidth]{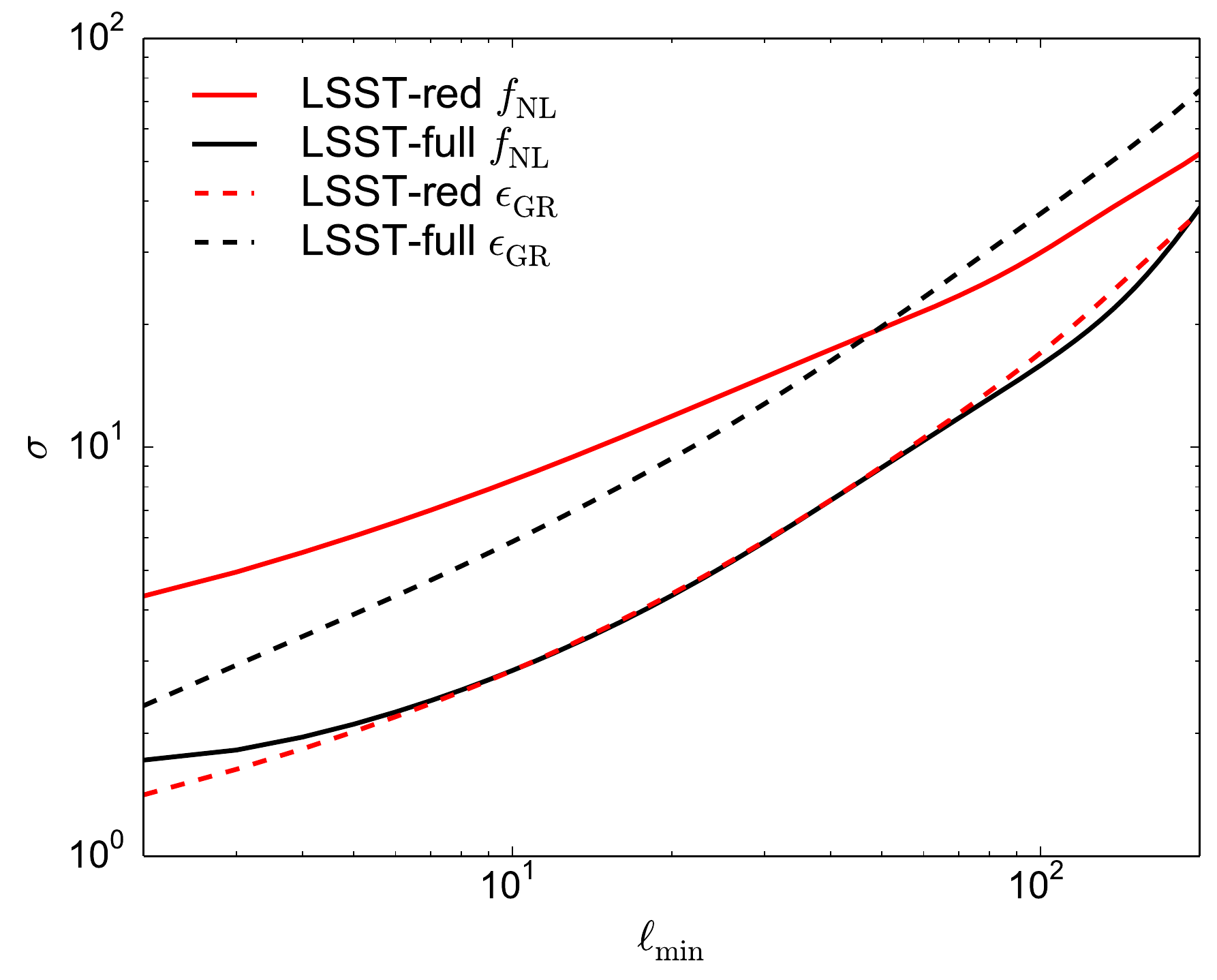}
 \caption{Dependence of the uncertainty on $\fnl$ (solid lines) and $\egr$ 
          (dashed lines) on the minimum scale probed by LSST for the two 
          samples considered here.}
 \label{fig:sigma_llim_lsst}
\end{figure}

Most of the sources of systematics that affect photometric redshift surveys are 
exactly the same as for their spectroscopic counterparts: galactic extinction, 
variations in sky brightness, seeing, and stellar contamination (both due to 
stars affecting the local observed number density of galaxies, and stars 
erroneously being included in the galaxy sample). All of these effects can 
potentially contaminate the signal measured on large angular scales. Figure 
\ref{fig:sigma_llim_lsst} shows the degradation in the constraints on $\fnl$ 
when the largest scales are omitted in the analysis. Limiting ourselves to 
scales $\ell \geq 10$ would increase our best-case error bars by over 50\%, to 
$\sigma(\fnl)\sim 2.8$.

The use of photometric redshifts also carries its own systematic effects. In 
order to obtain a reliable estimate of the power spectrum of the galaxy density 
field that we can use to constrain large scales, it is necessary to have a 
sufficiently accurate model of the window function for every redshift bin. 
Unless a spectroscopic subsample drawn from the same distribution as the 
photometric one is available, which is rarely the case, deriving a correct 
model for the true redshift distribution $\bar{N}(z)$ is a challenging task, 
although it has been noted that this issue could be ameliorated by 
cross-correlating the photometric sample with any spectroscopic survey 
\citep{2008ApJ...684...88N}. The presence of photometric redshift outliers can 
also modify the tails of the photo-$z$ distribution, which affects the shape of 
the redshift window functions. The level to which this effect is problematic 
will depend on how accurately the photo-$z$ pdf can be characterized.

%-------------------------------------------------------------------------------
\section{Discussion}
\label{sec:discussion}
\begin{table*}[t]
\centering{
{\renewcommand{\arraystretch}{1.6}
 \begin{tabular}{|p{3cm}|p{2.3cm}|c|c|c|}
 \hline
 {\bf Experiment type}& {\bf Experiment}     & $\sigma(\fnl)$ & $\sigma(\egr)$ & $\sigma(\ewl)$ \\
 \hline
 Intensity mapping    & SKA1-MID             & ~~~3.01~~~     & ~~~2.75~~~     & ---            \\
                      & ~~ (w. $1.5\times$ bias) & 0.90       & 1.90           & ---            \\
                      & CV-limited           & 1.68           & 1.97           & ---            \\
 \hline
 Continuum survey     & $S_{\rm cut}=10\uJy$ & 18.5           & 26.7           & 1.90           \\
                      & $S_{\rm cut}=5\uJy$  & 16.0           & 24.6           & 0.57           \\
                      & $S_{\rm cut}=1\uJy$  & 11.8           & 17.1           & 0.25           \\
 \hline
 Spectroscopic survey & H$\alpha$ survey     & 6.64           & 2.57           &  0.19 \\
                      & CV-limited           & 3.02           & 1.35           &  0.10 \\
 \hline
 Photometric survey   & LSST-red             & 4.32           & 1.41           & 0.14           \\
                      & LSST-full            & 1.71           & 2.33           & 0.04           \\
 \hline
 \end{tabular}} }
 \caption{Forecasted constraints on $\fnl$, $\egr$, and $\ewl$ for the different experiments
          explored in this work.}
 \label{tbl:constraints_summary}
\end{table*}

It has been argued that general relativistic corrections to the number density 
of galaxies should be observable with future cosmological surveys. In 
particular, ultra large-scale features in the power spectrum of density 
fluctuations could in principle be detectable with deep and wide surveys such 
as those that will be carried out by the next generation of experiments. In 
this paper we set out to systematically examine this claim for the most relevant
surveys planned for the next decade or so. At the same time, we performed 
forecasts for the expected constraints on the scale-dependent bias that arises 
from primordial non-Gaussianity, another key effect on ultra-large scales. Our
analysis uses a more rigorous formalism than is usually followed, based on
computing all possible angular cross-correlations between different redshift bins
instead of using an approximate 3-D power spectrum approach, and avoiding the flat-sky
and Limber approximations.

Our main conclusion is that, if we restrict ourselves to the single-tracer power 
spectrum of the density fluctuations, in either two or three dimensions,{\it all
previously-undetected contributions to the power spectrum of source number counts are
completely unobservable}. Note that we have labelled these terms as `GR effects' in
this work, thus excluding the lensing magnification term, which we have treated
separately due to its very different properties (see below).

In hindsight, this result is not surprising.\footnote{See e.g. \cite{2011PhRvD..84d3516C,
Jeong:2011as, Yoo:2012se} for similar statements using less quantitative analyses.} A number
of papers have previously attempted to forecast the optimal constraints on $f_{\rm NL}$ from a 
variety of surveys, with the general conclusion being that, at best, one can 
detect a value of $f_{\rm NL}\sim 1-2$ at one sigma. Our work has confirmed these 
results. Given that $\epsilon_{\rm GR}$ has a similar (although not identical) 
effect on the power spectrum as $\fnl \sim 1$, we expect the same level of 
sensitivity to relativistic effects. If we are to aspire to a statistically 
significant detection of the relativistic effects, we would need a sensitivity 
of $\sigma(\epsilon_{\rm GR})\sim 0.1-0.2$, which is clearly unachievable with 
any of the single-tracer survey techniques and strategies analysed here.

It is interesting to look at each of the survey techniques in turn to see why 
they fall short of our desired target:
\begin{itemize}[leftmargin=1em]
 \item On the face of it, \emph{intensity mapping} is a particularly promising 
 approach to efficiently surveying large volumes of the Universe -- it can 
 simultaneously produce very deep surveys and cover large areas. One would 
 expect this to be ideal for constraining both $\egr$ and $\fnl$. Relative to 
 the other techniques, intensity mapping suffers from the fact that one of the 
 substantial corrections on large scales -- the perturbation to angular 
 distances -- is absent. As was shown in Section \ref{ssec:th_egr}, because 
 one is measuring an intensity rather than source number density, there is 
 an exact cancellation of the lensing contributions to both the number density 
 and angular diameter distance corrections. This has a significant impact on 
 the size of the relativistic correction signal, and thus the detectability of 
 $\egr$.
 
 Intensity mapping can be used to obtain reasonably tight constraints on 
 $\fnl$, although there too it is placed at a disadvantage by the ``Gaussian''
 clustering bias, $b^G$, for neutral hydrogen being smaller than for other 
 types of probe. Note that while most halo-based models predict a low HI bias, 
 measurements of the clustering of damped Lyman-$\alpha$ systems carried out by 
 \cite{2012JCAP...11..059F} point towards a significantly higher value at 
 $z \gtrsim 2$. If we multiply our fiducial bias function by a factor of 
 $1.5\times$ (so that $b \approx 2$ at $z=2.2$), we obtain a significantly 
 better constraint of $\sigma(\fnl) = 0.9$ for an IM survey going out to 
 $z=3.5$. Our forecasts for $\fnl$ are very sensitive to the fiducial bias 
 model, and therefore could change significantly with better empirical 
 measurements.
 
 We must also point out that the survey specifications we assumed are such 
 that the measurements of the power spectra are only cosmic variance-limited on 
 the very largest scales. It might be possible to improve the constraints on 
 $\fnl$ (by up to 30\% in the most optimistic case) by reducing the noise (e.g. 
 by increasing the survey time).

 \item \emph{Radio continuum surveys}, while efficient at accessing large 
 volumes, are remarkably poor at constraining both $\epsilon_{\rm GR}$ and 
 $f_{\rm NL}$ in a single-tracer context. The loss of all radial information
 through projection over a wide redshift range has a significant toll on their 
 ability to discriminate between different effects in the power spectrum. If we 
 are to benefit from continuum surveys, a more sophisticated multi-tracer 
 approach is needed, as we will discuss below.

 \item We also looked at the ability of a \emph{spectroscopic survey} to 
 constrain large-scale features in the power spectra. The advantage of this 
 type of survey is ostensibly their ability to obtain high-resolution 
 measurements in both the angular and radial directions -- more so than in any 
 other type of survey considered here, although in practice this comes at a 
 price. Spectroscopic surveys are time-intensive, and so surveys are limited to 
 smaller areas of the sky and shallower depths. The average number density of 
 (usable) galaxies is also lower than for other surveys, and hence the 
 effective noise on large angular scales can be more substantial. As was the 
 case with intensity mapping, our forecasts for $\fnl$ depend crucially on the 
 fiducial clustering bias, which for H$\alpha$ emitters is also relatively low.

 \item Deep and wide \emph{photometric surveys} seem to be the most effective 
 method for probing large scales, for the parameters we considered here. The 
 loss of resolution along the radial direction (as compared to a spectroscopic 
 survey) is compensated by the significantly wider redshift coverage and larger 
 survey area. Photometric surveys are also a more rapid way of counting 
 galaxies, so source number densities are higher than for their spectroscopic 
 surveys contemporaries. Finally, the clustering bias for the sources that LSST 
 will observe is significantly higher than for HI and H$\alpha$ galaxies, 
 boosting its ability to detect $f_{\rm NL}$ substantially.
\end{itemize}

\begin{figure}[t]
 \centering
 \hspace{-0.5em}\includegraphics[width=0.49\textwidth]{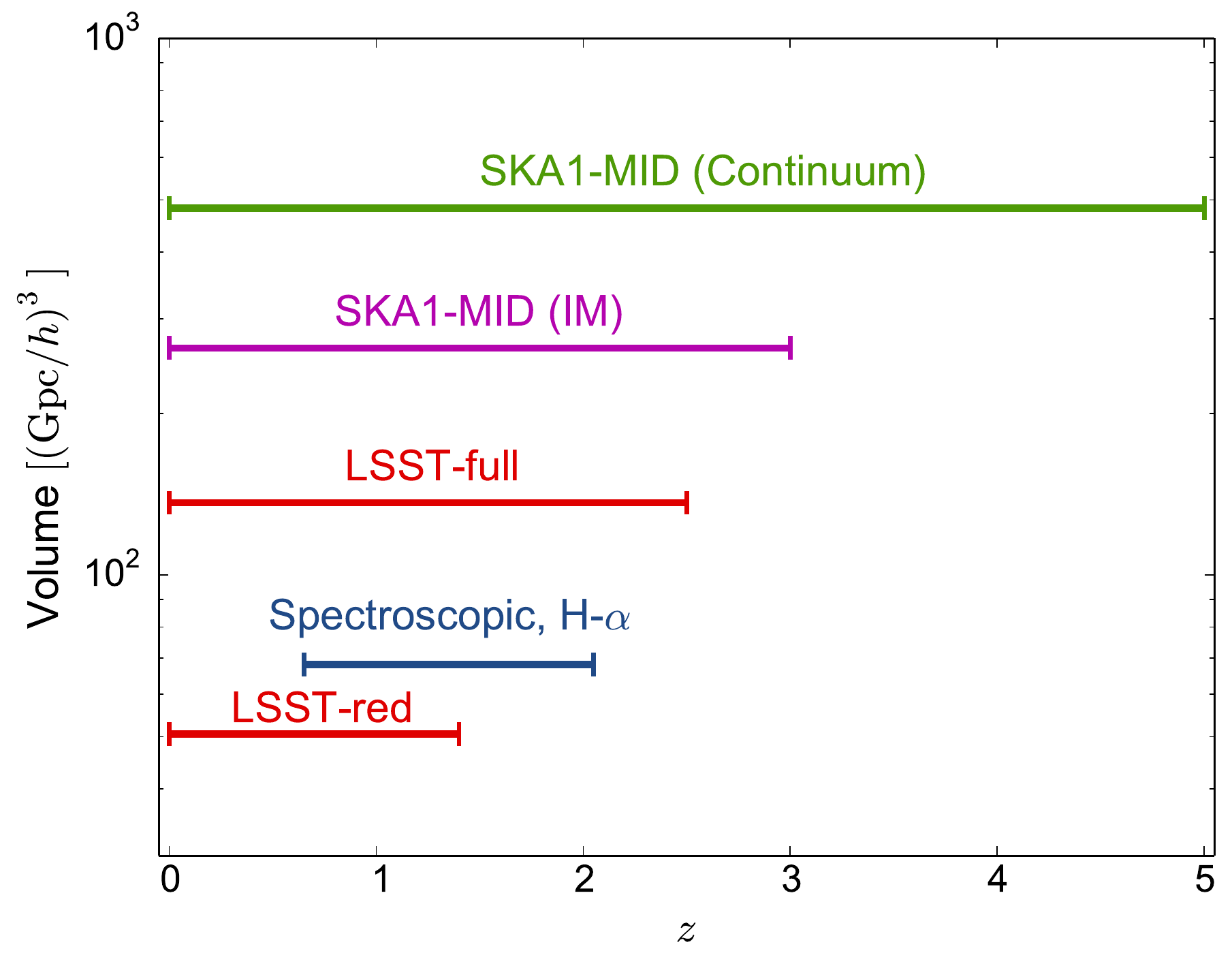}
 \caption{Comoving volume and redshift range covered by the experiments considered
          in this work.}
 \label{fig:volumes}
\end{figure}

Figure \ref{fig:volumes} compares the redshift range and comoving volume probed 
by the various experiments studied here. Measuring ultra large-scale 
observables depends critically on the ability to cover very large volumes, but 
we have seen here that this is far from the only factor. Even though an SKA1 
continuum survey should be able to access the largest volume of any of the 
surveys, its inability to use radial information prevents it from achieving a 
competitive measurement of $\fnl$ without the use of the multi-tracer 
technique.

\begin{figure}
 \centering
 \includegraphics[width=0.49\textwidth]{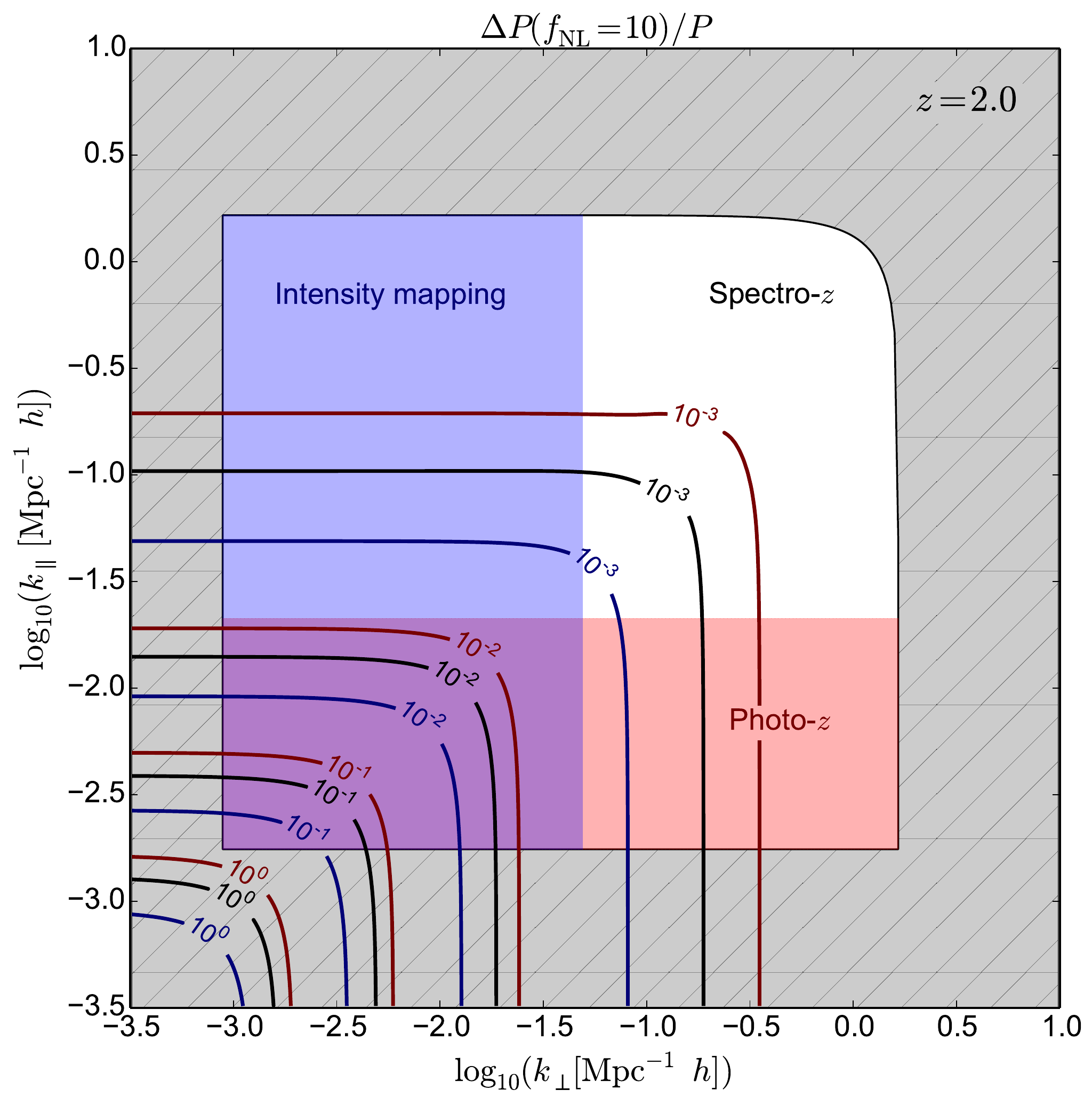}
 \caption{Regions in the space of transverse and parallel wavenumbers 
          ($k_\perp,\,k_\parallel$) accessible for various surveys.
          {\sl Red:} region accessible to a photometric redshift survey with 
          $\sigma_0=0.05$. The large $k_\parallel$-regime is lost due to the 
          inaccuracy of photo-$z$'s.
          {\sl Blue:} region accessible by an SKA1-MID IM survey in single-dish 
          mode ($D_{\rm dish}=15{\rm m}$). Small transverse scales are 
          inaccessible due to the beam width.
          {\sl White (containing red and blue regions):} region accessible to a 
          full-sky spectroscopic survey covering up to $z=2$. \protect\\
          {\sl Contours:} the relative amplitude of the contribution to the 
          three-dimensional power spectrum due to primordial non-Gaussianity, 
          with $\fnl=10$, for these three experiments, in black 
          (spectroscopic), red (photometric), and blue (intensity mapping). 
          Note that this amplitude depends on the clustering bias of each
          tracer.}
 \label{fig:k_regions}
\end{figure}

It is therefore also relevant to compare the range of radial and angular scales 
covered by each experiment. We do this in Figure \ref{fig:k_regions} for 
intensity mapping, photometric, and spectroscopic surveys covering redshifts 
$z<2$. The grey hashed region corresponds to inaccessible scales, either 
larger than the survey volume or smaller than the non-linear scale. While 
spectroscopic surveys have essentially complete access to the 
$k_\parallel-k_\perp$ plane, photometric surveys and intensity mapping are 
limited to only large radial or angular scales respectively.

This is not necessarily an important inconvenience for ultra large-scale 
observables. The coloured contours show the relative amplitude of the $\fnl$ 
signal for each of these surveys (spectroscopic in black, photometric in red 
and IM in blue) for $\fnl=10$. In the large-$k$ region that only spectroscopic 
surveys have access to, the amplitude of the signal is relatively low. Note 
also that radial and angular scales should not be treated equally in this plot; 
simply by a dimensionality argument (two angular dimensions vs. one radial), 
better angular coverage will usually be more advantageous.

\begin{figure}[t]
 %\centering
 \hspace{-1em}
 \includegraphics[width=0.49\textwidth]{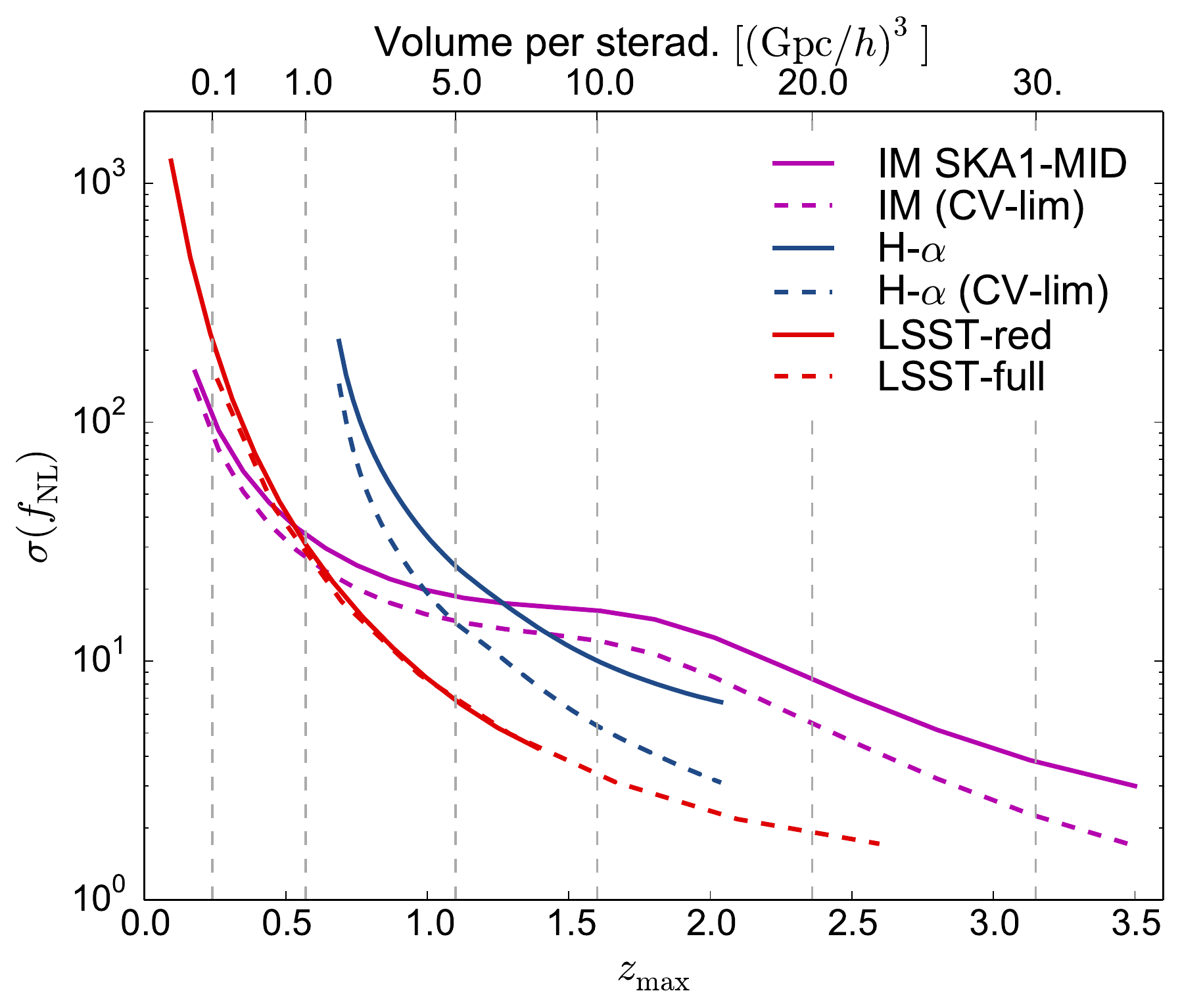}
 \caption{Constraints on $\fnl$ as a function of the maximum redshift covered
          by each of the surveys considered here. The comoving volume per
          steradian at each $z$ is indicated on the upper $x$-axis.}
 \label{fig:sfnl_z}
\end{figure}

Finally, our ability to measure any signal depends critically on its amplitude. 
In the case of primordial non-Gaussianity, this depends on the sample's 
clustering bias, since $\Delta b_{\rm NG} \propto (b - 1)$. This dependence can 
be observed in Figure \ref{fig:sfnl_z}, where we show the forecasted 
uncertainty on $\fnl$ as a function of the maximum redshift covered by each 
survey. Even though intensity mapping has the potential to cover the largest 
volume while also preserving radial information, the improvement of 
$\sigma(\fnl)$ with $z$ is significantly slower, in particular in the range 
$z\sim1-1.5$, where the HI bias is very close to unity.

We emphasise that all of our forecasts depend on a variety of astrophysical 
model assumptions. By this we mean that we have had to predict the number 
densities and biases for the surveys as a function of redshift, based on 
existing data and simulations. Getting these assumptions correct is key for 
accurate forecasting, and substantial work will have to be done -- for all 
types of surveys -- to better estimate these quantities. Nevertheless, we are 
confident that our models are sufficiently representative that our broad point 
is correct, and that the numbers we present here give a fair representation of 
what to expect from future surveys. On this point it is also worth noting that
it is in principle possible to further constrain the level of primordial
non-Gaussianity from measurements of higher-order correlations. Such measurements
are extremely challenging at present due to the large theoretical uncertainties
(e.g. in the form of the bias in the presence primordial non-Gaussianity,
the form of the bispectrum and its covariance in redshift space, the value of the
non-linear clustering bias etc.). Further studies of the three-point function might
eventually make such measurements possible, possibly superseding the forecasts
presented here.

Conservatively, we did not include the lensing magnification contribution to 
the power spectrum as one of the effects parametrized by $\egr$, even though it 
is a fully relativistic effect. This is because magnification has a significant 
amplitude on sub-horizon scales, and has in fact already been detected by 
cross-correlating pairs of distant tracers. This is qualitatively 
different to the situation for the other relativistic terms, which are 
significant only on ultra-large scales, and which have not been measured yet
in large-scale structure. Figure~\ref{fig:cl_terms} indicates that the lensing 
magnification contribution to the power spectrum can dominate the GR terms 
parametrized by $\egr$ at ultra-large scales, however. This large-scale 
contribution has not been detected in current small-volume surveys, but 
previous work has forecasted for its detectability \citep{2009PhRvD..79b3517Y,
2011PhRvD..83l3514N,2011MNRAS.415.3485Y,Yang:2013ceb,2015arXiv150601369M}. By
defining a parameter, $\ewl$, corresponding to the amplitude of the magnification
lensing term alone, we have quantitatively verified its detectability, showing 
that the large-scale lensing contribution should be detectable above $5\sigma$ 
in all relevant experiments. It has also been shown by
\cite{2011PhRvD..83l3514N,2012PhRvD..85b3511L} and \cite{2014arXiv1412.4781C}
that  omission of the lensing $+$ GR terms leads to a
bias in the recovered value of $f_{\rm NL}$. This was predicted to be at the
$\sim\! 3\sigma$ level for an SKA Phase 2 HI spectroscopic survey
\citep{2014arXiv1412.4781C}. Our analysis shows that $\egr$ cannot be responsible
for this bias, leading to the conclusion that it must be almost exclusively due 
to the ultra-large scale lensing term.

\begin{figure}[t]
 %\centering
 \vspace{1.7em}
 \hspace{-2em}\includegraphics[width=0.54\textwidth]{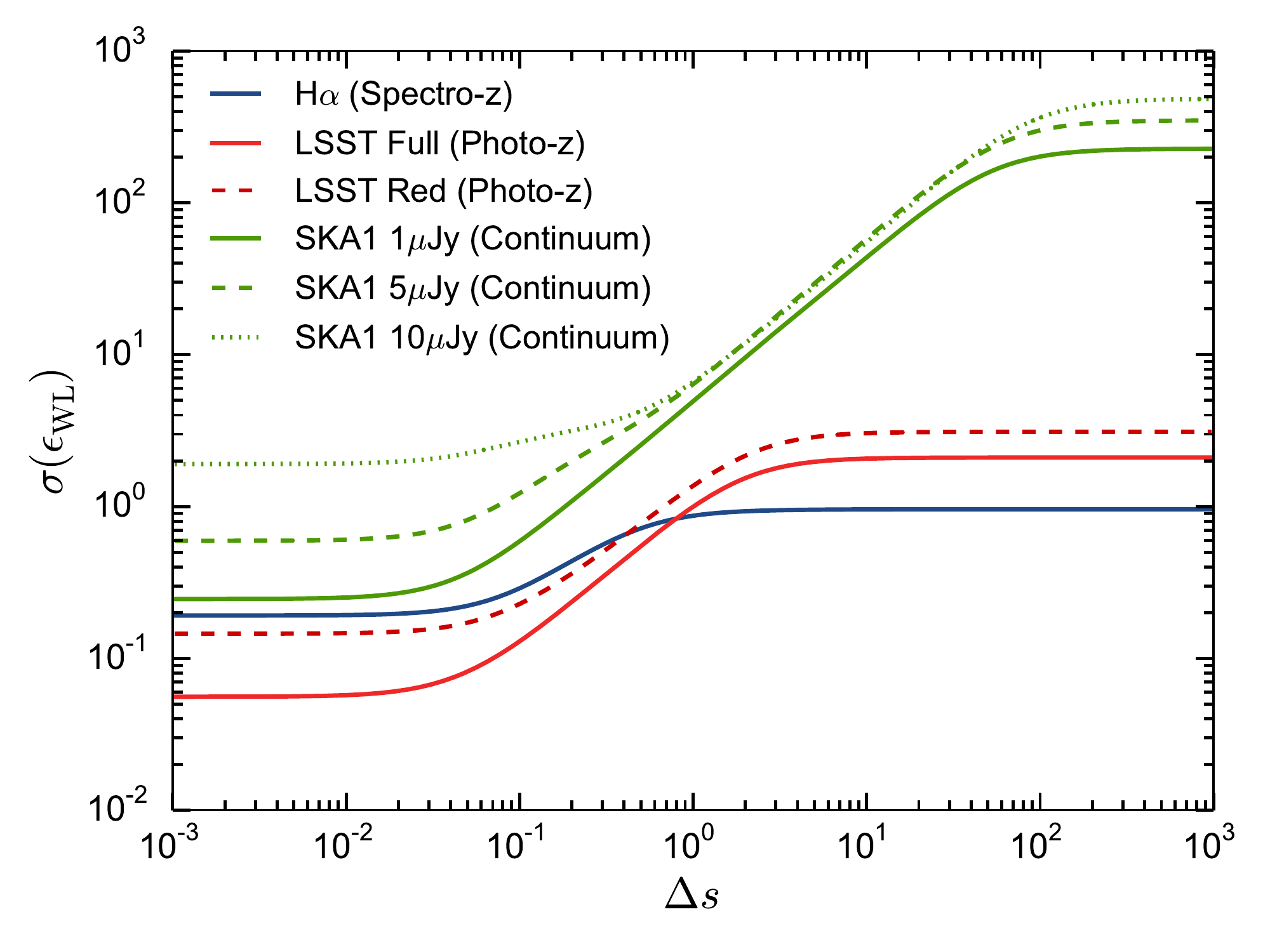}
 \caption{Constraints on $\ewl$ as a function of the prior on the magnification 
          bias. The same parameters, and priors for the other bias functions, 
          were used as in the $\fnl$ analysis, except for the continuum 
          survey, where the clustering bias was not marginalized.}
 \label{fig:ewl_sbias}
\end{figure}

We also tested the consistency of our results with the recent work by 
\cite{2015arXiv150601369M}, who find $\sim 1\%$ constraints on 
$\epsilon_{\rm WL}$ (their parameter $\beta$) for a Euclid photometric survey.
Marginalising over the same set of cosmological parameters as them (except the 
neutrino mass, $m_\nu$), and choosing $\ell_{\rm max} = 1000$, we obtain 
$\sigma(\ewl) = 0.06$ for the H$\alpha$ spectroscopic survey with $\Delta z = 0.1$ 
bins. This is consistent with their results to within a factor of a few, which 
is reasonable given the differences in survey specifications, fiducial 
magnification bias functions, and maximum $\ell$ used.

As a further test we then repeated this analysis, but restricted ourselves 
to only large-scale modes, $\ell \le 100$, and marginalized over the bias 
nuisance parameters as well. While one would likely fix the bias functions to 
their fiducial forms when attempting a first detection of the large-scale 
lensing effect, the bias uncertainties must be taken into account for 
precision measurements of $\ewl$ \citep[e.g. if used to test GR, as 
discussed in][]{2015arXiv150601369M}. We find a strong correlation between 
$\ewl$ and the magnification bias, especially at higher redshifts, which is to 
be expected given that both factors multiply the lensing term in the number 
count expression. Fig. \ref{fig:ewl_sbias} shows how the $\ewl$ constraint 
depends on the magnification bias prior for the various surveys; the 
spectroscopic and photometric surveys both require $\Delta s \lesssim 1$ to 
measure $\ewl$ to better than 100\%, and $\Delta s$ of order a few 
$\times 10^{-2}$ to 
reach their optimal constraints. Continuum surveys require a stronger prior 
of $\sim\! 0.1$ just to reach a 100\% constraint on $\ewl$ (for 1 and 5 $\mu$Jy 
flux limits), and are also subject to strong correlations with other bias 
parameters, as discussed in Section \ref{sec:continuum_fnl}. The continuum 
constraints shown in Fig. \ref{fig:ewl_sbias} were derived for fixed clustering 
bias, but if this is marginalized over (with a 10\% relative prior on $b$), 
$\sigma(\ewl)$ increases by a factor of $\sim 2$.

The fact that we are unable to detect $\epsilon_{\rm GR}$ from the 
single-tracer power spectrum should not at all lead us to give up hope of 
seeing the GR corrections, however. Indeed, this is just the first step in 
identifying the most effective observables for teasing out the ultra large-scale 
effects. In order to beat down the cosmic variance that is a fundamental 
barrier to single-tracer detectability, one must use multi-tracer techniques.
In some sense these divide out the stochastic part of the perturbation field, 
avoiding the effects of cosmic variance for certain (non-stochastic) 
quantities. By cross-correlating different tracers, with different bias 
functions, it is possible to isolate a number of terms from the scale 
dependence due to non-Gaussianity, relativistic effects, and growth of 
structure, in such a way as to obtain much tighter constraints than those from 
the overall power spectrum alone. It has been shown that multi-tracer 
techniques applied to continuum surveys can lead to almost an order of 
magnitude improvement in the detection of $f_{\rm NL}$, for example. We expect 
that the GR effects will also be detectable via this approach \citep{Yoo:2012se}.
A systematic analysis of multi-tracer techniques in the surveys described here 
is, therefore, an obvious next step in trying to identify robust methods for 
measuring ultra large-scale effects.

\emph{Note added ---} While this paper was being finalised, 
\cite{Raccanelli:2015vla} appeared, which discusses some related topics.

%-------------------------------------------------------------------------------
\section*{Acknowledgements}
\label{sec:acknowledgements}

We would like to thank Rachel Bean, Elisa Chisari, Enea Di Dio, Ruth Durrer, 
Matt Jarvis, Lance Miller, Francesco Montanari, and Eva-Maria M\"uller for 
their very valuable input on this paper. DA is supported by ERC grant 259505. 
PB is supported by ERC grant StG2010-257080. PGF acknowledges support from
STFC, BIPAC and the Oxford Martin School. RM and MS are supported by the 
South African Square Kilometre Array Project and the South African National 
Research Foundation. RM is also supported by the UK Science \& Technology 
Facilities Council, Grant No. ST/K0090X/1.

%-------------------------------------------------------------------------------
\appendix

%-------------------------------------------------------------------------------
\section{Modifications to CLASS}
\label{app:CLASS}

In order to compute the power spectra used for the Fisher forecasts, we used 
the public code {\tt CLASS} \citep{2011arXiv1104.2932L}. In its current 
version, {\tt CLASS} encompasses the extension {\tt CLASSgal} 
\citep{2013JCAP...11..044D}, 
which can be used to compute the transfer functions $\Delta^i$ in 
Eqs.~(\ref{eq:terms0}-\ref{eq:isw}). The public version of the code is easy to 
install, run, and modify, and we encourage its use; however, a number of 
modifications had to be implemented in order to make it usable for our work. 
We have made our modified version of the code publicly available at
\url{http://intensitymapping.physics.ox.ac.uk/codes.html}, and document the 
main changes here:
\begin{itemize}
 \item The terms $\Delta^{\rm L}$, $\Delta^{\rm P4}$ and $\Delta^{\rm ISW}$ in 
 Eqs.~(\ref{eq:terms0}-\ref{eq:isw}) can be computationally very demanding to 
 compute. The main reason for this is the wide redshift range covered by their 
 window functions -- these terms correspond to integrated effects along the 
 photon trajectory from the source to the observer. In the current version of 
 {\tt CLASS}, including these terms in the calculation of $C_\ell^{ij}$ for a 
 single redshift bin at $z=0.5$ takes about 10 minutes running on a modern 
 4-core computer. This makes including these terms for the large number of bins 
 used in this project (e.g. 100 bins for intensity mapping) prohibitively 
 expensive, so we invested some time in speeding up the calculation of these 
 terms.
 
 Two modifications were implemented. First of all, the code was parallelized 
 for distributed memory machines using MPI, so that each node computes the 
 transfer functions of a different set of redshift bins. Secondly, some 
 redundant calculations were circumvented by precomputing the window functions
 $W^{\rm L}(\eta)$, $W^{\rm P4}(\eta)$ and $W^{\rm ISW}(\eta)$ in 
 Eq.~(\ref{eq:windows}) and storing them in memory. This last modification speeds 
 up the computation of the integrated terms by a factor of $\sim4$, although it 
 requires more memory.
 
 \item We implemented the effect of primordial non-Gaussianity by including the 
 scale-dependent contribution to the bias in Eq.~(\ref{eq:db_fnl}).
 
 \item We modified the I/O system for the bias parameters $b(z)$, $s(z)$ and 
 $\fevo(z)$. These can now be supplied as tabulated $z$-dependent functions. 
 Furthermore, $\fevo$ must now be provided separately from the redshift 
 distribution, $\bar{N}(z)$.
 
 \item We implemented the possibility of adding an extra parameter for each 
 redshift bin that corresponds to the photo-$z$ uncertainty, so that the window 
 function in each bin can be computed as in Eq.~(\ref{eq:window_photoz}). In 
 doing this, we also modified the I/O system for defining the redshift bins. 
 The bin properties must now be supplied in different columns in a separate 
 text file. We believe this system is better suited for a large number of 
 redshift bins.
 
 \item Finally, we included the extra parameter $\egr$, used in this paper to 
 parametrize the amplitude of the relativistic corrections. This is not a 
 general-purpose modification.
\end{itemize}

%-------------------------------------------------------------------------------
\section{Survey specifications}\label{app:specs}

In this Appendix we provide the detailed specifications for all four of our
reference surveys. The codes used to generate the fiducial redshift
distributions and bias functions have been made available
online.\footnote{\url{http://intensitymapping.physics.ox.ac.uk/codes.html}}

%-------------------------------------------------------------------------------
\subsection{Intensity mapping}
\subsubsection{Noise model and redshift binning}

For an IM autocorrelation experiment, the simplest case is to assume that the 
noise is uncorrelated between different frequency channels, and has a white 
noise power spectrum:
\begin{equation}
  N^{ij}_\ell=\delta_{ij}\sigma_{\rm sr}^2.
\end{equation}
Here $\sigma_{\rm sr}^2$ is the noise variance per steradian, and can be 
calculated as follows: the noise per-pointing can be estimated as the rms 
temperature fluctuation of the system, $T_{\rm sys}$, scaled by the number of 
independent samples measured (given by $\delta\nu\,t_{\rm p}$, where 
$\delta\nu$ is the frequency channel width and $t_{\rm p}$ is the integration 
time per pointing). $t_{\rm p}$ can be approximated by $t_{\rm tot}\,
\Delta\Omega/(4\pi f_{\rm sky})$, where $f_{\rm sky}$ is the surveyed fraction 
of the sky, $\Delta\Omega$ is the solid angle covered in each pointing, and 
$t_{\rm tot}$ is the total survey time. Finally, scaling this by the total 
number of dishes in the experiment, we obtain the power spectrum
\begin{equation}
 N^{ij}_\ell=\delta_{ij}\frac{T_{\rm sys}^2(\nu_i)\,4\pi\,f_{\rm sky}}
 {\delta\nu\,t_{\rm tot}\,N_{\rm dish}}.
\end{equation}
Note that the variance per pointing has been multiplied by $\Delta\Omega$ to 
obtain the variance per steradian, which cancels the dependence on 
$\Delta\Omega$. The system temperature receives two contributions, 
$T_{\rm sys}=T_{\rm sky}+T_{\rm inst}$, due to atmospheric and background radio 
emission ($T_{\rm sky}\simeq 60\, {\rm K} \times (\nu/300\,{\rm MHz})^{-2.5}$) 
and instrumental noise ($T_{\rm inst}$).

Finally, it is worth noting that two different conventions have been adopted in 
the literature regarding the effect of the beam that defines the angular 
resolution of the experiment. The difference is in interpreting the beam as 
smoothing the signal on scales beyond the resolution, or as enhancing the noise 
at those same scales. We use the former, so that the model for the total 
observed power spectrum is:
\begin{equation}
 C^{ij}_\ell=C^{S,ij}_\ell\,B_\ell^i\,B_\ell^j+N^{ij}_\ell,
\end{equation}
where $B^i_\ell$ is the harmonic transform of the instrumental beam in the 
$i-$th frequency bin. We have assumed that the beams are Gaussian, 
$B^i_\ell=\exp(-\ell(\ell+1)\theta_B^2/2)$, where $\theta_B$ is related to the 
beam FWHM through $\theta_{\rm FWHM}=2\sqrt{2\ln2}\,\theta_B$. The beam width 
can be related to the dish diameter approximately as 
$\theta_{\rm FWHM}\simeq c/(\nu\,D_{\rm dish})$.

In our forecasts for SKA1-MID, we used the instrumental parameters 
$T_{\rm inst}=25\,K$, $f_{\rm sky}=0.75$, $t_{\rm total}=10^4\,{\rm h}$, 
$D_{\rm dish}=15\,{\rm m}$, and $N_{\rm dish}=254$, and assumed a minimum 
frequency of $350\,{\rm MHz}$, corresponding to $z_{\rm max}\simeq3$.

Modern radio receivers have very high frequency resolution (e.g. 
$\delta\nu\sim0.1\,{\rm MHz}$), so HI intensity mapping experiments should be 
able to resolve radial structures on scales much smaller than those relevant 
for cosmology. We are therefore free to choose the width and 
shape of the redshift bins used for the cosmological analysis. In order to 
avoid inhomogeneous coverage of radial scales, we divide the total frequency 
band into frequency bins of varying width $\Delta\nu(\nu)$ such that the 
corresponding comoving size $\Delta\chi$ is held constant. We estimated the 
minimum number of frequency bins needed for the constraints on $\fnl$ and 
$\egr$ to converge, finding that at least 100 bins were necessary. This 
corresponds to a radial width of $\Delta\chi\simeq44\,{\rm Mpc}/h$.

%-------------------------------------------------------------------------------
\subsubsection{Nuisance parameters and redshift evolution}

We model the clustering, magnification, and evolution biases for HI using an 
approach based on the halo model. We first assume that a one-to-one 
relationship exists between halo mass and HI mass, $M_{\rm HI}=M_{\rm HI}(M,z)$. 
The density and clustering bias can then be computed as
\begin{align}
 &\rho_{\rm HI}(z)=\int_{M_{\rm min}}^{M_{\rm max}}dM\,n(M,z)\,M_{\rm HI}(M,z),\\
 &b_{\rm HI}(z)=\int_{M_{\rm min}}^{M_{\rm max}}dM\,n(M,z)\,b(M,z)\,\frac{M_{\rm HI}(M,z)}
 {\rho_{\rm HI}(z)},
\end{align}
where $n(M,z)$ is the halo mass function (comoving number density per unit 
mass), and $b(M,z)$ is the halo bias. The background brightness temperature can 
then be computed in terms of $\rho_{\rm HI}$ using Eq.~(\ref{eq:im_bg}).

As described in Section \ref{sssec:imap_th}, the transverse distance perturbations cancel 
out for intensity mapping, so that $s_{\rm HI}(z)=2/5$ exactly. Furthermore, 
since we observe the emission from all of the HI in each patch of the sky, the 
evolution bias can be computed directly by differentiating $\rho_{\rm HI}$ with 
respect to $z$,
\begin{equation}
 \fevo=-\frac{d\,\log[\rho_{\rm HI}(z)]}{d\,\log(1+z)}.
\end{equation}
All that remains is to specify the function $M_{\rm HI}(M,z)$. As in
\cite{2015ApJ...803...21B}, we assume a power-law relation 
$M_{\rm HI}(M,z)\propto M^\alpha$ with an exponent $\alpha\simeq0.6$, and with 
the normalization set by constraints on $\Omega_{\rm HI}$ at $z=0.8$ from 
\cite{2013MNRAS.434L..46S}.

\subsection{Radio continuum surveys}
\subsubsection{Number counts, bias functions and noise model}
The models for the signal and noise power spectra are very simple compared with 
the other probes, as the galaxy sample is distributed in a single redshift bin, 
with a window function given by the redshift distribution,
\begin{equation}
 W(z)\propto \bar{N}(z).
\end{equation}
We estimate $\bar{N}(z)$ for the radio sources from empirical estimates of the 
luminosity functions of the main radio populations, since these also contain 
the necessary information to estimate the magnification and evolution biases. 
We consider four main radio galaxy types: star-forming galaxies (SF), 
starbursts (SB), radio-quiet quasars (RQQ), and Faranoff-Riley type I AGNs 
(FRI).\footnote{We also considered FRII galaxies, but their 
number density is so low that they contribute negligibly to the total number
counts.} The luminosity functions for each population were computed 
following the prescriptions of \cite{2008MNRAS.388.1335W}.

We will now outline the procedure used to calculate the redshift distribution, 
$s$, and $\fevo$ in all cases, and refer the reader to 
\cite{2008MNRAS.388.1335W} and the references \cite{2001ApJ...554..803Y} (SF 
and SB), \cite{2003ApJ...598..886U} (RQQ), and \cite{2001MNRAS.322..536W} (FRI) 
for details on the observations that the luminosity functions are based on.
These details are also summarized in Appendix \ref{app:lfradio}.
As with optical and IR surveys (see Sections \ref{sec:spectroscopic} and 
\ref{sec:photometric}), the $k$-correction to the flux measured in a given 
band is also needed in order to accurately estimate the observed number counts. 
This can be done for radio sources by assuming a particular SED for each 
population. For this, we again used the models from 
\cite{2008MNRAS.388.1335W}.

In its rest frame, a radio source has a luminosity per unit frequency given by
\begin{equation}
 L_{\nu}\equiv\frac{dE_e}{dt_ed\nu_e}=
 L_{\nu_*}\frac{\varphi(\nu)}{\varphi(\nu_*)},
\end{equation}
where $\nu_*$ is a pivot frequency and $\varphi(\nu)$ is the source SED. This 
is related to the flux per unit frequency measured by the observer by
\begin{equation}
 S_\nu\equiv\frac{dE_o}{dt_od\nu_odA_o}=
 \frac{L_{\nu(1+z)}}{4\pi\chi^2(z)(1+z)}(1-2\delta_\perp).
\end{equation}
Consider a radio survey in a given frequency band ($\nu\in[\nu_0,\nu_f]$). The 
average flux density measured for one source is defined as
\begin{equation}
 \bar{S}(\nu_0,\nu_f)=\int_{\nu_0}^{\nu_f}S_\nu\frac{d\nu}{\nu_f-\nu_0}.
\end{equation}
A given source will be detected if its average flux is above the detection limit 
$S_{\rm cut}$, and therefore all sources with a pivot luminosity above a 
minimum value $L_{\nu_*}^{\rm cut}\equiv\bar{L}_{\nu_*}^{{\rm cut}}(1+2\delta_\perp)$ 
will be included in the sample, where the average threshold luminosity is
\begin{equation}
 \bar{L}_{\nu_*}^{{\rm cut}}(z,S_{\rm cut})=
 \frac{4\pi S_{\rm cut}\chi^2(z)(1+z)\varphi(\nu_*)}
 {\bar{\varphi}(\nu_0(1+z),\nu_f(1+z))},
\end{equation}
and the average SED in the observed band is
\begin{equation}
 \bar{\varphi}(\nu_1,\nu_2)\equiv\int_{\nu_1}^{\nu_2}\varphi(\nu)\frac{d\nu}{\nu_2-\nu_1}.
\end{equation}

\begin{figure}[t]
 \centering
 \includegraphics[width=0.49\textwidth]{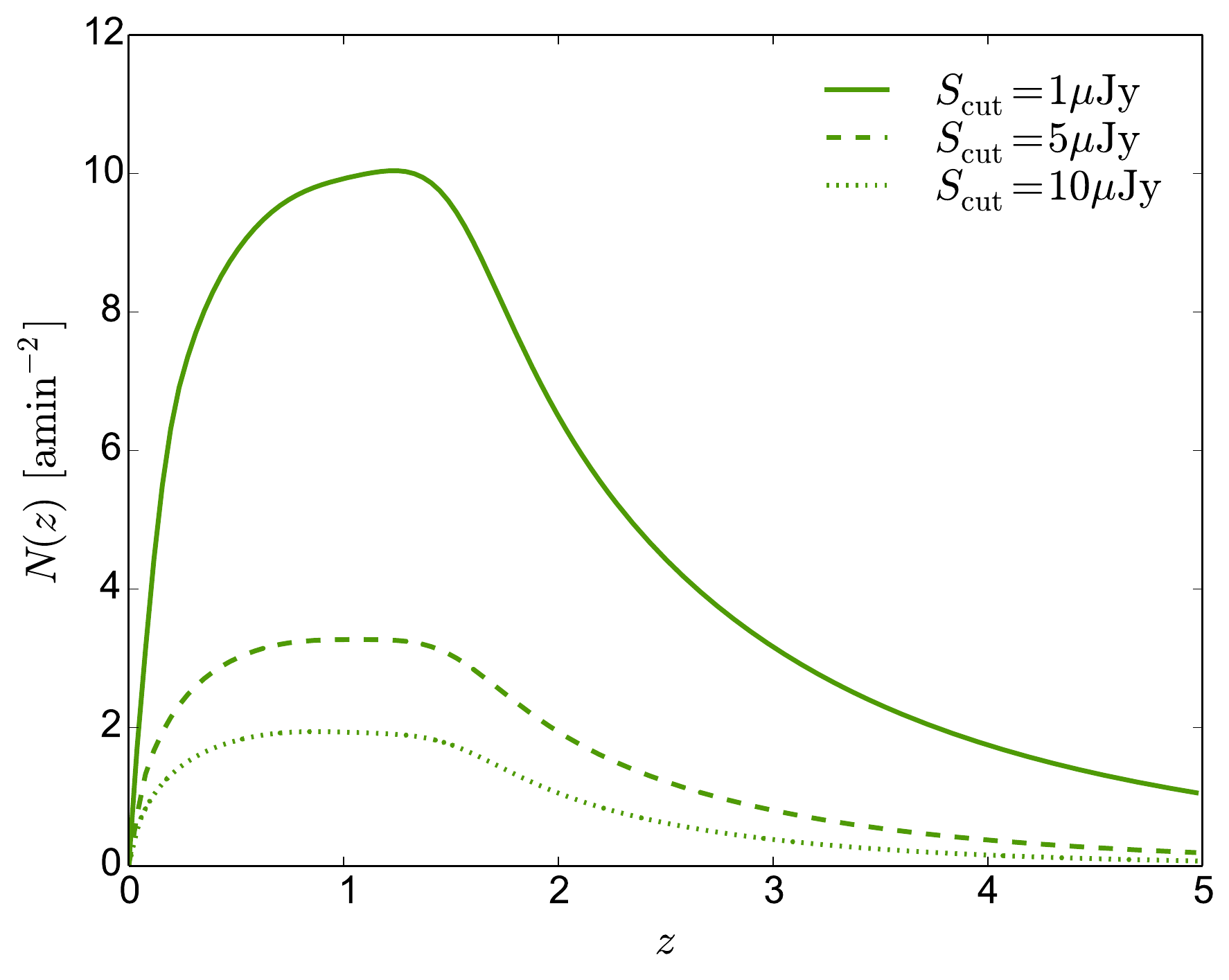}
 \caption{Angular number density of radio galaxies as a function of redshift 
          for the combined continuum sample considered here, and for different flux cuts.}
          \label{fig:nz_continuum}
\end{figure}

Given a pivot frequency $\nu_*$, the luminosity function 
$\bar{n}_s(z,\ln L_{\nu_*})$ at that frequency, and a characteristic SED 
$\varphi(\nu)$, the redshift distribution of sources can be computed as
\begin{equation}
  \bar{N}(z)=\frac{c\,\chi^2(z)}{(1+z)^3H(z)}
  \bar{\mathcal{N}}(z,>\bar{L}^{\rm cut}_{\nu_*}),
\end{equation}
where $\bar{\mathcal{N}}$ is defined as in Eq.~(\ref{eq:def_cumlum}). $s(z)$ and 
$\fevo(z)$ are then calculated from $\bar{\mathcal{N}}$ using 
Eqs.~(\ref{eq:fevo_def}) and (\ref{eq:s_def}).

For the clustering bias, we follow the same approach used in 
\cite{2008MNRAS.388.1335W}, and assign a fixed halo mass to each population. 
The corresponding bias is then found as the halo-model bias for that mass as a 
function of redshift. For this we parametrize the halo bias as in
\cite{1999MNRAS.308..119S}. We have also explored the approach followed in
\cite{2014MNRAS.442.2511F}, where each population is given a distribution of 
halo masses rather than a fixed one, and the bias is found by averaging over 
that distribution. No significant differences were found between the 
approaches, so we use the first, simpler one.

Once the redshift distribution and bias functions have been calculated for each 
population, we compute them for the combined sample as a weighed average of the 
individual ones,
\begin{align}
& \bar{N}_{\rm tot}(z)=\sum_a \bar{N}_a(z),\\
& (b,s,f_{\rm evo})_{\rm tot}(z)=\sum_a (b,s,f_{\rm evo})_a(z)\,
\frac{\bar{N}_a(z)}{\bar{N}_{\rm tot}(z)},
\end{align}
where $a$ labels the population. Separating the different populations is a very 
costly observational task. For our purposes, the main benefit of doing this is 
to allow the use of the multi-tracer technique to circumvent cosmic variance. 
Since we have postponed the multi-tracer analysis for future work, we will only 
report our forecasts here for the combined sample of radio sources. Figure 
\ref{fig:nz_continuum} shows the redshift distribution for the combined sample 
for the three different detection limits considered here. The luminosity 
functions, redshift distributions, and bias can be obtained using a code that 
we have made publicly available.\footnote{\url{http://intensitymapping.physics.ox.ac.uk/codes.html}} 
This provides an easy way to obtain number counts and power spectra without 
needing to query the full simulation of \cite{2008MNRAS.388.1335W}. Our 
results are consistent with the simulation except for the total number counts 
of star-forming galaxies, where our numbers are a factor of $2.5\times$ higher. 
This is consistent with what is described in \cite{2015arXiv150103825J}, and 
references therein, however.

The most relevant source of noise in a clustering analysis of discrete 
sources is shot noise, determined by the number density of sources. The angular 
number density of radio sources in the sample, $\bar{N}_\Omega$, is determined 
by integrating the redshift distribution
\begin{equation}
 \bar{N}_\Omega\equiv\int_0^\infty dz\, \bar{N}(z),
\end{equation}
and the noise power spectrum is given by
\be
N_\ell=\bar{N}_\Omega^{-1}.
\ee

\subsection{Spectroscopic redshift surveys}
\subsubsection{Noise and signal model}
Spectroscopic galaxy surveys measure Eq.~(\ref{eq:total_snc}), the perturbation 
to flux-limited number counts 
$\Delta_{\rm N}(z, \mathbf{\hat{n}}, >\log F_{\rm cut})$. This depends on the 
clustering, magnification, and evolution bias functions for the source 
population. The latter two can be derived from the background luminosity 
function of the sources, $n_s(z,\log L)$, given the flux limit and efficiency 
of the survey. Number counts are subject to a Poisson shot noise term that 
depends on the number density of sources, which can also be obtained from the 
luminosity function.

\begin{figure}[t]
 \centering
 \includegraphics[width=0.49\textwidth]{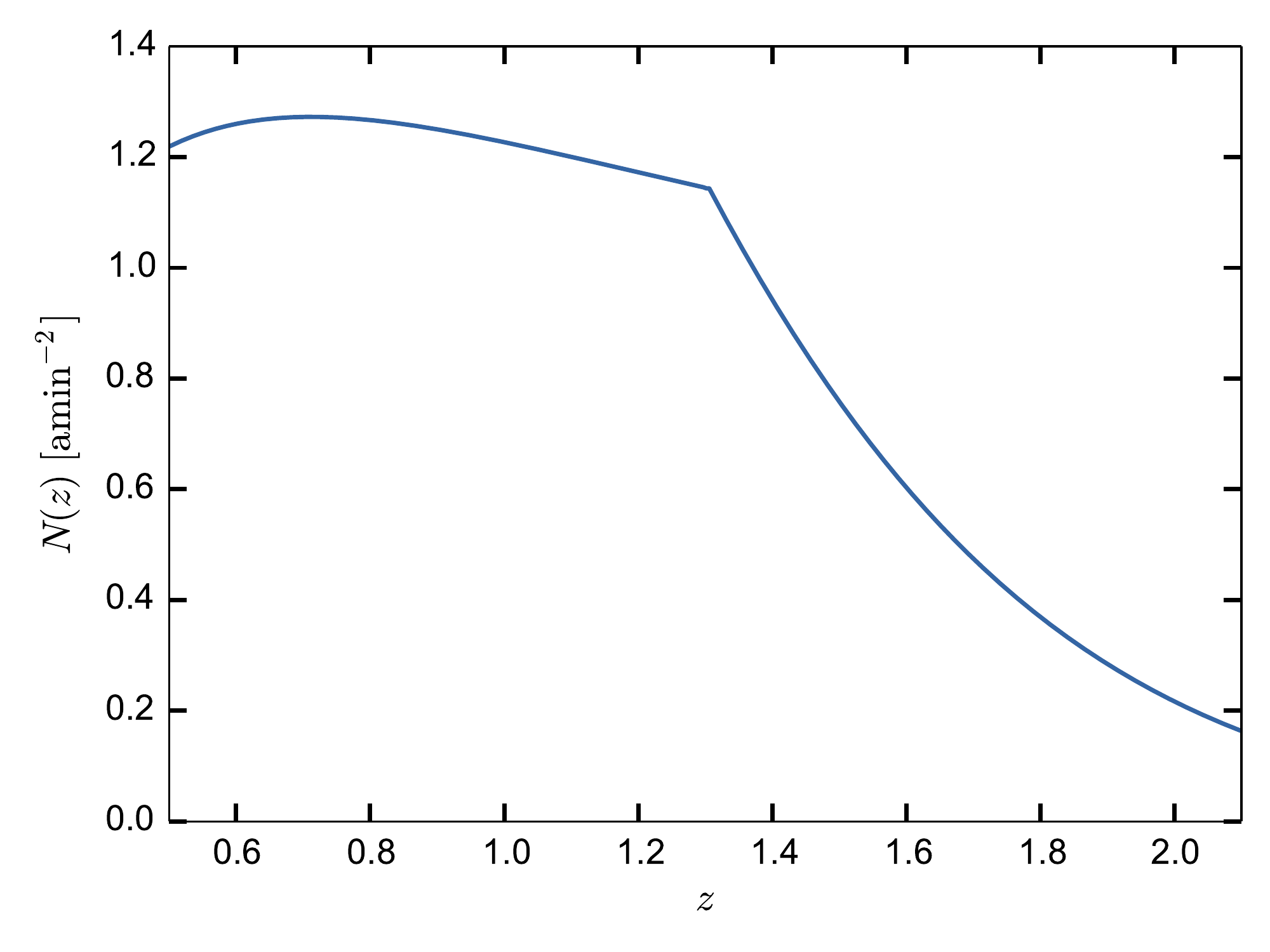}
 \caption{Predicted number density of galaxies as a function of redshift for 
 the H$\alpha$ spectroscopic survey.}
 \label{fig:spectro_nz}
\end{figure}

We follow the current set of public specifications for a realistic H$\alpha$ 
spectroscopic survey, described in \cite{2013LRR....16....6A}. The number 
counts presented in Table 1.3 of that paper can be approximately reproduced by 
using the best-fit H$\alpha$ Schechter luminosity function found by 
\cite{2010MNRAS.402.1330G},
\bea
n(z, L>L_{\rm cut}) = \int^\infty_{x_{\rm cut}} \epsilon\, \phi^* \, x^\alpha e^{-x}\, dx; ~~~ x
\equiv L / L^*(z), \nonumber\\
L^*(z) = \left\{
            \begin{array}{ll}
              5.1 \times 10^{41} (1 + z)^{3.1} \, & ~~(z < 1.3) \\
              6.8 \times 10^{42} \, & ~~(1.3 < z < 2.2)
            \end{array}
          \right. ~ {\rm erg/s},\nonumber
\eea
where $L$ is the H$\alpha$ line luminosity and we have assumed a flux limit of 
$F_{\rm cut} = 3 \times 10^{-16}$ erg/s/cm$^2$, an efficiency of 
$\epsilon = 0.45$, a faint-end slope of $\alpha = -1.35$, and comoving number 
density normalisation $\phi^* = 1.37 \times 10^{-3}$ Mpc$^{-3}$. The predicted 
galaxy number density as a function of redshift is shown in 
Fig.~\ref{fig:spectro_nz}. There are significant uncertainties in this model, 
which we account for by marginalizing over a set of bias function nuisance 
parameters, as explained in Sect. \ref{sec:fisher}. Finally, for the clustering 
bias we use the simplified prescription from \cite{2013LRR....16....6A}, 
$b(z) = \sqrt{1+z}$, which we also subject to the nuisance parametrisation.

The assumed bias functions for the H$\alpha$ survey were shown in 
Fig.~\ref{fig:bias}. The evolution bias is large and negative, and grows more 
negative with redshift until $z=1.3$, where there is a discontinuity in the 
$L^*(z)$ model, beyond which $L^*(z) = {\rm const.}$ and so $f_{\rm evo} = 0$. 
This is not particularly realistic, but is the best that can be done until 
updated constraints on the high-redshift H$\alpha$ luminosity function become 
available. The magnification bias grows rapidly with redshift, and deviates 
significantly from $s=0.4$ over most of the range, meaning that there is little 
or no cancellation of the transverse scale perturbation as there was in the
intensity mapping survey (although this is contingent on the uncertain 
behaviour of the luminosity function model for $z>1.3$). The clustering bias 
is relatively close to unity across the entire redshift range, slowly evolving 
from a minimum of $b \approx 1.3$ at $z=0.65$ to a maximum of $b \approx 1.8$ 
at $z \approx 2$.

We consider three constant-width redshift binnings for the selection function 
over the interval $0.65 \le z \le 2.05$: $\Delta z = (0.025, 0.05, 0.10)$, 
yielding (56, 28, 14) bins respectively. The target redshift uncertainty for 
(e.g.) Euclid is $\sigma_z \le 0.001\, (1+z)$ \citep{2011arXiv1110.3193L}, 
which is always significantly smaller than the narrowest bin width. As such, we 
assume a uniform (tophat) selection function, weighted by the source redshift
distribution, $\bar{N}(z)$.

Our noise model assumes that only shot noise is relevant, and that 
effects such as spectroscopy failures and point source masking have been taken 
into account in the survey efficiency, $\epsilon$. The shot noise angular power 
spectrum is
\be
\label{eq:shot_spec}
 N_\ell^{ij} = \delta_{ij} / n_i ; ~~~ n_i \equiv \int_{z_i}\bar{N}(z)\, dz.
\ee
with $n_i$ measured in units of steradians$^{-1}$. In our forecasts, we 
assumed that a wide range of multipoles can be recovered, $2 \le \ell \le 1000$, 
with no cuts at high or low $\ell$ due to systematics, non-linear effects and 
so on (this assumption was relaxed in Sect.~\ref{sec:spectro:systematics}).

\subsection{Photometric redshift surveys}

\subsubsection{Redshift distribution and bias parameters}

In order to compute $\bar{N}(z)$, $s(z)$, and $\fevo(z)$ for our two samples, 
we need an estimate of the luminosity function for both red and blue galaxies, 
preferably in the $r$-band, for which the LSST specifications are provided. We 
describe the method used for this task here.

For the red sample we follow a method similar to that used by 
\cite{2011A&A...527A..26J}. First, an estimate of the $B$-band luminosity 
function for red galaxies is obtained from \cite{2007ApJ...665..265F} as a 
Schechter function with constant slope $\alpha=-0.5$, and $z$-dependent $\phi_*$ 
and $M_*$, measured in a number of redshift bins in the interval $z\in(0.2,1.2)$. 
We extrapolate the luminosity function to higher/lower redshifts by fitting 
the values of these parameters, measured by \cite{2007ApJ...665..265F}, to the 
models:
\begin{align}
& M_*(z)=M_0+M_1\,z,\\
& \phi_*(z)=\frac{\phi_0}{1+(z/z_0)^a}\,[10^{-3}\,{\rm Mpc}^{-3}]
\end{align}
with $M_0=-20.6$, $M_1=-0.49$, $\phi_0=1.82$, $z_0=1.04$ and $a=7.17$. In order 
to translate this into an $r$-band luminosity function, we use $B-r=1.32$
\citep{1995PASP..107..945F} and assume that $B-r$ does not evolve significantly 
for the red sample in the redshift range under study.

For the full sample, we use the $r'$-band luminosity function found by 
\cite{2006A&A...448..101G}, and approximate $r'\simeq r$. This is again given 
as a Schechter function with constant slope $\alpha=-1.33$ and $z$-dependent 
$M_*$ and $\phi_*$, for which we have used the following parametrizations:
\begin{align}
& M_*(z)=M_0+a\,\ln(1+z),\\
& \phi_*(z)=(\phi_0+\phi_1\,z+\phi_2\,z^2)\,[10^{-3}\,{\rm Mpc}^{-3}]
\end{align}
with $M_0=-21.49$, $a=-1.25$, $\phi_0=2.59$, $\phi_1=-0.136$, and 
$\phi_2=-0.081$. The luminosity function for blue galaxies is then estimated as 
the difference between those of the full and red samples.

An absolute magnitude, $M$, measured in a given rest-frame band for a galaxy at
redshift $z$ is related to the apparent magnitude in the 
observer-frame band, $m$, by 
\begin{equation}\label{eq:apptoabs}
  M=m-25-5\,\log_{10}\left[\frac{d_L(z)}{1\,{\rm Mpc}\,h^{-1}}\right]+
  \log_{10} h-k(z),
\end{equation}
where $d_L(z)$ is the luminosity distance and $k(z)$ is the $k$-correction
corresponding to that galaxy's SED redshifted to $z$. We estimated $k(z)$ for
red and blue galaxies by running the code {\tt kcorrect} 
\citep{2007AJ....133..734B} on the
spectra of an elliptical galaxy and a barred spiral galaxy (Sbc) respectively, 
as measured by \cite{1980ApJS...43..393C}. We approximate and extrapolate these
$k$-corrections as $k_{\rm red}(z)\sim2.5\,z$ and $k_{\rm blue}\sim1.5\,z$. We 
verified that these parametrizations are compatible with the $k$-corrections 
shown in \cite{1995PASP..107..945F} for both types, and also that the final 
redshift distributions did not vary significantly when other functional forms 
are used. For a given magnitude limit $m_{\rm cut}$, we use Eq.~(\ref{eq:apptoabs})
to compute the corresponding luminosity cut $\bar{L}_{\rm cut}$ as a function of
redshift. The redshift distribution, magnification bias, and evolution bias
for each population are then estimated using Eqs.~(\ref{eq:nz_def}), 
(\ref{eq:s_def}), and (\ref{eq:fevo_def}) respectively, with the luminosity 
functions described above.

Regarding the clustering bias, for the full sample we use the parametrization
$b_{\rm full}(z)=1+0.84\,z$, based on the simulations of \cite{2004ApJ...601....1W}
and quoted in the LSST science book \citep{2009arXiv0912.0201L}. Red galaxies 
should have a larger bias, which we parametrize as $b_{\rm red}(z)\simeq1+z$. 
This parametrization is compatible with bias measurements at redshifts $z<1$ 
\citep[e.g.][]{2008ApJ...672..153C}. Since the red population dies off at 
$z\sim1.4$, this extrapolation should not significantly influence our final 
result, especially as we ultimately marginalize over $b(z)$.

The redshift distributions for the two samples considered here are shown in 
Figure \ref{fig:nz_LSST}. For this figure, as well as in our forecasts, we
assume a magnitude limit of $i=25.3$, corresponding to $r\sim26$ 
for typical galaxy colours, as quoted in \cite{2009arXiv0912.0201L}. According 
to the models used here, the total number of galaxies 
observed by LSST should be $\sim$40 per arcmin$^2$, in qualitative agreement 
with previous results \citep{2006A&A...457..841I,2009arXiv0912.0201L}.

\subsubsection{Photometric redshifts, binning and shot noise}

The quality of a given photo-$z$ algorithm is normally quoted in terms 
of its rms error, $\sigma_z^2\equiv\langle(z_{\rm photo}-z_{\rm true})^2\rangle$,
which is typically parametrized as
\begin{equation}
 \sigma_z(z)=\sigma_0\,(1+z).
\end{equation}
The photometric redshift requirement for the LSST gold sample, as quoted in
\cite{2009arXiv0912.0201L}, is $\sigma_0<0.05$, with a goal of 0.02. As 
described above, since the spectral properties of red galaxies make their 
photometric redshifts more accurate than those of the blue population, we have 
assumed the following photometric redshift uncertainties for the two samples:
\begin{equation}
\sigma_0^{\rm red}=0.02,\hspace{12pt}\sigma_0^{\rm full}=0.05.
\end{equation}

We have also assumed that the data will be analysed by dividing the sample into 
a number of photo-$z$ bins. Let $z_0^i$ and $z_f^i$ be the limits of the $i$-th
bin. The window function in this bin must trace the true $z$ distribution of
galaxies within it, and is therefore given by the product of the overall
redshift distribution and the photo-$z$ probability distribution, integrated
over the bin:
\begin{equation}\label{eq:window_photoz}
 W^i(z)\propto\bar{N}(z)\,w^i(z),
\end{equation}
\begin{equation}
 w^i(z)=\int_{z^i_0}^{z^i_f}dz_p\,p(z_p|z),
\end{equation}
where $p(z_p|z)$ is the photo-$z$ pdf. Assuming a Gaussian photo-$z$ 
distribution, we can write the window functions $w^i(z)$ analytically, as
\begin{equation}
 w^i(z)=\frac{1}{2}
 \left[{\rm erf}\left(\frac{z-z^i_0}{\sqrt{2}\sigma_z}\right)-
 {\rm erf}\left(\frac{z-z^i_f}{\sqrt{2}\sigma_z}\right)\right].
\end{equation}

\begin{figure*}[t]
 \centering
 \includegraphics[width=0.49\textwidth]{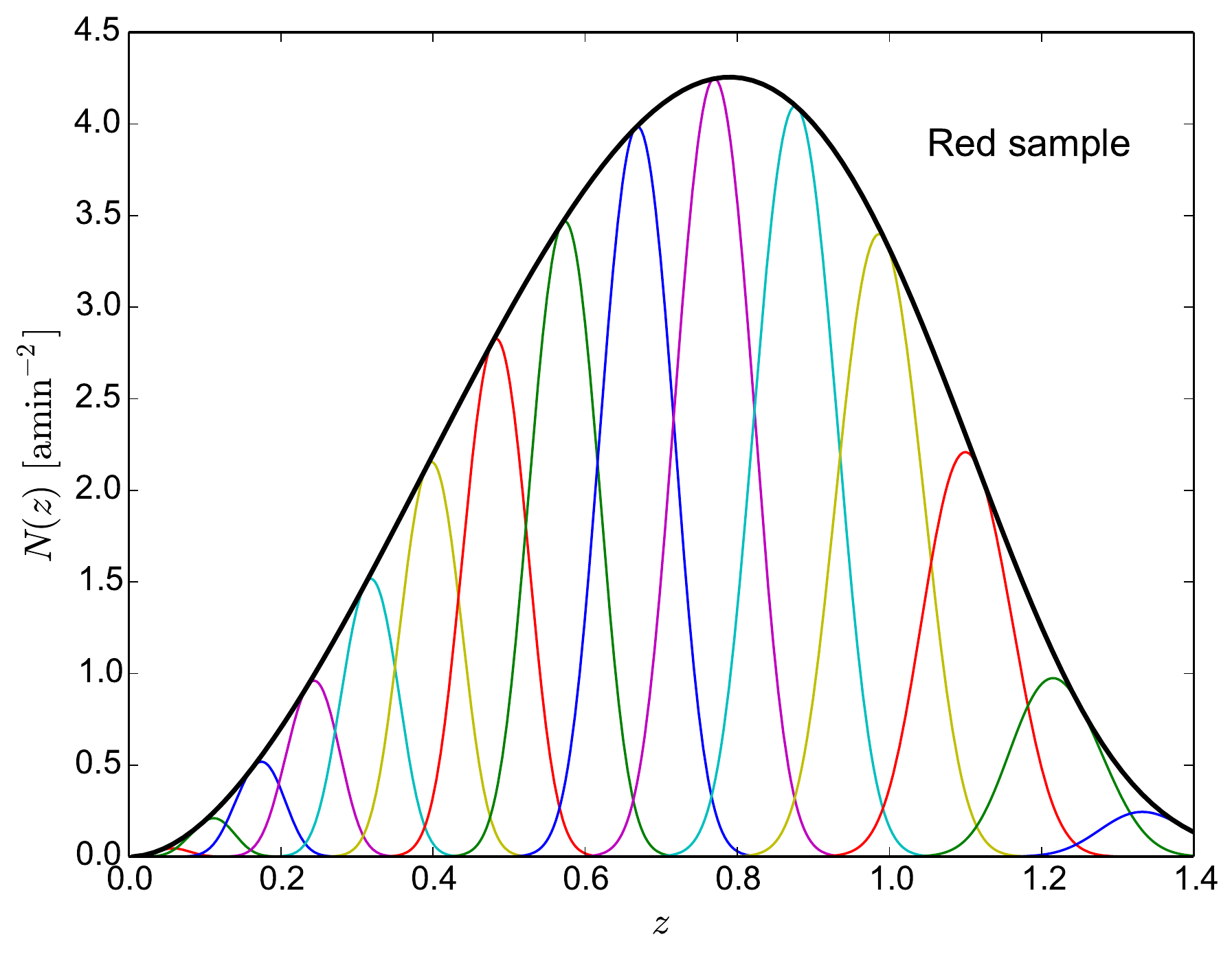}
 \includegraphics[width=0.49\textwidth]{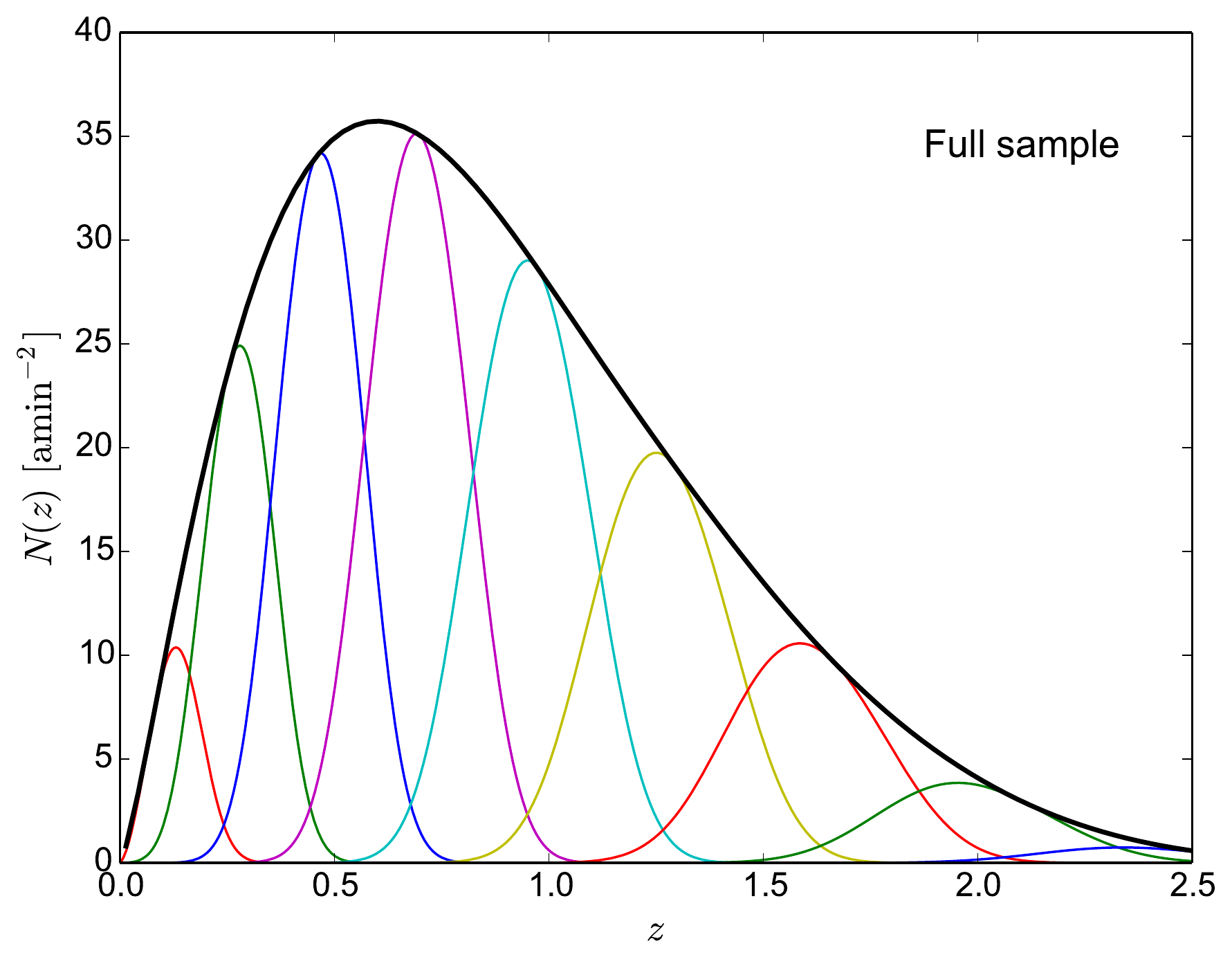}
 \caption{Angular number density of galaxies as a function of redshift for the
          LSST ``red'' (left plot) and ``full'' (right plot) samples. The 
          colored lines show the window functions of the different redshift
          bins considered here.}
          \label{fig:nz_LSST}\vspace{1.5em}
\end{figure*}

The tails of $w^i(z)$ correlate different redshift bins to a much larger degree
than the intrinsic correlations due to gravitational clustering do, which is
an expression of the loss of information on radial scales. In order to reduce
this correlation and avoid redundant calculations, the width of the redshift
bins are usually defined to be of the order of $\sigma_z$. In this work we
have chosen to define the width of our bins to be three times the photo-$z$
dispersion at the bin centre. For the values of $\sigma_z$ assumed here, this
results in 15 bins for the red sample and 9 bins for the full sample. The
window functions for the bins are shown in Figure \ref{fig:nz_LSST}.

Finally, as in the case of spectroscopic and continuum surveys, the main
source of statistical noise in the measurement of clustering anisotropies is
shot noise. The noise bias term for photometric surveys is thus given by
\begin{equation}
 N^{ij}_\ell=\frac{\delta_{ij}}{n^i\,n^j},
\end{equation}
where
\begin{align}
 &n^i\equiv\int_0^\infty dz\, \bar{N}(z)\,w^i(z),\\
\end{align}
Note that this reduces to Eq.~(\ref{eq:shot_spec}) in the case of top-hat windows (i.e.
$\sigma_z\rightarrow0$).

\section{Luminosity function of radio sources}
\label{app:lfradio}
We will summarize here the steps that were followed to compute the luminosity 
functions of the different radio sources that would be observable with a 
continuum survey. We follow almost exactly what was done by 
\cite{2008MNRAS.388.1335W} to simulate the distribution of radio galaxies, but 
we would like to describe the details of the calculation here for the benefit of
the potential users of the codes used in this paper.

We will discuss five main galaxy populations: star-forming galaxies, starbursts, 
radio-quiet AGNs and radio-loud AGNs of types FRI and FRII. We note that we use 
the convention for the luminosity function used in Section \ref{ssec:th_egr}, 
where $n_s(z,L)$ is the physical (not comoving) number density of objects per 
unit $\ln L$.

\subsection{Star-forming galaxies and starbursts}
The luminosity functions of normal star-forming galaxies and starbursts used 
here are based on the luminosity function derived from the IRAS 2 $\mu$Jy sample 
by \cite{2001ApJ...554..803Y} at $1.4\,{\rm GHz}$. This is modelled as a sum of 
two Schechter functions, which we identify with two populations: normal and 
starburst galaxies respectively. Thus, for both normal star-forming galaxies 
and starbursts, the $z=0$, $1.4\,{\rm GHz}$ luminosity function is parametrized 
as
\begin{equation}
 n_s(z=0,L)=n_*\,\left(\frac{L}{L_*}\right)^\alpha\exp\left[-\frac{L}{L_*}\right],
\end{equation}
where in each case:
\begin{align}
 &n_*=\begin{cases} 
      3.2\times10^{-4}\,{\rm Mpc}^{-3}\,\,\,\,\text{for normal galaxies} \\
      8.3\times10^{-6}\,{\rm Mpc}^{-3}\,\,\,\,\text{for starbursts} \\
   \end{cases}\\
 &L_*=\begin{cases} 
      2.10\times10^{22} \,[{\rm W/Hz/sr}]\,\,\,\,\text{for normal galaxies} \\
      1.44\times10^{23}\,[{\rm W/Hz/sr}]\,\,\,\,\text{for starbursts} \\
   \end{cases}
\end{align}
and $\alpha=-0.633$ in both cases. This parametrization is valid for luminosities
above $\log_{10}L/[{\rm W/Hz/sr}]=19.6\,$, and it is assumed to be constant for
lower values of $L$, in agreement with \cite{2007MNRAS.375..931M}.

In agreement with \cite{1993MNRAS.263..123R}, we assume a pure luminosity
evolution of this luminosity function,
\begin{equation}
  n_s(z,L)=(1+z)^3\,n_s(z=0,L\,f(z)),
\end{equation}
\begin{equation}
 f(z)=\begin{cases} 
      (1+z)^{3.1}  \,\,\,\,\,\,{\rm for}\,\,\,\,z<z_0\\
      (1+z_0)^{3.1}\,\,\,\,{\rm for}\,\,\,\,z\ge z_0
      \end{cases},
\end{equation}
with $z_0=1.5$. This parametrization was obtained assuming a cosmological model
$(\Omega_M,\Omega_\Lambda)=(1,0)$. We adapt it to our fiducial
cosmology by scaling the number densities and luminosities by the
ratios of comoving volumes and luminosity distancies in both models,
\begin{equation}\label{eq:coslf}
  n^{(1)}_s(z,L|\Omega_1)=
  n^{(2)}_s\left(z,L\,\frac{\chi_{(2)}^2(z)}{\chi_{(1)}^2(z)}\right)
  \frac{\chi_{(2)}^2(z)\,H_{(1)}(z)}{\chi_{(1)}^2(z)\,H_{(2)}(z)},
\end{equation}
where the indices $(1)$ and $(2)$ indicate quantities computed in
two different cosmological models.

The SED for both normal and starburst galaxies was assumed to be
\begin{equation}
 \varphi(\nu)\propto
   \nu_{\rm GHz}^2\,(1-e^{-\tau})\left(1+10\nu_{\rm GHz}^{-0.65}\right)
\end{equation}
with $\nu_{\rm GHz}\equiv\nu/(1\,{\rm GHz})$ and
$\tau=(\nu_*/\nu_{\rm GHz})^{2.1}$,
with $\nu_*=0.005$ for normal galaxies and $\nu_*=1$ for starbursts.
This corresponds to a combination of thermal free-free emission and
non-thermal synchrotron from supernovae. We did not include a dust
component in this SED, as mentioned in \cite{2008MNRAS.388.1335W},
which should be irrelevant for the range of redshifts and frequencies
studied here.

\subsection{Radio-loud AGNs}
The luminosity function for FRI and FRII radio-loud AGNs was based
on model ``C'' of the luminosity function at $151\,{\rm MHz}$ derived by
\cite{2001MNRAS.322..536W}. This luminosity function consists of
low-luminosity and high-luminosity components, which we identify with
FRI and FRII sources respectively.

For FRI galaxies, the $z=0$ luminosity function takes the form of
a Schechter function
\begin{equation}
 n_s(z=0,L)=\frac{n_*}{\ln10}\left(\frac{L}{L_*}\right)^\alpha
 \exp\left[-\frac{L}{L_*}\right],
\end{equation}
with $n_*=10^{-7.12}\,({\rm Mpc})^{-3}$,
$\log_{10}L_*/(1\,{\rm W/Hz/{\rm sr}})=26.1$ and $\alpha=-0.539$.
The model also assumes a pure density evolution,
\begin{equation}
 n_s(z,L)=(1+z)^3\,f(z)\,n_s(z=0,L)
\end{equation}
\begin{equation}
  f(z)=\begin{cases} 
      (1+z)^{4.3}  \,\,\,\,{\rm for}\,\,\,\,z<z_0\\
      (1+z_0)^{4.3}\,\,\,\,{\rm for}\,\,\,\,z\ge z_0
      \end{cases},
\end{equation}
where $z_0=0.706$.

For FRII galaxies, the luminosity function takes the form of a
Schechter function with an inverted exponential term,
\begin{equation}
 n_s(z=0,L)=\frac{n_*}{\ln10}\left(\frac{L}{L_*}\right)^\alpha
 \exp\left[-\frac{L_*}{L}\right], 
\end{equation}
with $n_*=10^{-6.196}\,({\rm Mpc})^{-3}$,
$\log_{10}L_*/(1\,{\rm W/Hz/{\rm sr}})=26.95$, and $\alpha=-2.27$.
As for FRI, a pure density evolution is assumed, with
\begin{equation}
 f(z)=\exp\left(-\frac{(z-z_0)^2}{2\,\sigma_*^2(z)}\right)
\end{equation}
\begin{equation}
  \sigma_*(z)=\begin{cases} 
      0.559 \,\,\,\,{\rm for}\,\,\,\,z<z_0\\
      1.378 \,\,\,\,{\rm for}\,\,\,\,z\ge z_0
      \end{cases},
\end{equation}
where $z_0=1.91$.
As in the case of star-forming galaxies, these luminosity functions
were derived for an Einstein-de Sitter background, so had to be
adapted to our fiducial cosmology using Eq.~(\ref{eq:coslf}).

A power-law SED with $\varphi(\nu)\propto\nu^{-0.75}$ was assumed
for both types of radio-loud AGN.

\subsection{Radio-quiet AGN}
Radio-quiet AGN make up the majority of the total AGN population, as is
observed from the hard X-ray luminosity function. This can be combined
with the relation between hard X-ray and radio luminosities
\citep{2000A&A...356..445B},
\begin{equation}
 \log_{10}(L_{2-10\,{\rm keV}}/({\rm erg}\,{\rm s}^{-1}))=
 1.012\,\log_{10}(L_{1.4\,{\rm GHz}}/({\rm W/Hz/sr}))+21.3,
\end{equation}
to derive the $1.4\,{\rm GHz}$ luminosity function.

For this we use the X-ray luminosity function of \cite{2003ApJ...598..886U}
in the $2-10\,{\rm keV}$ band, parametrized at $z=0$ as
\begin{equation}
 n_s(z=0,L_X)=\frac{A}{\ln10}\left[(L_X/L_*)^{\gamma_1}+
 (L_X/L_*)^{\gamma_2}\right]^{-1},
\end{equation}
where $L_X$ is the X-ray luminosity, $A=5.04\times10^{-6}\,{\rm Mpc}^{-3}$,
$\gamma_1=0.86$, $\gamma_2=2.23$, and
$\log_{10}(L_*/({\rm erg}\,{\rm s}^{-1}))=43.94$.

The evolution with redshift is parametrized as a luminosity-dependent
density evolution,
\begin{equation}
 n_s(z,L_X)=(1+z)^3\,f(z,L_X)\,n_s(z=0,L_X),
\end{equation}
with
\begin{equation}
 f(z,L_X)=\begin{cases} 
      (1+z)^{4.23}\,\,\,\,{\rm for}\,\,\,\,z<z_0\\
      (1+z_0)^{4.23}\,\left(\frac{1+z}{1+z_0}\right)^{-1.5}
      \,\,\,\,{\rm for}\,\,\,\,z\ge z_0,
      \end{cases},
\end{equation}
where $z_0$ is a luminosity-dependent function
\begin{equation}
 z_0(L_X)=\begin{cases} 
      1.9\,(L_X/L_1)^{0.335} \,\,\,\,{\rm for}\,\,\,\,L_X<L_1\\
      1.9 \,\,\,\,{\rm for}\,\,\,\,L_X\ge L_1,
      \end{cases}
\end{equation}
with $\log_{\rm 10}(L_1/({\rm erg}\,{\rm s}^{-1}))=44.6$.

A power-law SED with $\varphi(\nu)\propto\nu^{-0.7}$ was assumed for radio-quiet AGN.

%-------------------------------------------------------------------------------
\bibliography{bibliography}
\bibliographystyle{hapj}

\end{document}